\documentclass[11pt]{article}
\usepackage{harvard,a4,epsfig}
\usepackage[germanb,english]{babel}
\newtheorem{theorem}{Lemma}

\newcommand{\be}{\begin{equation}}
\newcommand{\ee}{\end{equation}}
\newcommand{\BE}{\begin{eqnarray}}
\newcommand{\EE}{\end{eqnarray}}
\newcommand{\forget}[1]{}
\newcommand{\LY}{Lieb and Yngvason}
\newcommand{\TD}{thermodynamics}
\newcommand{\TEA}{Ehren\-fest-Afanas\-sjewa}

\newcommand{\C}{Cara\-th\'eo\-do\-ry{}}
\newcommand{\statechange}[1]{\stackrel{\cal #1 }{\displaystyle
\longrightarrow}}
\newcommand{\statechangehat}[1]{\stackrel{\hat{\cal #1 }}{\displaystyle
\longrightarrow}}
\newcommand{\nprec}{%
\mbox{%
$\mbox{}\,\displaystyle\prec\hspace{-.75em} \displaystyle\not
\hspace{.75 em}
\,\mbox{}$}
}

\begin{document}
\newcommand{\dstreep}{d \mkern-5mu \rule[1.2 ex]{.7 ex}{.08ex}}
\title{%
Bluff your way in the Second Law of Thermodynamics
}
\author{%
Jos Uffink\\Department of History and
Foundations of Science\\
Utrecht University, P.O.Box 80.000, 3508 TA Utrecht, The
Netherlands\\
e-mail:  uffink@phys.uu.nl }
\maketitle

\begin{abstract}
 The aim of this article is to analyse the relation between the second law
of thermodynamics and the so-called arrow of time. For this purpose, a
number of different aspects in this arrow of time are distinguished, in
particular those of time-(a)symmetry and of (ir)reversibility. Next I
review versions of the second law in the work of Carnot, Clausius, Kelvin,
Planck, Gibbs, Carath\'eodory and Lieb and Yngvason, and investigate their
connection with these aspects of the arrow of time.
  It is shown that this connection varies a great deal along with these
formulations of the second law.  According to the famous formulation by
Planck, the second law expresses the irreversibility of natural processes.
But in many other formulations irreversibility or even time-asymmetry
plays no role.
 I therefore argue for the view that the second law has nothing to do
with the arrow of time.
  \end{abstract}
 \section{Introduction}

There is a famous lecture by the British physicist/novelist C.~P.
Snow about the cultural abyss between two types of intellectuals:
those who have been educated in literary arts and  those in  the
exact sciences. This lecture, \emph{the Two Cultures} (1959),
characterises the lack of
 mutual respect between them in a passage:
 \begin{quote}\small
A good many times I have been present at gatherings of
people who, by the standards of the traditional culture, are
thought highly educated and who have with considerable gusto
been expressing their incredulity at the illiteracy of
scientists. Once or twice I have been provoked and have asked
the company how many of them could describe the Second Law of
Thermodynamics. The response was cold: it was also negative.
Yet I was asking something which is about the equivalent of:
{\em have you read a work of
Shakespeare\/}?\nocite{Snow}\end{quote}
 Snow stands up for the view that exact science is, in its own right, an
essential part of civilisation, and should not merely be valued
for its technological applications. Anyone who does not know the
Second Law of Thermodynamics, and is proud of it too, exposes
oneself as a Philistine.

Snow's plea will strike a chord with every physicist who has ever attended
a birthday party.  But his call for cultural recognition creates
obligations too. Before one can claim that acquaintance
with the Second Law is  as indispensable to a cultural education as
{\em Macbeth\/} or {\em Hamlet}, it should obviously be clear what this
law states.  This question is surprisingly difficult.

The Second Law made its appearance in physics around 1850, but a half
century later it was already surrounded by so much confusion that the
{\em British Association for the Advancement of Science\/} decided to
appoint a special committee with the task of providing clarity about the
meaning of this law. However, its final report~\cite{Larmor}
did not settle the issue. Half a century later, the physicist/philosopher
Bridgman still complained that there are almost as many formulations of
the second law as there have been discussions of
it~\cite[p.~116]{Bridgman}.
And even today, the Second Law remains so obscure that it
continues to attract new efforts at clarification. A recent
example is the work of \citeasnoun{LY}.

This manifest inability of the physical community to reach consensus about
the formulation and meaning of a respectable physical law is truly
remarkable. If Snow's question had been: `Can you describe the Second Law
of Newtonian Mechanics?' physicists would not have any problem in
producing a unanimous answer.  The idea of installing a committee for this
purpose would be just ridiculous.

A common and preliminary description of the Second Law is that it
guarantees that all physical systems in thermal equilibrium can be
characterized by a quantity called entropy, and that this entropy cannot
decrease in any process in which the system remains adiabatically
isolated, i.e.\ shielded from heat exchange with its environment.
 But the law has many faces and interpretations;  the comparison to a work
of Shakespeare is, in this respect, not inappropriate.\footnote{%
      Actually, in the second edition of \emph{The Two Cultures},
      Snow expressed regret for comparing the Second Law to a work of
      Shakespeare, due to the formidable conceptual problems connected
      with the former.}
 One of the most frequently discussed aspects of the
Second Law is its relation with the `{arrow of time}'.
 In fact, in many texts in philosophy of physics the Second Law figures as
an emblem of this arrow.
  The idea is, roughly, that typical thermodynamical processes are
irreversible, i.e.\ they can only occur in one sense only, and that this
is relevant for the distinction between past and future.

At first sight, the Second Law is indeed  relevant for this arrow.  If
the entropy can only increase during a thermodynamical
process, then obviously, a reversal of this process is not possible.
 Many authors believe this is a crucial feature, if not the very
essence of the Second Law.  Planck, for example, claimed that, were it not
for the existence of irreversible processes, `the entire edifice of the
second law would crumble [\ldots] and theoretical work would have to start
from the beginning.' \cite[ \S 113]{Planck}, and  viewed entropy increase
as a `universal measure of irreversibility' (ibid.~\S 134). A similar
view
is expressed by Sklar in his recent book on the foundations of statistical
mechanics (1993, p.~21): `The crucial fact needed to justify the
introduction of [\ldots] a definite entropy value is the irreversibility of
physical processes.' \nocite{Sklar}

 In this respect, thermodynamics seems to stand in sharp contrast with the
rest of classical physics, in particular with mechanics which, at least in
Hamilton's formulation, is symmetric under time reversal.  The problem of
reconciling this thermodynamical arrow of time with a mechanical world
picture is usually seen as the most profound problem in the foundations of
thermal and statistical physics; see~\citeasnoun{Davies},
\citeasnoun{Mackey}, \citeasnoun{Zeh}, \citeasnoun{Sklar} and
\citeasnoun{Price}.

However, this is only one of many problems awaiting a student of the
Second Law. There are also authors expressing the opposite viewpoint.
Bridgman writes:
 \begin{quote}\small
 It is almost always emphasized that thermodynamics is
   concerned with reversible processes and equilibrium states and that it
   can have nothing to do with irreversible processes or systems out of
   equilibrium \ldots \cite[p.~133]{Bridgman} \label{Br}
 \end{quote}
 It is not easy to square this view, ---and the fact that Bridgman
presents
it as prevailing among thermodynamicists--- with the idea that
irreversibility is essential to the Second Law.

Indeed, one can find other authors  maintaining that the
Second Law has little to do with irreversibility or the arrow of time;
 in particular  Ehrenfest-Afanassjewa,
(1925, 1956, 1959),\nocite{TEA25,TEA56,TEA59}
\nocite{Landsberg56,Jauch}
Landsberg (1956) and Jauch (1972, 1975).  For them,
the conflict between the irreversibility of thermodynamics and the
reversible character of the rest of physics is merely illusory, due
to a careless confusion of the meaning of terms.  For example, Landsberg
remarks that the meaning of the term `reversible' in thermodynamics
has nothing to do with the meaning of this term in classical mechanics.
However, a fundamental and consistent discussion of the meaning of these
concepts is rare.

Another problem is that there are indeed many aspects and
formulations of the Second Law, which differ more or less from the
preliminary circumscription offered above. For example, consider
the so-called `approach to equilibrium'.  It is a basic assumption
of thermodynamics that all systems which are left to themselves,
i.e.\ isolated from all external influences, eventually evolve
towards a state of equilibrium, where no further changes occur.
One often regards this behaviour as a consequence of the Second
Law. This view is also suggested by the well-known fact that
equilibrium states can be characterised by an entropy maximum.
 However, this view is problematic. In thermodynamics,
entropy is not defined for arbitrary states out of equilibrium.  So how
can the assumption that such states evolve towards equilibrium be a
consequence of this law? \forget{Thus, although the approach to
equilibrium does provide a clear-cut thermodynamical arrow of time, its
relation to the Second Law itself remains unclear.  }

 Even deliberate attempts at careful formulation of the Second Law
sometimes end up in a paradox.
 One sometimes finds a formulation which admits that thermodynamics aims
only at the description of systems in equilibrium states, and that,
strictly speaking, a system does not always have an entropy during a
process.  The Second Law, in this view, refers to processes of an isolated
system that begin and end in equilibrium states and says that the entropy
of the final state is never less than that of the initial
state~\cite[p.~381]{Sklar74}.  The problem is here that, by definition,
states of equilibrium remain unchanged in the course of time, unless the
system is acted upon. Thus, an increase of entropy occurs only if the
system is disturbed, i.e.\ when it is not isolated.

It appears then that it is not unanimously established what the Second Law
actually says and what kind of relationship it has with the arrow of time.
 The aim of the present paper is to chart this amazing and confusing
multifariousness of the Second Law; if only to help prevent
embarrassment when, at a birthday party, the reader is faced with
the obvious counter-question by literary companions. Or, if the
reader wishes to be counted as a person of literary culture, and
guard against arrogant physicists, one can also read this article
as a guide to how to bluff your way in the Second Law of
Thermodynamics.

The organization of the article is as follows. In section~\ref{status}, I
will describe a few general characteristics of thermodynamics, and its
status within physics.  Section~\ref{pijl} is devoted to the distinction
between several meanings one can attribute to the arrow of time.  Next, in
sections~\ref{Carnot},~\ref{CK} and~\ref{terug}, I will trace the
historical development of the orthodox versions of the Second Law,
focussing at each stage on its relation to the arrow of time.
  This historical development finds its climax in the intricate arguments
of Planck, which I review in section~\ref{Planck}.

Then I address two less orthodox but perhaps more vital versions of the
Second Law, due to Gibbs (section~\ref{Gibbs})  and Carath\'eodory
(section~\ref{Car}). I will argue that
these versions do not carry
implications for an arrow of time (with a slight qualification for
\C).
  In section~\ref{debat}, I discuss the debate in the 1920's between
Born, Planck and \TEA{}, which was triggered by the work of \C.

Despite a number of original defects, the approach pioneered by
\C{} has in recent years turned out to be the most promising route
to obtain a clear formulation of the Second Law.  Section~\ref{LY}
is devoted to the work of Lieb and Yngvason, which forms the most
recent major contribution to this approach. Finally, in
section~\ref{moral}, I will discuss some conclusions. In
particular, I will discuss the prospects  of
 giving up the idea that the arrow of time  is  crucially
related to the  Second Law.

 \section{The status of thermodynamics \label{status}}

Classical \TD{} can be described as the study of phenomena
involved in the production of work by means of heat; or, more
abstractly, of the interplay of thermal and mechanical energy
transformations.  The theory is characterised by a purely
empirical (often called `phenomenological')  approach.  It avoids
speculative assumptions about the microscopic constitution or
dynamics of the considered systems. Instead, a physical system is
regarded as a `black box' and one starts from a number of
fundamental laws (\emph{Haupts\"atze}), i.e.\ generally formulated
empirical principles that deny the possibility of certain
conceivable phenomena, in particular various kinds of perpetual
motion.
 The goal is then to introduce all specific thermodynamical
quantities and their general properties
by means of these laws. This is the approach to the theory  taken by
Carnot,
Clausius, Kelvin, and Planck, and with some differences also by Gibbs and
\C.

Anyone who studies classical thermodynamics today will encounter a wide
range of views on its status. In the eyes of many modern physicists, the
theory has acquired a somewhat dubious status.  They regard classical
thermodynamics as a relic from a bygone era. In particular the refusal to
adopt the atomic hypothesis is seen as typical nineteenth century cold
feet.
 Also, one often reads that thermodynamics is really a subject for
engineers and therefore(?) not an appropriate vehicle for fundamental
knowledge about nature. Further, the `negative' character of its laws,
i.e.\ the fact that they state what is impossible rather than what is
possible, seems offensive to many authors.\footnote{%
         This objection, expressed again and again by many commentaries,
          is not easy to comprehend.  Elsewhere in physics one also
          finds `negatively' formulated principles (like the light postulate, the
         uncertainty principle, etc.)
          but one never encounters this reproach.
         Besides it is obvious that every negative lawlike statement,
      can also be rephrased  positively
       by a change of vocabulary.
        The issue is
        therefore only cosmetic.  A more substantial problem
         that probably lurks in the background
         is  ambivalence in the meaning of  `possible'.
 }

Indeed, the view that \TD{} is obsolete is so common that many
physicists use the phrase `Second Law of Thermodynamics' to denote
some counterpart of this law in the kinetic theory of gases or in
statistical mechanics.  However, I will not embrace this
manoeuvre.  In this article, the term `Second Law of
Thermodynamics'  refers to an ingredient of classical
thermodynamics, and not some other theory.

On the other hand, even in the  twentieth century one can find
prominent physicists who appreciated  \TD.
  Einstein, whose earliest publications were devoted to the foundations of
the Second Law,  remained  convinced throughout his life  that
thermodynamics is the only universal physical theory that will never be
overthrown.
 He recommended this remark
`for the special attention of those who are skeptics on principle'
\cite[p.~33]{Einstein48}.
 Other important physicists who devoted part
of their work to thermodynamics are Sommerfeld, Born, Schr\"odinger,
Pauli, Chandrasekhar and Wightman.
 Planck in particular reported  (1948, p.~7)  \nocite{Planck48}
how strongly he was impressed by the
universal and irrefutable validity of thermodynamics.  As a schoolboy, he
already experienced his introduction to the First Law of Thermodynamics as
an evangelical revelation (`wie eine Heilsbotschaft').
 The Second Law became, mainly through his own contributions, a
proposition of comparable stature; (see e.g.\  \citeasnoun[ \S 136]{Planck}).

Similar veneration is expressed in a famous
quotation by Eddington:
   \begin{quote}\small
 The law that entropy always increases, ---the second law
of thermodynamics---holds, I think, the supreme position among the
laws of Nature. If someone points out to you that your pet theory
of the universe is in disagreement with Maxwell's equations---then
so much the worse for Maxwell's equations. If it is found to be
contradicted by observation---well, these experimentalists bungle
things sometimes. But if your theory is found to be against the
second law of thermodynamics I can give you no hope; there is
nothing for it but to collapse in deepest humiliation \label{Edd}
\cite[p.~81]{Eddington}. \end{quote} Apparently there is, apart
from the view that \TD{} is obsolete, also a widespread belief
among physicists in its absolute authority.

Apart from its authority, thermodynamics is also often praised for
its clear and rigorous formulation. Maxwell (1877) regarded the
theory as a `a science with secure foundations, clear definitions
and distinct boundaries'.\nocite{Maxwell77} Sommerfeld (1952)
called it a `{\em Musterbeispiel\/}' of an axiomatised
theory\nocite{Sommerfeld}.
 It is also well-known  that Einstein drew inspiration from \TD{} when he
formulated the theory of relativity and that he intended to construct this
theory in a similar fashion, starting from similar empirical principles of
impossibility \cite{Klein}.

  But there are also voices of dissent on this issue of clarity and
rigour. The historian Brush notes:
  \begin{quote}\small
As anyone who has taken a course in thermodynamics is well
aware, the mathematics used in proving Clausius' theorem
[i.e.\ the Second Law] is of a very special kind, having
only the most tenuous relation to that known to
mathematicians\nocite{Brush}
(1976, Vol.~1, p.~581).
\end{quote}
He was recently joined by the mathematician
 Arnold (1990, p.~163):
\begin{quote}\small
Every mathematician knows it is impossible to
understand an elementary course in thermodynamics.
\nocite{Arnold}\end{quote}
  Von Neumann once remarked that whoever uses the term `entropy' in a
discussion always wins:
 \begin{quote}\small
\ldots no one knows what entropy really is, so in a debate you
will always have the advantage
 (cited by\nocite{Tribus}
Tribus and McIntire, 1971, p.~180).\end{quote}
an invaluable piece of advice for the  true bluffer!

The historian of science and mathematician Truesdell made a
detailed study of the historical development of thermodynamics in
the period 1822--1854. He characterises the theory, even in its
present state, as
 `a dismal swamp of obscurity'\nocite{Truesdell}
(1980, p.~6)
and
  `a prime example to show that physicists are not exempt from
  the madness of crowds' (ibid.~p.~8).
He is outright cynical about the respect with which non-mathematicians
treat the Second Law:
 \begin{quote}\small
Clausius' verbal statement of the second law makes no sense
[\ldots]. All that remains is a Mosaic prohibition; a century
of philosophers and journalists have acclaimed this
commandment; a century of mathematicians have shuddered and
averted their eyes from the unclean.
(ibid.\ p.~333).

Seven  times in the past thirty years have I tried to
follow the argument Clausius offers
[\ldots]  and seven times  has it blanked and  gravelled me.
[\ldots] I cannot explain what I cannot
understand (ibid.\ p.~335).
\end{quote}
From this anthology it emerges that although many prominent physicists
are firmly convinced  of, and express  admiration for the Second Law,
there are also serious complaints, especially from mathematicians,
about a lack of clarity  and rigour in its  formulation.\footnote{%
   But  here  too there are dissidents:  `Clausius' \ldots definition [of
   entropy] \ldots appeals to the mathematician only.' \cite{Callendar}.}
 At the very least one can say that the Second Law suffers from an image
problem: its alleged eminence and venerability is not perceived by
everyone who has been exposed to it.
 What is it that makes this physical law so obstreperous that every
attempt at a clear formulation seems to have failed? Is it just the usual
sloppiness of physicists? Or is there a deeper problem? And what exactly
is the connection with the arrow of time and irreversibility? Could it be
that this is also just based on bluff?

Perhaps readers will shrug their shoulders over these questions.
Thermodynamics is obsolete; for a better understanding of the
problem we should turn to more recent, statistical theories. But
even then the questions we are about to study have more than a
purely historical importance.  The problem of reproducing the
Second Law, perhaps in an adapted version, remains one of the
toughest, and controversial problems in statistical physics. It is
hard to make progress on this issue as long as it remains unclear
what the Second Law says; i.e.\ what it is that one wishes to
reproduce.  I will argue, in the last section, using the example
of the work of Boltzmann, how much statistical mechanics suffered
from this confusion.

Since there is no clear-cut uncontroversial starting point, the only way
to approach our problem is by studying the historical development of the
Second Law.  I will further assume that respect ought to be earned and
from now on write the second law without capitals.

\section{Possibility, irreversibility, time-asymmetry,  arrows and
ravages\label{pijl}}

In order to investigate the second law in more detail, it is
necessary to get a tighter grip on some of the philosophical
issues involved, in particular the topic of the arrow of time
itself. But first there is an even more general issue which needs
spelling out. As we have seen, the basis of the second law is a
claim that certain processes are impossible. But there are various
senses in which one can understand the term `possible' or related
dispositional terms.  At least three of these are relevant to our
enterprise.

    (i) `Possible' may mean: `allowed by some given theory'. That is,
  the criterium for calling a process possible is whether one can specify
a model of the theory in which it occurs.  This is the sense which
is favoured by modern philosophers of science, and it also seems
to be the most fruitful way of analysing this notion. \forget{, I
will adopt it here as a default.  } However, thermodynamics has a
history of more than 150 years in which it did not always have the
insights of modern philosophy of science at hand to guide it. So,
one should be prepared to meet other construals of this term in
the work of our main protagonists.

(ii) The term `possible' may be taken to mean: `available in the
actual world' (or in `Nature').  This is the view that Planck and
many other nineteenth century physicists adopted. For them, e.g.\
the statement that it is possible to build a system which exhibits
a particular kind of perpetual motion means that we can actually
build one.
 \forget{
Similarly, in the question of whether the recovery of an initial
state is possible, one wishes to obtain this recovery in our
actual world.  The idea, e.g., that a return to our youth would be
permitted by a particular theory, (i.e.\ the theory allows a
possible world in which it occurs) is a meager solace for those
who would like to see the effects of ageing undone in the actual
world. In this reading, the notion of theoretically allowed models
plays no role.}

An important aspect of reading `possibility' in this way is that
 the question of whether a process is possible or not, is not decided by
the theory, but by `the furniture of the world', i.e.\ the kinds
of systems and interactions there actually are. This includes the
systems and forms of interactions which we have not even
discovered and for which we lack an appropriate theory. So, the
claim that such a process is impossible, becomes a statement that
transcends theoretical boundaries. It is not a claim to be judged
by a theory, but a constraint on all physical theories, even those
to be developed in the future. Clearly, the idea that the second
law is such a claim helps explaining why it inspired such feelings
of awe.

(iii) A third sense of `possible' is `available to us'. \forget{
Although we will not be concerned much with this view,}  We shall
see that some authors, in particular Kelvin, were concerned with
the loss of motive power `available to man'.  This reading makes
the notion dependent on the human condition.  This is generally
considered as a drawback. If the second law would be merely a
statement  expressing a human lack of skills or knowledge, it
would cease to be interesting,
both physically and philosophically.\footnote{%
    However, this not to say that   the viewpoint is unviable. In
    fact, it seems to be very close to Maxwell's views. He often
    emphasized the importance of  the human condition
    (as opposed to the `demonic condition') in writings like: `\ldots
    the notion of
    dissipated energy would not occur to a being who \ldots could
    trace the motion of every molecule and seize it at the right
    moment. It is only to a being in the intermediate stage, who can
    lay hold of some forms of energy while others elude his grasp,
    that energy appears to be passing inevitably from the available to
    the dissipated state.' \cite{Maxwell78}.  Planck
    strongly opposed this view; see \cite[\S 136]{Planck}.}

Next we consider the arrow of time. What exactly does one mean
by this and related terms?  For this question one can consult the relevant
literature on the philosophy of time
\cite{Reichenbach,Grunbaum,Earman,Kroes,Horwich}.

An important aspect of time that is distinguished in this
literature is the idea of the \emph{flow} or progress of time.
Human experience comprises the sensation that time moves on, that
the present is forever shifting towards the future, and away from
the past.  This idea is often illustrated by means of the famous
two scales of McTaggart.  Scale $B$ is a one-dimensional continuum
in which all events are ordered by means of a date.
 Scale $A$ is a similar one-dimensional continuous ordering for the same
events, employing terms like `now', `yesterday', `next week', etc. This
scale shifts along scale $B$ as in a slide rule.

Another common way of picturing this idea is by attributing a
different ontological status to the events in the past, present
and future. Present events are the only ones which are `real' or
`actual'. The past is gone, and forever fixed. The future is no
more actual than the past but still `open', etc. The flow of time
is then regarded as a special ontological transition:  the
creation or actualisation of events. This process is often called
\emph{becoming}. In short, this viewpoint says that grammatical
temporal tenses have counterparts in reality.

Is this idea of a flow of time related to thermodynamics?
 Many authors have indeed claimed that the second law provides a
physical foundation for this aspect of our
experience~\cite{Eddington,Reichenbach,Prigogine}.
  But according to contemporary understanding, this view is
untenable~\cite{Grunbaum,Kroes}. In fact the concept of  time flow
hardly ever enters in any physical theory.\footnote{%
       Newton's conception of  absolute time which
      `flows equably and of itself' seems the only exception.}
In a physical description of a process, it never makes any difference
whether it occurs in the past, present or future.
 Thus, scale $B$ is always sufficient for the formulation of physical
theory~\footnote{%
      This statement holds strictly speaking only for non-relativistic
      theories.  Nevertheless, for special-relativistic theories
      an analogous statement is valid,
      when the one-dimensional scales of
      McTaggart are   replaced by
      partial orderings~\cite{Dieks,Muller}.} and the
above-mentioned ontological distinctions only play a metaphysical
role.  Thermodynamics is no exception to this, and therefore
unable to shed any light on this particular theme.

A second theme, which is much closer to the debate on the second law, is
that of \emph{symmetry} under time reversal.  Suppose we record some
process on film and play it backwards.  Does the inverted sequence look
the same? If it does, e.g.\ a full period of a harmonic oscillator, we
call the process time-symmetric. But such processes are not in
themselves very remarkable.
 A more interesting question concerns physical laws or theories.  We call
a theory or law time-symmetric if the class of processes that it allows is
time-symmetric. This does not mean that all allowed processes have a
palindromic form like the harmonic oscillator, but rather that a censor,
charged with the task of banning all films containing scenes which violate
the law, issues a verdict which is the same for either
direction of playing the film.

More formally, the criterion can be phrased as follows.
Many theories employ  a state space
 $\Gamma$ which contains all possible states of a system.  The
instantaneous state is thus  represented as a point  $s$ in
$\Gamma$ and a process as a parametrised curve:
 \[ {\cal P}  = \{s_t \in \Gamma \;:\: t_i \leq t\leq t_f\}\]The laws
of the theory only  allow a definite class of
processes (e.g.\ the solutions of the equations of motion).
Call this class $\cal W$, the set of all possible worlds (according to
this theory).
  Let now  $R$  be a transformation  that turns a state $s$  into its
`time reversal' $Rs$.  It is always assumed that $RRs=s$ (i.e.\
$R$ is an involution).
 In classical mechanics, for example, $R$ is the transformation which
reverses the sign of all momenta and magnetic fields. In a theory
like classical thermodynamics, in which the state does not contain
velocity-like parameters, one may simply take $R$ to be the
identity transformation.

  Further, the time reversal ${\cal P}^*$
of a process ${\cal P}$ is
defined as:
  \[ {\cal P}^*  =  \{ (Rs)_{-t} \;:\:  -t_f \leq  t  \leq -t_i \}.
 \]
The theory  is called time-symmetric if the class $\cal W$ of possible
worlds is closed under time reversal, i.e.\ if the following holds:
 \be \mbox{If } {\cal P} \in {\cal W}\: \mbox{~then~}\: {\cal P}^* \in
{\cal W}. \label{traf} \ee

 Note that this criterion is formulated without recourse to
metaphysical notions like `becoming' etc.
The mathematical form of the laws themselves (and a given choice for $R$)
determines whether the theory is time-symmetric or not.
 Note also that the term `time-reversal' is not meant literally. That is
to say, we consider \emph{processes} whose reversal is or is not allowed
by a physical law, not a reversal of time itself. The prefix is only
intended to distinguish the term from a spatial reversal.
 Furthermore, note that we have  taken `possibility' here in
sense (i) above; that is, it is not relevant here whether the
reversed processes ${\cal P}^*$ occur in the actual world. It is
sufficient that the theory allows them.  Thus, the fact that the
sun never rises in the west is no obstacle to celestial
mechanics qualifying as time-symmetric.\footnote{%
    Of course one may also
    develop notions of time-(a)symmetry in other senses. It is
    interesting to mention, in this context, the distinction between
    Loschmidt's and Kelvin's arguments for
    the time-symmetry of classical
    mechanics, i.e.\ their versions of the \emph{Umkehreinwand}.
    Loschmidt observed that for every mechanical process $\cal P$ a
    time reversed process is also a model allowed by classical
    mechanics.
    This is possibility in sense (i). Kelvin, on the other hand, discussed
    the issue  of actually obtaining the time reversal of a given
    molecular motion, by means of a physical intervention, namely by
        collisions with `molecular cricket bats'. This is closer to sense
(ii).}

Is this theme of time-(a)symmetry related to the second law?  Even though
the criterion is unambiguous, its application to \TD{} is not a matter of
routine.  In contrast to mechanics, \TD{} does not possess equations of
motion.  This, in turn, is due to the fact that thermodynamical processes
only take place after an external intervention on the system.  (Such as:
removing a partition, establishing thermal contact with a heat bath,
pushing a piston, etc.) They do not correspond to the autonomous behaviour
of a free system.
 This is not to say that time plays no role.  Classical \TD{} in the
formulation of Clausius, Kelvin or Planck is concerned with processes
occurring in the course of time, and its second law does allow only  a
subclass of possible
worlds, which is indeed time-asymmetric. However, in the formulations by
Gibbs and \C{} this is much less clear.
 We shall return in due course to the question of whether \TD{} in these
versions is time-asymmetric.

As a side remark, I note that the discussion about the relation
between the second law and time-asymmetry is often characterized
by a larger ambition.  Some authors are not satisfied with the
mere observation that a theory like \TD{} is time-asymmetric, but
claim that this theory can be held responsible, or gives a
physical foundation, for the distinction between past and future.
This claim has been advanced in particular by Reichenbach.  He
argued that by definition we could identify our concept of
`future' with the direction of time in which entropy increases.

 Reichenbach's claim  has been criticized by
\cite{SklarUp}. The  main objections, in my opinion, are
that the claim would entail that all other forms of time-asymmetry which
might be found in other physical theories (such as cosmology,
elementary particles physics, etc.)  should also be characterizable in
terms of a thermodynamical asymmetry.
  The question whether this is really the case, has often been discussed
 \cite{Landsberg,Savitt} but an affirmative answer is not yet established.
Another objection is that even if humans are placed in a local
environment in which entropy decreases, e.g.\ in a
refrigerator cell, this does not seem to affect their sense of temporal
orientation.
  More important perhaps is the objection that the programme to
  define the distinction between past and future by means of the second
 law is only sensible if it turns out to be possible to introduce the
 second law itself without presupposing  this distinction.
   The classical formulations of the second law certainly do not meet this
criterion.

Another theme concerns `\emph{irreversibility}'.  This term is
usually attributed to processes rather than theories. In the
philosophy of science literature the concept is however intimately
connected with time-asymmetry of theories.  More precisely, one
calls a process $\cal P$ allowed by a given theory irreversible if
the reversed process ${\cal P}^*$ is excluded by this theory.
Obviously, such a process $\cal P$ exists only if the theory in
question is time-asymmetric. Conversely, every time-asymmetric
theory does admit irreversible processes in this sense. These
processes constitute the hallmark of time-asymmetry and,
therefore, discussions about irreversibility and time-asymmetry in
the philosophy of science coincide for the most part. However, in
thermodynamics, the term is commonly employed with other meanings.
Therefore, in an attempt to avoid confusion, I will \forget{
continue to refer to this aspect as time-asymmetry, and} not use
the term `(ir)reversibility' in this sense.

 In the thermodynamics literature one often uses the term
`irreversibility' to denote a different aspect of our experience
which, for want of a better word, one might also call
\emph{irrecoverability}. Our experience suggests that in many
cases the transition from an initial state $s_i$ to a final state
$s_f$, obtained during a process, cannot be fully undone, once the
process has taken place.  Ageing and dying, wear and tear, erosion
and corruption are the obvious examples.  In all such cases, there
is no process which starts off from the final state $s_f$ and
restores the initial state $s_i$ completely. As we shall see in
more detail in Section~\ref{Planck}, this is the sense of
irreversibility that Planck intended, when he called it the
essence of the second law.

Many writers have followed Planck's lead  and emphasised this
theme of irrecoverability in connection with the second law.
 Indeed, Eddington introduced his famous phrase of  `the arrow of time'
in a general discussion of the `running-down of the universe', and
illustrated it with many examples of processes involving
`irrevocable changes', including the nursery rhyme example of
Humpty-Dumpty who, allegedly, could not be put together again
after his great fall.
 In retrospect, one might perhaps say that a better expression for this
theme is the \emph{ravages} of time rather than its arrow.

This present concept of irreversibility is different from that of
time-asymmetry in at least three respects. In the first place, for a
`recovery' the only thing that counts is the retrieval of the initial
state. It is not necessary that one specifies a process $\cal P^*$ in
which the original process is retraced step by step in the reverse order.
In this respect, the criterion for reversibility is weaker than that for
time-symmetry, and irreversibility is a logically stronger notion than
time-asymmetry.

A second
difference is that in the present concept, one is concerned with a
\emph{complete} recovery.
  As we shall see, Planck repeatedly emphasised that the criterium for a
`complete recovery' of the initial state involves, not only the
system itself, but also its environment, in particular all
auxiliary systems with which it interacted.

\forget{By imposing such additional desiderata,
the condition for reversibility becomes stronger than time-symmetry, and,
in this respect, irreversibility is weaker than time-asymmetry.}

This reference to states of the environment of a system already lends a
peculiar twist to classical thermodynamics that we do not meet in other
theories of physics.\footnote{%
    The reason for this is,  again, that in thermodynamics
    processes are due to an external intervention on the system;
    whereas in mechanics it is natural ---or at least always
    possible--- to study autonomous processes of a system which is
    isolated from
    its environment.  That is to say: even in those cases where
    interactions with external environment occur it is in principle
    possible to include their mechanical behaviour into the
    description, in order to obtain a
    larger, isolated system.}
    The problem is that the theory aims at stating conditions which
    allow the introduction of the notions temperature, entropy and
    energy, which are needed to characterise the thermodynamical state
    of a system. This entails that one cannot assume ---on pains of
    circularity--- that (auxiliary systems in) the environment already
    possess a thermodynamical state.  We will meet several instances
    where this problem raises its head.

However, assuming for the moment that it makes sense to attribute, at
least formally, a state $Z$ to the environment, one may give a formal
 criterion for the present concept of reversibility as follows. Since we
are not interested in the intermediate stages of a process here, we adopt
an abbreviated representation. Let $\cal P$ be a process that produces the
transition:
 \[
\langle s_i,Z_i\rangle
\statechange{P} 
\langle s_f,Z_f\rangle. \]
(Such an abbreviated representation of a process is often called a
 `change of state'.)
Then  $\cal P$ is reversible iff another process $\cal P'$ is
possible which produces the state change
\[ \langle s_f,Z_f\rangle  \statechange{P'}
\langle s_i,Z_i\rangle.  \]

 The third respect in which Planck's concept of irreversibility differs
from time-asymmetry concerns the notion of `possible'.  As
we shall see, Planck insisted that the `recovery process' $\cal P'$ is
available in our actual world, not merely in some model of the
theory.  That is,
in the question of whether the recovery of an initial state is possible,
one wishes to obtain this recovery in our actual world.  The idea, e.g.,
that a return to our youth would be permitted by a particular theory,
(i.e.\ the theory allows a possible world in which it occurs) is a
too meager
solace for those who would like to see the effects of ageing undone in the
actual world.  In this reading, the notion of theoretically allowed models
plays no role.
In this respect, recoverability is stronger than time-symmetry. Taking
the first and third respect   together, we see that
 (ir)recoverability does not imply, and is not implied
by time-(a)symmetry.  See Figure~1 for illustrations.

\begin{figure}[t]
 \begin{center}
\begin{picture}(288,190)(10,20)
 \epsfig{file=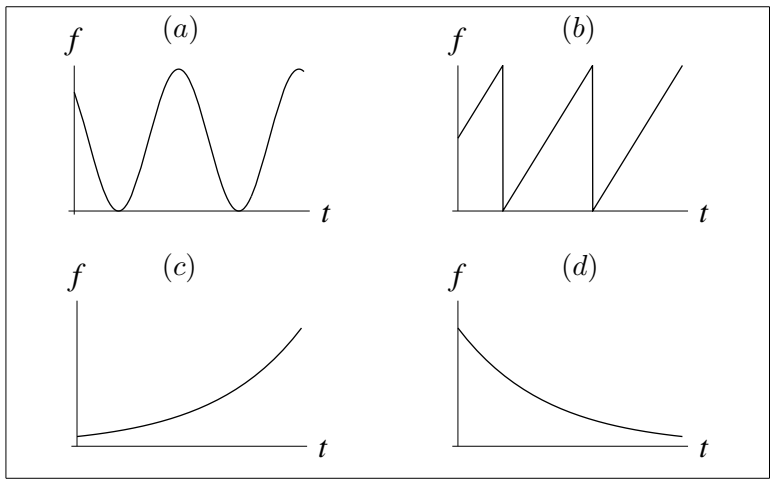}
\put(-230,167){$(a)$}
\put(-79,167){$(b)$}
\put(-230,77){$(c)$}
\put(- 79, 77){$(d)$}
\end{picture}
  \end{center} \caption{%
 \small \it    Illustrating the distinction between time-(a)symmetry
         and
    irreversibility in the sense of irrecoverability. Consider some
     processes of a fictional system in  an otherwise empty universe,
     characterised by the time evolution of some quantity $f$.
    Process (a) is time-symmetric and recoverable; (b) is
    time-asymmetric and recoverable.
    The processes (c) and (d) are both time-asymmetric and
       irrecoverable.
    A theory allowing only these two irrecoverable processes is
    time-symmetric.
  }
 \end{figure}

However, the thermodynamical literature  also uses the term `reversible'
 in yet another meaning, which is not  straightforwardly connected with
the arrow of time at all. It is used to denote processes which proceed so
delicately and slowly that the system remains close to an equilibrium
state during the entire process. This is comparable to, say, moving a cup
of tea filled to the brim, without spilling.
    We shall see in section~6 that this is the meaning embraced by
Clausius. Actually, it seems to be the most common meaning of the term, at
least in the physical and chemical literature; see e.g.\
\cite{HZ,Denbigh89}. A  modern and more apt name for this kind of
processes is
\emph{quasi-static}.\footnote{%
     Yet another term employed for this type of process is
    \emph{adiabatic}. This terminology probably originates from
  the so-called `adiabatic theorem' of
     P.~Ehrenfest (1916). \nocite{Ehrenfest1916}%
       This usage is highly confusing, because
     in the terminology of \TD{} (which is adopted here) a
     process is called adiabatic if it takes place without heat exchange
    between the system and its environment.
    I note that Ehrenfest formulated  his theorem for what  he at  first
    called `adiabatic-reversible changes'  (with reversible
in the sense of quasi-static). A few pages later he dropped the adjective
`reversible'  as being superfluous when  the process is periodic or
quasi-periodic.  (Obviously, he was now interpreting reversible in the
sense of
recoverable.) As a result,  `adiabatic' became the stand-in
terminology for  quasi-static.
 }

The present concept makes no direct reference to a direction of time.
Indeed, the concept is neutral with respect to time reversal, because the
time reversal of a (non-)quasi-static process is obviously again a
(non-)quasi-static process.
 Still, one can easily see, at least roughly, how the terminology arose.
 Indeed, for simple systems, thermodynamics allows all quasi-static
processes. That is to say, for any two equilibrium states $s$ and
$t$, and every smooth curve in the space of equilibrium states
which connects them, there always is a quasi-static process (in an
appropriate environment, of course) which always remains close to
this curve, but also one which closely follows the curve in the
opposite direction (i.e.\ from $t$ to $s$). In this sense, the
time reversal of a quasi-static process is
allowed by the theory.\footnote{%
    However, one should be careful
    not to take this statement literally.  The `reversal' of the
    quasi-static process is generally not the exact time reversal of
    the original process, but remains close to it within a small
error.}
  This, clearly, is why quasi-static processes are
traditionally called reversible.
This conclusion does not hold generally for non-quasi-static
curves,  and, therefore, they are considered as irreversible.

Note that in the present sense, a process is `irreversible' if and
only if it involves non-negligible deviations from equilibrium. Obvious
examples are turbulence and life processes.  It is well-known that
classical thermodynamics is inadequate to give a detailed description of
such processes (or, indeed, of their time reversal).  From this
perspective, Bridgman's view that thermodynamics has little to say about
irreversible processes becomes less puzzling.

In fact, Bridgman  is one of few authors who recognise a
distinction between notions of (ir)reversibility:
\begin{quote}\small
[\ldots] reversible engines and reversible processes play an
important
role in the conventional thermodynamical expositions. I would like to
make the comparatively minor point that the emphasis on reversibility is
somewhat
misplaced.  [\ldots]
\forget{In stead of reversible processes one might better speak of
processes of maximum efficiency.}
It is not the reversibility of the process that is of primary importance;
the importance of reversibility arises because when we have reversibility,
we also have \emph{recoverability}. It is the recovery of the original
situation that is important, not the detailed reversal of steps which led
to the original departure from the initial situation. \cite[p.~122]
{Bridgman}
\end{quote}
 Note, however, that at the same instant at which he makes the distinction
Bridgman also draws a close connection between the two: he claims that
reversibility implies recoverability. We shall see later that this
inference is incorrect.

The above discussion of meanings of irreversibility is not exhaustive.
Part of the physical literature on (ir)reversibility employs the term in
order to denote (in)determinism of the evolutions allowed by a theory.  A
deterministic process is then called reversible because the evolution
$U_t:  s \rightarrow U_t(s) =s_t$ is an invertible mapping
\cite{Landauer,Mackey}.  Indeterministic evolutions arise in classical
mechanics in the description of open systems, i.e.\ systems that form part
of a larger whole, whose degrees of freedom are not included in the state
description.

Indeed, one important approach to the foundations of thermal and
statistical physics aims at explaining irreversibility by an appeal to
open systems~\cite{BL,E.Davies,Lindblad,RR}.
 In this article,
however, this view is not discussed.\footnote{%
     Yet it is somewhat ironic that, whereas some of the  above
      formulations of the second law  pertains
    exclusively to isolated systems,
    this view regards it as essential condition for irreversible
    processes that the system is interacting with an environment.}
 Other ramifications, even farther removed from our subject, can be found
in the literature on the so-called `principle of microscopic
reversibility'.

 In his  \emph{the Nature of the Physical World} Eddington  introduced
 the catch phrase `the arrow of time'. Actually
\forget{it is not clear at all
what he meant by this. He} he employed the term as a metaphor that could
cover the whole array of themes discussed above. It is perhaps best to
follow him and use the `arrow of time' as a neutral term encompassing both
time-asymmetry  and irreversibility.

\section{The prehistory of thermodynamics: Carnot \label{Carnot}}

Sadi Carnot's \emph{R\'eflexions sur
la Puissance motrice du Feu}  appeared in 1824.
 It was this work which eventually led to the birth of
thermodynamics. Still, Carnot's memoir itself does not belong to
what we presently understand as thermodynamics. It was written
from the point of view that heat is an indestructible substance,
the so-called \emph{calorique}.
 This assumption proved to be in conflict with empirical evidence.\footnote{%
      After his death,
      papers were found in which Carnot expressed doubt
      about the conservation of heat.
      These notes were only published in 1878 and
      did not influence the factual development of the theory.}
However, Carnot's main theorem  did agree with experience.
 Classical thermodynamics was born from the attempts around 1850 to save
Carnot's theorem by placing it on a different footing.

Even though Carnot's work does not belong to thermodynamics proper, it is
important for our purpose for three reasons.
 First, his theorem became, once encapsulated in that later theory, the
first version of a second law.  Secondly, the distinction between
reversible and irreversible processes can already be traced back
to his work. And most importantly, many commentators have claimed
that Carnot's work already entails an arrow of time.

Carnot was concerned with heat engines: devices which operate in a
cycle and  produce work by absorbing heat from one heat reservoir
(the `furnace')  with high temperature $\theta_1$ and ejecting
heat in another (the `refrigerator') with a lower temperature
$\theta_2$. Both  reservoirs are assumed to be so large that their
state is unaltered by their heat exchange with the engine. The
engine is then capable of repeating the cycle over and over again.

The operation of such an engine is comparable to that of a water mill: its
power to produce work results from the transport of heat from high to low
temperature, just as a  mill works by transporting water from a
higher to a lower level.  And just as the mill does not consume water, no
more does the heat engine reduce the amount of \emph{calorique}.

Naturally, Carnot was interested in the efficiency of such heat
engines: i.e., the ratio of the total work produced during a cycle
and the amount of heat transported from the furnace to the refrigerator.
He obtained a celebrated result, which in a modern formulation, can be
stated as follows:\footnote{%
     An elaborate analysis and reconstruction of his
     argumentation has been  given
by~\citeasnoun{TruesdellBharatha},\citeasnoun{Truesdell}.
    This reconstruction differs, however, from the formulation adopted
here.
See also footnote~\ref{truesdellovercarnot} below.
}
\begin{quote}\small
{\sc Carnot's Theorem:}
 Let the furnace  and refrigerator temperatures  $\theta_1$ and
$\theta_2$  be given (with $\theta_1 > \theta_2$).
 Then:\\
(\emph{i})  all heat engines operating in a {\em reversible\/}
cycle have the same efficiency. Their efficiency is therefore a
universal expression depending only on the temperatures $\theta_1$
and $\theta_2$.
\\
(\emph{ii}) all other heat engines have an efficiency  which is
less than or equal to that obtained by reversible cycles.
\end{quote} Thus, the  efficiency obtained by the reversible cycle
is  a universal function,  often called  the \emph{Carnot
function}: $C(\theta_1, \theta_2)$.

 Carnot obtained his theorem from a principle that states the
impossibility of a \emph{perpetuum mobile}: it is not possible to
build an apparatus that produces an unlimited amount of work
without consumption of \emph{calorique} or other resources.\footnote{%
        Carnot's principle states the impossibility of what is today
        called the `\emph{perpetuum mobile} of the \emph{first} kind'.
         This fact is remarkable because it has often been claimed
        that this principle immediately entails the {\em first\/} law of
        thermodynamics (Joule's principle of equivalence of work and
        heat),
     e.g.\ by Von Helmholtz~(1847)\nocite{Helmholtz} and
         \citeasnoun{Planck}.
    The caloric theory employed by
        Carnot, which  violates
        the first law, is a manifest counterexample for this claim.
 }

His argument is a well-known reductio ad absurdum:  If there were a heat
engine $A$, performing a reversible cyclic process between the reservoirs
with temperatures $\theta_1$ and $\theta_2$, having less efficiency than
some other engine $B$, which also performs a cycle between these two
reservoirs, then we should be able to combine them in a composite cycle in
which the reversible engine $A$ is employed backwards, pumping the same
amount of heat from the refrigerator back into the furnace, that $B$ had
used in producing work.
 But since $A$ is assumed to have a lower efficiency, it needs less work
to restore the heat to the furnace than produced by $B$. In other words,
we would obtain a surplus of work, which can be used for any purpose we
like.
   Moreover, this composite process is cyclic, because both engines and
heat reservoirs return to their initial states. It is can thus be repeated
as often as we like, and we would have constructed a \emph{perpetuum
mobile}.

The most striking point about this  theorem, at least for Carnot himself
and for those who continued his work, was the implication that the maximum
efficiency should be independent of the medium used in the heat engine.
It remains the same, whether the engine employs steam, air, alcohol or
ether vapour, etc. This was not at all obvious to his contemporaries.

The obvious next question is then to determine the function
$C(\theta_1, \theta_2)$. Because the efficiency of a reversible
cyclic process is independent of the construction of the engine or
details of the process, one may restrict attention to the simplest
version:  the Carnot cycle. This is a reversible cyclic process
consisting of four steps:  two isothermal steps, where heat
exchange takes place with the reservoirs of temperature $\theta_1$
and $\theta_2$, alternating with two steps in which the system is
adiabatically isolated.

Carnot realised that the analogy with the water mill could not be
stretched too far.  Thus, while the maximal efficiency of a mill
depends only on the difference in the height of the levels, we
have no grounds for assuming that the efficiency of a Carnot cycle
simplifies to a function of $\theta_1 - \theta_2$ alone.

To avoid this difficulty, Carnot specialised his consideration to
Carnot cycles where the heat reservoirs have infinitesimally
different temperatures $\theta$ and $\theta + d\theta$. Let the
efficiency of such a cycle be $\mu(\theta)d\theta$ where
   \be  \label{mu}
\mu(\theta) :=  \left. \frac{\partial C(\theta, \theta_2)}{\partial
\theta_2}
\right|_{\theta_2 = \theta}. \ee
Assuming  the cycle is performed on an ideal gas,
he  obtained  the result:
 \be \mu (\theta)  = \frac{R}{Q} \log \frac{V_2}{V_1},
\label{ex}\ee
where  $Q$ is the amount of heat absorbed by the gas when its volume is
expanded from $V_1$ to $V_2$  at constant temperature $\theta$,
and $R$ is the gas constant. Thus, $\mu(\theta)$ can be determined from
experimentally accessible quantities. This is the closest Carnot got to
the  actual determination of $C(\theta_1, \theta_2)$.

The main question for my purpose is now to what extent the work of Carnot
is connected with the arrow of time.  I cannot find any passage in his
work in which he mentions such a connection. But it is true that his
theorem introduced the dichotomy between two types of cycles, which are
today called `reversible' and `irreversible'. However, he does not
actually use these terms.\footnote{%
    The name `reversible' is apparently due to Kelvin (1851).}
 So we should be cautious about the meaning of this dichotomy in this
context.

In actual fact, Carnot's own discussion starts by giving an explicit
description of a Carnot cycle for steam. In passing, he mentions that:
 `The operations we have just described might have been performed in an
inverse direction and order (Mendoza, 1960, p.11).' Next, he
formulates his theorem by claiming
 \forget{
that there exists no means to obtain motive power by transfer of
caloric, preferable to the method just indicated (i.e.\ the Carnot
cycle for steam). He concludes} that `the maximum of motive power
resulting from the employment of steam is also the maximum of
motive power realizable by any means whatever.' (ibid.\ p.~12)

However, he realised that a more precise formulation of this
conclusion was desirable.  He continued:
\begin{quote}\small
We have a right to ask, for the proposition just enunciated, the
following questions: what is the sense of the word \emph{maximum}?
By what sign can it be known that this maximum is attained? By
what sign can it be known whether the steam is employed  at the
greatest possible advantage in the production of motive power?
(ibid.~p.~12)
\end{quote}
\forget{
   A reversible cycle is simply a process that can be run both forwards
and backwards: instead of doing work while transporting heat from
the furnace to the refrigerator, we can employ the engine as a
heat pump: when work is done on the device, it will transport heat
from the refrigerator to the furnace.  Implicitly, he assumed that
the efficiency remains the same in both modes of operation. } In
answer to this question, he  proposes a necessary and sufficient
criterion~\cite[p.~13]{Mendoza}: it should be avoided that bodies
of different temperature come into direct thermal contact, because
this would cause a spontaneous flow of heat.
 In reality, he says, this criterion cannot be met exactly: in order to
exchange heat with a reservoir the temperature of the system needs to be
slightly different from that of the reservoir. But this difference may be
as small as we wish, and therefore we can neglect it. In modern terms: the
condition is that the process should be quasi-static at all stages which
involve heat exchange.

 Carnot explicitly mentions cases where the condition is
not met (p.~12,14), and argues that the spontaneous heat flow occurring
there is unaccompanied by the production of work, and that thus motive
power is lost, just as in a mill that spills its water.

Accordingly, even at this early stage, there are two plausible options for
a definition of the `reversible cycle'.  Either we focus on the property
of the Carnot cycle that it can also be run backwards, and use this as a
definition.  This is the option later chosen by Kelvin in 1851.  Of
course, this is a natural choice, since this property is essential to the
proof of the theorem.
 Or else, one can view the necessary and sufficient condition which Carnot
offers as a definition of reversibility.  As we shall see, this is more or
less the option followed by Clausius in 1864. In that case, a cyclic
process is by definition irreversible if, and only if, it involves a
direct heat exchange between bodies with different temperature.

However this may be, let me come back to the main question: does Carnot's
work imply an arrow of time, either in the sense of time-asymmetry, or in
the sense of irrecoverability? Let us take these questions one by one.

Is Carnot's theory time-asymmetric? That is: does it allow the
existence of processes while prohibiting their time reversal? And
more precisely, are the above irreversible cycles examples of such
processes? The answer to the first question is easy. Carnot's
principle forbids devices which produce work without consuming
some kind of resources. It has no qualms about their time
reversals, i.e.\ devices that consume work without producing any
effect, or leaving  any trace on  other resources. We conclude
that the theory is time-asymmetric.

The answer to the last-mentioned question, however, is less
obvious. Still, I think it is affirmative.  Consider some cycle
$\cal C$ which violates Carnot's criterion, and may therefore be
called `irreversible'. This cycle has less than maximum
efficiency.
  Now, suppose that its time reversal ${\cal C}^*$ is also allowed.  Does
this entail a contradiction? We may assume that the efficiency of the
reverse process ${\cal C}^*$ is the same as that of $\cal C$. After all,
 process ${\cal C}^*$ would not be considered as the reversal of $\cal
C$, unless it requires the same amount of work to transport the
same amount of heat back to the furnace.  Its efficiency is
therefore also less than maximal. Clearly, the supposition that
${\cal C}^*$ exists does not by itself violate Carnot's theorem.
However, we do obtain a contradiction by a very similar argument.
Indeed, the cycle ${\cal C}^*$ operates as a heat pump.  Thus,
`less than maximum efficiency' means that it requires \emph{less}
work to transport a given amount of heat from the refrigerator
into the furnace than a Carnot cycle.  Hence, combining the heat
pump ${\cal C}^*$ with a Carnot cycle in ordinary mode one obtains
a \emph{perpetuum mobile} of the first kind.
 Thus  ${\cal C}^*$
is not allowed by   this theory.\footnote{%
    \label{truesdellovercarnot}
    I note, however, that \citeasnoun{TruesdellBharatha}
and \citeasnoun{Truesdell}
     reach the opposite conclusion.  They argue that Carnot
    implicitly relied on a background theory of calorimetry, which
     involves standard differential calculus for heat and work
    and in  which  processes are always representable as
    differentiable curves in some state space.
    They  call this  `the doctrine of specific and latent
    heat'.
    Truesdell points out that all processes which can be handled by this
    approach are by definition reversible (by which he means that the
    curve can be traversed in either direction).  Thus,
        their reconstruction of Carnot's theory is completely time
    symmetric.   As a consequence, Truesdell
    denies that the dichotomy between cycles with maximum
    efficiency and those with less than maximal efficiency
        should be identified
    with  that between reversible and irreversible processes.  His
    interpretation of Carnot's theorem is rather that it states that
    Carnot cycles attain the maximum efficiency among all those
    reversible cycles where $\theta_1$ and $\theta_2$ are the
    \emph{extreme} temperatures;  see
    \cite[p.~117,168,303]{Truesdell} for details.
 }

Yet it seems to me that Carnot's work gives no indications of an
irreversibility of physical processes, in the sense of
irrecoverability discussed in section~\ref{pijl}. In the first
place, Carnot's theory does not imply the existence of
irreversible processes: his principle and theorem would remain
equally valid in a world were all cyclic processes have  maximum
efficiency. However, this is clearly not the world we live in.
Carnot explicitly acknowledged, that as a matter of fact,
irreversible cycles do exist, and that, moreover, it is rather the
reversible cycle which is an ideal that cannot be constructed in
reality.

Even so, even if we accept this fact,  it is not evident whether
these irreversible processes bring about irrecoverable changes.
 Carnot is concerned only with cycles. At the end of a cycle, all changes
in the system have become undone, even if the cyclic process was (in his
sense) irreversible. There is no question of a quantity of the system that
can only increase. The only option for find irreversible (irrecoverable)
changes must then lie in the environment, i.e.\ in the heat reservoirs
employed. But these are conceived of as buffers of fixed temperature,
\forget{s infinite heat capacity,} whose states do not change as a result
of the working of the engine.

I admit that it is possible to adopt a more liberal reading of the link
between Carnot's work and irreversibility.  The spontaneous flow of heat,
arising when two bodies of different temperatures come in thermal contact
constitutes, in his words, a `loss of motive power' (p.13, 14). One may
think that this denotes a loss in the course of time;  i.e.\ that during
the operation of an irreversible cycle, motive power somehow disappears.
In this reading ---adopted e.g.\ by Kelvin, and also by later commentators
e.g.~\citeasnoun{Brush}--- the power of the reservoirs to produce useful
work is decreased by irreversible cycles.  Irreversible cyclic processes
thus bring about irrecoverable changes: a `degradation' of energy from
useful to less useful forms.\footnote{%
    The problem is here, that  one would like to see this change in
    the environment reflected in the state of the heat reservoirs.
    There are specific cases were this is conceivable. For example,
    consider a case where the heat reservoirs are systems  of two
    phases, say a water/ice mixture and a liquid/solid paraffine
    mixture. These heat reservoirs maintain their fixed temperature,
    while at the same time, one can register the heat absorbed or lost
    by the heat reservoir by a shift of the boundary surface between
    the two phases. However, it is not easy to generalise this to
    arbitrary heat reservoirs.}

Apart from the fact that it is hard to make this reading precise, in view
of the construal of the reservoirs as unchanging buffers, there is to my
eyes a more natural explanation of these passages.  One can understand the
term `loss' as expressing only the counterfactual that if an ideal,
reversible machine
had been employed, a higher efficiency would have been achieved.
 There is only loss in an irreversible cyclic process in the sense that
the potential of the heat reservoirs to produce work has not been fully
exploited. We are then concerned with a comparison of the actual
irreversible cycle and another reversible cycle in a possible world, not
with irretrievable changes in this world.

But even if one accepts the liberal view, we still cannot say, in
my opinion, that this irreversibility is a {\em consequence\/} of
the theorem of Carnot.  Maybe a comparison with mechanics
clarifies the point. The first law of Newtonian mechanics states
that a free body persists in a state of uniform rectilinear
motion.  But free bodies are, just like the reversible cyclic
process of Carnot, only an idealisation.
 `Real' bodies, as is often said, always experience friction and do not
persevere in a state of uniform motion.  In fact, in the long run, they
lose their speed.  Here too, if one so desires, one can discern an
irreversibility or one-sided tendency of nature.\footnote{%
        One can even adduce the authority of Newton himself for this point
        of view: `Motion is much more apt to be lost than got and is
        always on the decay'; (cited by \cite[p. 23]{Price}).}
 But even so, it is clear that this view is an addition to, and not a
consequence of, Newton's first law. Similarly, the idea that the
reversible cycle is only an idealisation, and all actual cycles
are irreversible, is an addition to and not a consequence of
Carnot's theorem.

Another argument to the same effect is the following.  If Carnot's theory
implies irreversibility then this should also be the case when we actually
apply it to water mills. The theorem that all reversible water mills
operating between two given water levels have the same efficiency (and
that this efficiency is larger than that of any irreversible mill) can be
obtained by an analogous argument. But there are few authors willing to
draw the conclusion that there is an arrow of time in purely
mechanical/hydrodynamical systems; even if such a hydrodynamical arrow is
also not excluded by this theorem
 (e.g.\ the principle: `water always seeks the lowest level').

\section{Clausius and Kelvin \label{CK}}

 \subsection{The introduction of the second law}

 The main contributions towards the development of
thermodynamics are those by Kelvin (W. Thomson)  and
Clausius.\footnote{%
    Of course
    the work of several other authors was also highly
    significant, such as
    Rankine, Reech and Clapeyron.  And although
    I agree with those
    historians who argue that the role of these lesser-known authors
    is commonly underestimated in the traditional historiography of
    thermodynamics, I will not attempt to do justice to them.}
 Kelvin had noted in 1848 that Carnot's theorem allows the design of an
absolute scale for temperature, i.e.\ a scale that does not depend
on the properties of some special substance (water, mercury,
alcohol, the ideal gas). But at this time, he was still convinced
of the caloric view of heat which Carnot had adopted.

 The birth of the second law, or indeed of thermodynamics itself, is
usually located in an article by~\citeasnoun{Clausius50}.  In this
work one finds, for the first time, a clear rejection of the
conservation of heat, while the validity of Carnot's theorem is
maintained. Clausius showed that this theorem could also be
derived from another argument, in which the conservation of heat
was replaced by the equivalence principle of Mayer and Joule,
stating that from work heat can be produced, and vice versa, with
a universal conversion rate ($J$ =4.2 Nm/Cal). This is the `first
law' of thermodynamics.

In order to obtain Carnot's theorem the argument employing the
perpetuum mobile had to be adapted. Clausius' reasoning assumed
the impossibility of what we today call the {\em perpetuum mobile
of the second kind\/}: a periodically operating machine producing
no other effect but the transport of heat from a lower to a higher
temperature.

 The argument rests, just like Carnot's, on a reductio ad absurdum. If
Carnot's theorem were false, Clausius argues, we could build a combined
machine that works in a cycle and whose only effect would be that heat is
transported from a cold to a hot reservoir. But this would be absurd,
says Clausius, because:
 \begin{quote}\small \selectlanguage{german}
[\ldots] das widerspricht dem sonstigen Verhalten der W\"arme, indem sie
\"uberall das Bestreben zeigt, vorkommende Temperaturdifferenzen
auszugleichen und also aus den \emph{w\"armeren} K\"orpern in die
\emph{kaltern} uberzugehen\footnote{%
    `[\ldots]this contradicts the further behaviour of heat, since it
    everywhere shows a tendency to smoothen  any occurring temperature
    differences and therefore to pass from {\em hotter\/} to {\em
    colder\/} bodies.' }
\cite[p.~50]{Clausius64b}.
\forget{\footnote{`Dann w\"urden am Schlusse beiden K\"orper wieder in
ihrem
urspr\"unglichen Zustande sein; ferner w\"urden die erzeugte und die
verbrauchte Arbeit sich gerade abgehoben haben, und somit k\"onnte auch
nach dem fr\"uheren Grundsatze die Quantit\"at der W\"arme sich weder
vermehrt noch vermindert haben.  Nur im Bezug auf die \emph{Vertheilung}
der W\"arme w\"are ein Unterschied eingetreten, indem mehr W\"arme von $B$
nach $A$ als von $A$ nach $B$ gebracht w\"are, und somit im Ganzen ein
Uebergang von $B$ nach $A$ stattgefunden h\"atte.  Durch Wiederholung
dieser beiden abwechselnden Processe k\"onnte man also, ohne irgend eine
Kraftaufwand oder eine andere Ver\"anderungen, beliebig viel W\"arme aus
einem \emph{kalten} K\"orper in einen \emph{warmen} schaffen, und
das
widerspricht dem sonstigen Verhalten der W\"arme, indem sie \"uberall das
Bestreben zeigt, vorkommende Temperaturdifferenzen auszugleichen und also
aus den \emph{w\"armeren} K\"orpern in die \emph{kaltern} uberzugehen.'
}}
\end{quote}
 This particular statement of Clausius is often regarded as the first
formulation of the second law.  But, remarkably, Clausius offers the
statement more or less  \emph{en passant}, as if it were obvious,
and not as a new principle or law in the
theory.\footnote{%
       The passage is apparently so inconspicious that in a recent
       compilation of historical papers on the second law \cite{Kestin}
    this article by Clausius is abridged before
       the author had a chance to state his seminal contribution.
              }
 According to the view of this paper, there are indeed  two fundamental
laws (\emph{Grunds\"atzen}) for  the theory. But they are: (i) the
Joule-Mayer principle and (ii) a (somewhat obscure) formulation of
what he takes to be  Carnot's theorem: \begin{quote}\small Der
Erzeugung von Arbeit [entspricht] als Aequivalent ein blosser
Uebergang von W\"arme aus einem
warmen in einen kalten K\"orper\footnote{%
    The production of work [has]  as  its equivalent a mere
    transition of heat from a warm into a cold body.'}
(\cite[p.~48]{Clausius64b}).
\end{quote}
 In the context, he makes clear that this equivalence is intended to refer
to the maximum amount of work that can be produced in a cycle by a heat
transfer between two reservoirs of given temperatures. The previous
statement about the natural behaviour of heat is only an element in his
argument to establish this `\emph{zweiten Grundsatz}'.

Note that although Clausius' argument in order to establish this
theorem only deals with cyclic processes, his statement about the
natural behaviour of heat flow does not explicitly mention this
restriction (and nor does  his version of Carnot's theorem).  This
is our first indication that the second law might develop into
something more general.

  One year later, Kelvin (1851)  also accepted the validity of the first
law, and similarly sought to put Carnot's theorem on this new footing. In
his article \emph{On the Dynamical Theory of Heat} he paraphrased
Clausius' argument, and raised his incidental remark to an axiom:
 \begin{quote}\small
It is impossible for a self-acting machine, unaided by
any external agency, to convey heat from one body to another at a higher
temperature\cite[p.~181]{Kelvin}. \end{quote}
 He also formulated a
variant by means of which Carnot's theorem could likewise be obtained:
\begin{quote}\small
      It is impossible, by means of inanimate material agency, to
        derive mechanical effect from any portion of matter by
        cooling it below the temperature of the coldest of the
       surrounding objects (ibid.~p.~179).\end{quote}
 \forget{\footnote{%
       The new absolute temperature scale $T$ is related to the previous
       scale $\tau$ as $ \tau (\theta)= \ln T(\theta)$.}}

Either of these axioms allows one to derive what Kelvin calls
`the second fundamental  proposition' of the theory:
\begin{quote}\small
{\sc Prop.~II.} (Carnot and Clausius)   If an engine be such that, when
it is worked backwards, the physical and mechanical agencies in every
part of its motions are all reversed, it produces as much mechanical
effect as can be produced by any thermo-dynamic engine, with the same
temperatures  of source and refrigerator, from a given quantity of heat
(ibid.~p.178).
\end{quote}
This is a clear formulation of the first part of Carnot's theorem,
i.e.\ the part  pertaining to reversible cycles.\footnote{%
     Although Kelvin does not explicitly
     mention the restriction to cyclic processes, this restriction was
     intended.
     At the beginning of the article he writes: `Whenever in what
       follows, \emph{the work done or the
     mechanical effect produced} by a thermo-dynamic engine is mentioned
     without qualification, it must be understood that the mechanical
     effect produced, either in a non-varying machine, or in
      a complete cycle, or any number of complete cycles of a periodical
     engine, is meant.' \cite[p.~177]{Kelvin51}.\label{fnk}}
 In fact, Kelvin introduces this term here, referring to the condition mentioned above as the `condition of
complete reversibility'.

Kelvin then applies this proposition to an infinitesimal Carnot
cycle performed on an arbitrary fluid, where the temperature
varies between $\theta$ and $\theta + d\theta$, and the volume
between $V$ and $V+ dV$. He shows that the function (\ref{mu}) can
be written as
 \be \mu(\theta) =
\frac{1}{M(V,\theta)} \frac{\partial p(V, \theta)}{\partial\theta},
\label{latent} \ee
 where $M$ is the latent heat capacity.\footnote{%
    That is,  $M(V,\theta) dV$ is the amount of heat
    the system takes in when
    its volume is changed from $V$ to $V+dV$ at
    constant temperature $\theta$.}
 He calls this result the `complete expression' of `the second
    fundamental proposition' (ibid.~p.~187) and
 emphasises the remarkable fact that the right-hand side of (\ref{latent})
is the same for all substances at the same temperature.

Next, Kelvin considers Carnot cycles with a finite range of variation for
temperature and volume. He analyses these cycles into an infinite number
of cycles operating in an infinitesimal temperature range. Integrating the
above result, he obtains the following expression for the ratio of the
work produced by the engine and the heat supplied by the source (i.e.\ the
Carnot function):
 \[ C(\theta_1, \theta_2)= \frac {W}{Q} = J\left( 1 - \exp\left
(-\frac{1}{J}
\int_{\theta_1}^{\theta_2}
\mu(\theta) d\theta\right)\right). \]
Choosing the  absolute  temperature scale $T(\theta)$ such that
 \[
T(\theta) = \exp\frac{1}{J} \int^\theta_{\theta_0} \mu(\theta') d\theta'
\]
 (a step only taken by  Kelvin in 1854)
 and units such that $J=1$, the result takes the simpler and more
familiar form:
 \begin{equation}
  \frac {W}{Q} =   1 -  \frac{T(\theta_2)}{T(\theta_1)} =
  1-   \frac{T_2}{T_1}. \label{0}
\end{equation}
 The rest of his article is mainly devoted to an attempt to determine the
 values of  $\int \! \mu(\theta) d\theta $ from the steam tables collected
in the experiments by Regnault.

Thus, for Kelvin too, the `second fundamental proposition' of the
theory is still the Carnot theorem, or its  corollaries
(\ref{latent}) and (\ref{0}) for Carnot cycles.
\forget{\footnote{to be exact, Kelvin's favorite version of the
 second law is the differential version of (\ref{0}):
$\frac{dp}{d\theta} = JM\mu$, where $M$ is \ldots. }}
  The axioms only serve to derive these propositions.
 But today nomenclature has shifted. The two axioms are usually themselves
seen as versions of `the second law'. They are commonly presented as
follows
     (see e.g.\
        \cite{Born21,Zemansky1,Buchdahl}).
 \begin{quote}\small
 {\sc
Clausius' Principle:}
 It is impossible  to perform a  cyclic process
which has no other result than that  heat is absorbed from a reservoir
with a low temperature and emitted into a reservoir with a higher
temperature.

{\sc Kelvin's
Principle:} It is impossible to perform a  cyclic process
with no other result than  that heat  is absorbed from
a reservoir, and   work is performed.
\forget{The
        observation that this principle might fail
       is by Maxwell.
      Clausius introduces in 1850
        his version by in the argument from the absurdum
        assuming  nemen that two bodies, die both a  cyclic process
        doormaken  gebruikmakend of two heat reservoirs a different
        efficiency would possess. They would then elkaar aan
        kunnen drive,  and then:
        `At the end
       of the operations both bodies
       are in their original
       condition; further, the work produced will have exactly
       counterbalanced the work done, and therefore [\ldots]
         the quantity of heat can have neither increased nor
     diminished. The only change will occur in the {\em
      distribution\/} of heat, since more heat will be transferred
      from $B$ to $A$ [i.e.\ the two heat reservoirs] than from $A$ to
     $B$, and so on the whole
      heat will be transferred form $B$ to $A$. By repeating these two
      processes alternatively
     it would be possible to transfer as much heat as we please
     from a {\em cold\/} to a {\em hot\/} body, and this is not in
     accord with the other relations [intended is: behaviour (Verhalten)]
     of
     heat [\ldots]  since it
     always shows a tendency to equalize temperature differences
     and therefore to pass from {\em hotter\/} to {\em colder\/}
     bodies.'
     (Mendoza 1960, p.~134.)%
}
\end{quote}
 The most striking difference from the original formulation is obviously
that the explicit exclusion by Kelvin of living creatures has been
dropped. Another important point is that they are concerned only
with cyclic processes.
 It is not hard to devise examples in which heat is transmitted from a
lower to a higher temperature, or used up as work,
 when the condition that the system returns to its original state is
dropped.
  A further noteworthy point is the clause about `no other result'.
 A precise definition of this clause has always remained a difficult
issue, as we shall see in later sections.
Another question is the definition of a
heat reservoir.\footnote{Hatsopoulos and Keenan (1965, p.~xxv)
        \nocite{Keenan}
        argue that the definition of the concept  of a heat
    reservoir can only be given such a content that the second
        law becomes a tautology. Although I have doubts about this claim I
      agree that the question of the
        definition  of a heat reservoir is not trivial. The
       most natural conception seems to be
       that a heat reservoir is a system in thermal
       equilibrium  which can take in  or give off a finite
       amount of heat without
       changing its temperature or volume. This means that it must have
    an infinite  heat capacity.
        The question is then whether the thermodynamical  state of
        such a system changes if it absorbs or emits  heat,
        and how this  can be represented theoretically.
        That is, if an infinite heat reservoir exchanges
        a finite quantity of heat,  does its own state change
    or not?  \label{cap}
}

Kelvin already claimed that these two formulations of the second law were
 logically equivalent. An argument to this effect can be found in almost
all text books:  one shows that violation of one principle would
lead to the violation of the other, and vice versa. It took
three-quarters of a century before Ehrenfest-Afanassjewa
(1925,1956) \nocite{TEA25,TEA56} noticed that the two formulations
only become equivalent when we add an extra axiom to
thermodynamics, namely that all temperatures  have the same sign.
When we allow systems with negative absolute temperature ---and
there is no law in thermodynamics that disallows that--- one can
distinguish between these two formulations.  Her observation
became less academic when Ramsey (1956)\nocite{Ramsey} gave
concrete examples of physical
systems with negative absolute temperatures.\footnote{%
    \TEA{} argued that when we allow systems with both positive and
    negative temperatures the principle of Clausius, but not that of
    Kelvin is violated. At present, common opinion seems to be the
    opposite~\cite{Ramsey,Marvan}.}

With hindsight, it is easy to see that the two formulations are
not equivalent.  Clausius' principle makes recourse to the
distinction between low and high temperature. That is to say, his
formulation makes use of the idea that temperatures are ordered,
and it is therefore sensitive to our conventions about this
ordering. If, for example we replace $T$ by $-T$ the statement is
no longer true.  The modern formulation of Kelvin's principle on
the other hand only mentions the withdrawal of heat from a
reservoir and does not rely on the ordering of temperatures. This
principle is thus invariant under a change of conventions on this
topic.

How is it possible that so many books prove the equivalence of these two
formulations?  A short look at the argumentation makes clear where the
weak spot lies: one argues e.g.\ that the violation of  Clausius'
formulation implies the violation of Kelvin's formulation by coupling
the `anti-Clausius engine' to a normal Carnot cycle. Such a coupling is
assumed to be always possible without restriction.  The idea is
apparently that everything which has not been said to be impossible must
be possible. But in an unusual application (such as a
world in which negative temperatures occur) such an assumption is not at
all evident. \forget{Small wonder that mathematicians and other logically
educated readers confess problems with thermodynamical reasoning. }

However this may be, let us return to the main theme of our essay. What
are the implications of the second law for the arrow of time in the early
papers of Clausius and Kelvin?  If we consider their own original
statements (the `\emph{zweite Grundsatz}' of 1850 or the `second
fundamental proposition' of 1851), there is none.  For these are just
statements of the part of Carnot's theorem  concerning reversible
cycles. This part is time-symmetric.

But what if we take the more modern point of view that their formulation
of the second law is to be identified with Clausius' and Kelvin's
principle? We can largely repeat the earlier conclusions about the work of
Carnot. Both are explicitly time-asymmetric: they forbid the occurrence
of cyclic processes of which the time reversals are allowed. It is much
harder to connect them to the idea of irreversibility.
 Both versions refer exclusively to cyclic processes in which
there occur no irrecoverable changes in the system.
 The only option for finding such changes must lie in the
environment. But also in the work of Kelvin and Clausius it is not clear
how the environment can be described in thermodynamical terms.\footnote{%
       The conceptual problem that is created when
       the properties of the environment (a heat reservoir or
       perhaps the  whole universe) play a role  in the
       argument is ---with some sense of drama---  expressed by
       Truesdell:
     `This kind of argument
       [requires that ] \forget{does not provide a proof unless}
     properties of the
    environment are specified along with the properties of the
    bodies on which it acts. Here the environment is not
    described by [the theory], so there is no place in the
    formal argument  where such a proof [\ldots] could start.
        \forget{
        Arguments of this kind are common in presentations of
        classical  thermodynamics even today.} [\ldots] Mathematicians
        instinctively reject  such arguments, because they stand above
        logic. \forget{Earlier theories of physics [\ldots ] made no appeal
        to properties of a system larger than the one being treated.
        In classical mechanics, for example, nobody ever suggested
        that something could not happen to a body because otherwise
         the environment of that body might suffer!} [\ldots]
         This is the point in history where mathematics and physics,
        which had come together in the sixteenth century, began to
         part company'
 \cite[p.~98]{Truesdell}\label{tr}.}
 A connection with this aspect of the arrow of time is therefore simply
not present at this stage of the development of the second law.

Also the negative character of both formulations  gives rise
to this  conclusion. Brush observes that
 `it is clear that both
[Kelvin's and Clausius' principles] are negative statements and do
not assert any tendency toward irreversibility'
\cite[p.~571]{Brush}.
 The objection is here that these  versions of the law  would also be
valid in a world in  which all  cyclic  processes were reversible.
 \forget{Obviously here too is
 a somewhat more liberal reading.  The
Kelvin/Clausius-formulation  of the  second law expresses the
impossibility  of certain cyclic  processes.  The obvious
thought is that time reversed  cyclic  processes, where as only
result heat  from high to low  temperature  transported  {\em is
\/} possible.  But  here too the asymmetry strictly speaking
 follows from a addition to the second law. }

\section{From
the steam engine to the universe (and back again)\label{terug}}
\subsection{Universal dissipation}
 After the original introduction in 1850/1851 by Clausius and Kelvin the
second law underwent a number of transformations before it was
given the form in which we recognise it today, i.e.\ as the
entropy principle.
 A development which, indeed, is no less impressive than the psychological
development of Macbeth, where the loyal and rather credulous general
evolves into a suspicious and cruel tyrant. Here too, the metamorphosis
starts with the prophecy of a foul future.

In 1852 Kelvin proposed the view that there exists
a one-sided directedness in physical phenomena,
namely a `universal tendency in nature to the dissipation of mechanical
energy', and argued that this is a necessary consequence of his axiom.
 He expressed this tendency in the following words:
 \begin{quote}\small
I\@. When heat is created by a reversible
process (so that the mechanical energy thus spent may be \emph{restored}
to its primitive condition), there is also a transference from a cold body
to a hot body of a quantity of heat bearing to the quantity created a
definite proportion depending on the temperatures of the two bodies.

II\@. When heat is created by an unreversible process (such as
friction) there is a dissipation of mechanical energy, and a
full  \emph{restoration} of it to its primitive condition is
impossible.

III\@. When heat is diffused by \emph{conduction}, there is a
\emph{dissipation} of mechanical energy, and perfect
restoration is impossible.

IV\@. When radiant heat or light is absorbed, otherwise than
in vegetation, or in chemical action, there is a \emph{dissipation} of
mechanical energy, and perfect restoration
is impossible~\cite{Kelvin52}.
  \end{quote}
 He then considers the question how much energy is dissipated by
friction when steam is compressed in a narrow pipe, and estimates that
even in the best steam engines no less than 3/4 of the available motive
power is wasted. He draws from this and other unspecified `known facts
with reference to the mechanics of animal and vegetable bodies' the
conclusions:
\begin{quote}\small
Any {\em restoration\/} of mechanical energy, without more
than an equivalent of dissipation is impossible in inanimate material, and
probably never effected by organized matter, either endowed with vegetable
life or subjected to the will of an animated creature.

Within a finite period of time past, the earth must have been, and within
a finite time to come the earth must again be, unfit for the habitation of
man as presently constituted, unless operations have been, or are to be
performed, which are impossible under the laws to which the known
operations going on at present in the material world are
subject.
\end{quote}
 Here a number of important themes in the debate on the thermodynamical
arrow of time meet.  It is the first time in the history of
thermodynamics that a universal tendency of natural processes is
mentioned,  and attributed to the second law. Thus this law
obtains a cosmic validity and eschatological implication: the
universe is heading for what later became known as the `heat
death'.\footnote{%
     Parenthetically it may be remarked that Kelvin presented his
    conclusion in time-symmetric form:
     `\ldots must have been
     \ldots and must again be\ldots'. The idea is here probably
      that the temperature differences on earth were
       too large in the past and will be too small in the
     future to sustain life.}
 All but one of the aspects that make the second law so fascinating and
puzzling are present in this short paper, the only exception being the
concept of entropy.

At the same time the logic of Kelvin's argumentation is astonishing. Many
commentators have expressed their surprise at his far-reaching conclusions
about the fate of humankind immediately following his consideration of the
steam pipe.
 Further, his claim that the universal tendency towards dissipation
would be a `necessary consequence' of his axiom, is not supported with
any argument whatsoever.\footnote{%
         The only explanation for this omission I can think of is that
    Kelvin thought that the implication had already
        been demonstrated.  Indeed one finds
        in his earlier article of 1851 (\S 22) in discussing the case
        of a non-ideal machine the remark that the heat  is only
         partly used for a useful purpose, `the remainder
        being irrecoverably lost to man, and therefore
       ``wasted,'' although not {\em annihilated}.'
         Here the idea of irrecoverable dissipation is
        apparently already present.  A draft of this article is even more
    explicit about his belief in the universal directedness:
      `Everything in the material world is progressive'
        (\cite{Wise}).
        But here he does not connect this opinion with the second law.
       See also the passage in Kestin (p.~64 = Kelvin 1849).
    Recent historical work  suggests that Kelvin's
     view on dissipation is to be explained by his religious
    convictions \forget{rather than by
    scientific argument } \cite{OU,NW}. }
 Instead, he simply reinterprets Carnot's theorem as `Carnot's
proposition that there is an absolute waste of mechanical energy
available to man when heat is allowed to pass from one body to
another at a lower temperature, by any means not fulfilling his
criterion of a ``perfect thermo-dynamic engine'' '. Kelvin thus
apparently adopts the `liberal' reading of Carnot that we
discussed in section 4. His addition of the phrase `available to
man' blocks an otherwise reasonable reading of `waste' in terms of
a comparison between possible worlds.

  Note that  Kelvin now  uses the  terms
`reversible/unreversible'
in a sense which is  completely different from
 that of the `condition of reversibility' in his 1851 paper.
He does not consider cyclic processes  but instead processes in
which the final state differs from the initial state.  Such a
process is `unreversible' if the initial state  cannot be
completely recovered.
 A cyclic process is therefore by definition  reversible in the
present sense,  even if it is irreversible in the sense of Carnot.
 Obviously the necessary and sufficient criterion of Carnot for
reversibility is no longer applicable to Kelvin's 1852 usage of the
term.\footnote{%
     Note too that Kelvin does not consider the recovery of the state of
      the system but rather of the form of energy. The idea that
      irreversibility is a characteristic aspect of
      energy remained alive for a long time, e.g.\ in the shape of the
      principle of `degradation' of energy. It was more or less
    extinguished by  Planck.}

\subsection{The second law in mathematical, modified,
 analytical and extended form}
 In 1854 Kelvin\nocite{Kelvin54} published another instalment of
his \emph{Dynamical Theory of Heat}.
 Here he adopts the absolute temperature scale defined in terms of the
Carnot function, leading to the result (\ref{0})  for the Carnot process.
 If $Q_1$ and $Q_2$ denote the quantities of heat exchanged with the heat
reservoirs, with
$Q_1 = W+ Q_2$, we can write this  as
\begin{equation} \frac{Q_1}{T_1} - \frac{Q_2}{T_2} = 0 \label{1}
 \end{equation}
 Adopting the convention to take the sign of heat positive when heat
is taken in
by the system, and negative when it is emitted, this becomes
 \be \frac{Q_1}{T_1} + \frac{Q_2}{T_2} =0.\ee
 He then expands  the consideration to a
reversible cyclic
process of a system which can exchange heat  with an arbitrary number
of heat reservoirs, and obtains the result: \be
 \sum_i \frac{Q_i}{T_i} =0\label{som}.\ee
He concludes:\begin{quote}\small
   This equation may be regarded as the mathematical expression of
the second fundamental law of the dynamical theory of heat (p.~237).
\end{quote}
In a following instalment (part VII, from 1878)
he even calls the result
 (\ref{som}):
`the full expression of the Second Thermodynamic Law' \cite[p.~295]{Kelvin}.
\forget{
 [\ldots] is that if $Q_T$, $Q_{T'}$, \&c., denote the quantities of heat
emitted from a body when at temperatures $T$, $T'$ respectively, during
operations changing its physical state in any way, the sum $\sum
\frac{Q_T}{T} $ must vanish for any cycle of changes if each is of a
perfectly reversible character, and if at the end of all the body is
brought back to its primitive state in every respect.}

 Thus, once again, the second law in Kelvin's formulation, remains a
time-symmetric statement, that only pertains to reversible cycles. His
doctrine of universal dissipation apparently plays no role whatsoever!

  Clausius too developed his point of view. Also in 1854, he presented a
`modified version' of the second law (now called: \emph{Hauptsatz}).
  He put it in the form:
 \begin{quote}\small  \selectlanguage{german}
Es kann nie W\"arme aus einem k\"alteren in einen w\"armeren K\"orper
\"ubergehen, wenn nicht gleichzeitig eine andere damit zusammenh\"angende
Aenderung eintritt.\footnote{%
    `Heat cannot
    of itself pass from a colder to a hotter body without
    some other change, connected herewith, occurring at the same
    time.'}%
 \cite[p.~134]{Clausius64b}\end{quote}%
\forget{
 He considers the Carnot process for an ideal gas and shows that with
Kelvin's absolute temperature scale
the  universal efficiency  can be  written
 as
 \[ C(T_1, T_2)  = \frac{W}{Q_{\mbox{1}}}
= \frac{Q_{\mbox{1}} -  Q_{\mbox{2}}}
{Q_{\mbox{1}}} =\frac{T_1 - T_2} {T_1} \]
where  $Q_1$ and $Q_2$ are
 the absorbed and ejected amounts of heat and $T_1$, $T_2$ the
temperatures of the two heat reservoirs ($T_1>T_2$).
}
     The fact that Clausius offers this statement, which is closely
related to what he had already written in 1850, as
     a \emph{modified} formulation of the second law
     underlines that at that time he had not regarded this as a law.
Nevertheless, his present formulation is indeed modified: instead of a
sweeping but vague statement about the natural tendency of heat to flow
from a hot to a cold body, he now says that heat never flows from cold to
hot unless there is some accompanying change. Unfortunately, it remains
unclear what one should understand by such changes.

 He then considers, just like Kelvin, reversible cycles in which a system
exchanges heat with an arbitrary finite number of heat reservoirs of different
temperatures $T_i$ and obtains the equation (\ref{som})
by an analogous argument.\footnote{%
     The only distinction between Clausius and Kelvin is that
     at this time the former did not accept
     Kelvin's definition for absolute temperature and therefore uses an
     indefinite function $f(\theta)$ instead of $ T$.
     He restricts himself to a cyclic process for an ideal gas, and
     expresses as a conjecture that $f(\theta) \propto pV$.}
 Clausius calls $Q_i/T_i$ the `equivalence value'
(`\emph{Aequivalenzwerth}'), of the heat exchange, and he reads the
equation (\ref{som}) as expressing that the heat absorbed and ejected in a
Carnot process possess equal equivalence value.

 Clausius also discusses the case in which the heat reservoirs undergo a
temperature change during the cyclic process.  In this case he replaces
the sum by an integral:
 \begin{equation} \oint \frac{\dstreep Q}{T} = 0\label{2}\end{equation}
This is  his  `analytical expression' of  the  second law  for
reversible (\emph{umkehrbare})
cyclic processes.\footnote{%
    Clausius had no special notation for
    cyclic integrals or non-exact differentials  and wrote
    the left-hand side of (\ref{2}) as $\int \frac{dQ}{T}$.}

In this formulation, $T$ stands for the temperature of the heat reservoir
with which the system exchanges the heat $\dstreep Q$. But, because of
Carnot's criterion, the cyclic process is reversible if and only if the
heat reservoir and system have the same temperature during the exchange.
Thus, if the system has a uniform temperature, the integral can be
considered as referring to the system by itself, and no longer to
properties of the heat reservoirs.  \label{sys}

 At the end of this paper~\cite[p.~151]{Clausius64b}
 Clausius gives a brief treatment of irreversible (\emph{nicht
umkehrbare})  cyclic processes, for which case he obtains the
equation
 \begin{equation} \oint
\frac{\dstreep Q}{T}
\leq 0\label{3}. \end{equation}
        His argument is as follows:
        for an \emph{umkehrbar} cyclic process  the
        result (\ref{2})  rests on the argument that according to the
        modified version of the second law the integral
        cannot be positive. The reversed cyclic process, where the
     integral has the opposite sign, must also satisfy this condition,
     and the integral is therefore also not negative.
     Therefore it must vanish. In the case of the \emph{nicht umkehrbar}
    cyclic process the second part of this argument is
     not applicable, but the first part remains valid. Hence we obtain
(\ref{3}).

A further paper~\cite{Clausius62} presents what in his collected work is
referred to as the `extended form' of the second law.  Here, he studies
processes where the final state of the system differs from the initial
state. For convenience I will call these `open processes'. For this
purpose Clausius needs a number of assumptions about the possible change
of states of that system, and hence about its internal constitution. He
characterizes the state of the system by introducing two abstruse
quantities:
 the `\emph{vorhandene W\"arme}' $H$ and the `\emph{Disgregation}' $Z$.
The definition of these quantities is not very clear (Clausius merely
remarks about the \emph{Disgregation} that it represents a `degree of
distribution', which is related to
 the ordering of the molecules)  and for our purpose actually not very
important.\footnote{%
    Apparently, Clausius was inspired by, and aimed to
    improve upon, Rankine's 1853 formulation of thermodynamics, which
    adopted the quantity $H$
     (`actual heat') and a quantity $F$ which was intimately related
    to $Z$, known as the `heat potential' (see
        Hutchison, 1973). \nocite{Hutchison73}
        Unlike Clausius, however, Rankine employed an elaborate
    microscopic picture of molecular
    vortices in terms of which these functions could be defined.
    Nevertheless,
    the idea of separating entropy into two distinct quantities was
    not so weird as it may seem to modern
eyes. See~\citeasnoun{Klein69}
        for a clear exposition.
\forget{For the ideal gas e.g.\ we
    today write $\dstreep Q= TdS = c_V dT + R dV /V$. This can  be
    separated as $dH= c_V dT$; $dZ = R d \log V$.}}

I only mention that Clausius here considers infinitesimal pieces of an
open process and formulates the second law as:
 \be
 \frac{\dstreep Q + d H}{T} + dZ= 0 \ee
for\emph{umkehrbar}  and \be
 \frac{\dstreep Q + d H}{T} + dZ\geq 0 \ee for \emph{nicht umkehrbar}
processes.
He emphasizes  \cite[p.~244]{Clausius64b} that this extension of the
second
law rests on additional assumptions and does not follow from the earlier
versions.

More important for our purpose is that, now the limitation to cyclic
processes is dropped, Clausius has to be more explicit than before in
stating the criterion for what he means by the term
`\emph{umkehrbar}'.
 \begin{quote}\small
\selectlanguage{german}
\label{x}
Wenn die Anordnungs\"anderung in der Weise stattfindet, dass
dabei Kraft und Ge\-gen\-kraft gleich sind, so kann unter dem Einflusse
derselben Kr\"afte die Aenderung auch im umgekehrten Sinne geschehen.
Wenn aber eine Aenderung so stattfindet, dass dabei die \"uberwindende
Kraft gr\"osser ist als die \"uberwundene, so kann unter dem Einflusse
derselben Kr\"afte die Ver\"anderung nicht im umgekehrten Sinne geschehen.
Im ersteren Falle sagen wir, die Ver\-\"an\-de\-rung habe in
\emph{umkehrbarer}
Weise stattgefunden, im letzeren, sie habe in \emph{nicht umkehrbarer}
Weise stattgefunden.

Streng genommen muss die \"uberwindende Kraft immer st\"arker sein, als
die \"uberwundene; da aber die Kraft\"uberschuss keine bestimmte Gr\"osse
zu haben braucht,  so kann man ihn sich immer kleiner und kleiner
werdend denken,
so dass er sich dem Werthe  Null bis zu jedem beliebigen Grade n\"ahert.
Mann sieht daraus, dass der Fall, wo die Ver\"anderung in umkehrbarer
Weise stattfindet, ein Gr\"anzfall ist, den man zwar nie vollst\"andig
erreichen, dem man sich aber beliebig n\"ahern  kann\footnote{%
   `When a change of arrangement takes place in such a way  that
    force and counterforce are equal, the change can take place in the
    reverse direction also under the influence of the same forces.  But if
    a change takes place  in such a way that the overcoming force is greater
    than that which is
    overcome, the transformation cannot take place in the opposite
    direction under the influence of the same forces. We may say that the
    transformation has occurred in the first case in a {\em reversible\/}
    manner, and in the second case in an \emph{irreversible} manner. \\
    Strictly speaking, the overcoming force must always be more powerful
    than the force which it overcomes;  but as the excess of force is
    not required to have any assignable value, we may think of it as
    becoming continually smaller and smaller, so that its value may
    approach to nought as nearly as
    we please. Hence it may be seen that the case in which the
    transformation takes place reversibly is a limit which in reality is
    never reached but to which we can approach as nearly as we please.'
\forget{(Clausius, (Phil.\ Mag\.\  {\bf 24} 81-97 (1862); Kestin
p..~140.) } }
 \cite[p.~251]{Clausius64b}.
\end{quote}

  This definition is clearly related to, and in a certain sense a
sharpening of, the necessary and sufficient criterion of Carnot.
   For both authors the reversible process may be regarded as a limit of a
series of processes in which the  disturbance from the equilibrium
state become smaller and smaller. But Clausius' condition is more
stringent.
  Whereas Carnot only demanded equality of temperature for all bodies in
thermal contact,  Clausius demands equality for all kinds of `forces'.
 (Note that Clausius' concept of `force' is more or less Aristotelian. It
denotes any cause of change and includes temperature gradients).
Thus, his criterion demands also, e.g.\ in a compression process,
that the piston is pushed very gently, with a force which nearly
balances the pressure exerted by the gas.  Thus a Carnot process
is not necessarily reversible in Clausius' sense. Indeed, in an
experimental realisation of a Carnot process, adiabaticity of the
two adiabatic stages of the cycle is often secured
  by making them so  {\em fast\/} that the system has no chance
  to exchange heat with its environment.  The main difference with Carnot
is, however, that Clausius applies the criterion to open processes.

More importantly, Clausius' definition differs considerably from
 Kelvin's 1852 notion of reversibility. For Clausius, a process is called
reversible when it proceeds very gently. This is very close to what we
today call `quasi-static'.  Whether the initial state of such a process is
recoverable is another matter.  We shall return to this distinction
between Kelvin's notion of `reversible' and Clausius' \emph{`umkehrbar'}
below.

In his \cite{Clausius64a}, he embraced the idea that the second
law has implications for the direction of natural processes.  For
this occasion he adopts a more positive reading of this law:
 heat transport from bodies with high to bodies with low temperature
 can occur `by itself', but is not possible from low to high temperature
`without compensation'.
 These rather vague clauses are intended to express the same idea as the
phrase
 `without other associated changes' from 1854.
(See in particular the footnote   p.~134-5 in \cite{Clausius64b}.)
  The clause serves to exclude both the possibility of changes of states
in the environment as well as in the system itself (when it does not
perform a complete cycle).

He then proposes the view that his present formulation  of the
second law  expresses a universal tendency, that will end in the
heat death of the universe:
 \begin{quote}\selectlanguage{german}
 In diesen S\"atzen [\ldots] dr\"uckt sich eine allgemein in der Natur
obwaltende Tendenz zu Ver\"anderungen in einem bestimmten Sinne
aus. Wendet mann dieses auf das Weltall im ganzen an, so gelangt
man zu einer eigenth\"umlichen Schlu{\ss}folgerung, auf welche
zuerst W.Thomson aufmerksam machte, nachdem er [\ldots] sich
meiner Auffassung des zweiten Hauptsatzes angeschlossen hatte.
Wenn n\"amlich im Weltall [\ldots] die W\"arme stets das Bestreben
zeigt, ihre Ver\-theilung in der Weise zu \"andern da{\ss} dadurch
die bestehenden
 Temperaturdifferenzen ausgeglichen werden, so mu{\ss} sich das Weltall
allm\"ahlich mehr und mehr zu dem Zustand n\"ahern, wo die Kr\"afte keine
neuen Bewegungen mehr hervorbringen k\"onnen, und keine
Temperaturdifferenzen mehr existiren.''%
\footnote{%
    `These statements [\ldots] express a generally prevailing
    tendency
    in Nature towards changes in a definite sense. If one applies
    this to the universe in total, one reaches a remarkable
    conclusion, which was first pointed out by W. Thomson, after
    [\ldots] he had accepted my view of the second law. Namely, if,
    in the universe, heat always shows the endeavour to
    change its
    distribution in such a way that  existing temperature
    differences are thereby  smoothened, then the universe must continually get
    closer and closer to the state, where the forces cannot produce
    any new motions, and no further temperature differences exist.'
 }%
 \cite[p.~323]{Clausius64b}\end{quote}

 In his next paper~\cite{Clausius65} introduces the  concept of entropy.
  Again, he considers cyclic as
well as open processes. But this time, he does not resort to
hypothetical physical quantities. Instead he starts from the observation
that the relation (\ref{2}) implies that for an open \emph{umkehrbar}
process,
say from state $s_i$ to $s_f$, the integral
 \[ \int_{s_i}^{s_f} \frac{\dstreep Q}{T}  \]
 is independent of the integration path, i.e.\
depends only on the initial and final state.
 By a standard argument, one can show that this implies the existence
of a state function $S$ such that
 \begin{equation}
 \int_{s_i}^{s_f} \frac{\dstreep Q}{T} = S(s_f) - S(s_i) \label{3b}.
 \end{equation}
Thus, the equivalence value of a transformation can be determined
as the change of {entropy} between initial and final state.

Now Clausius considers a \emph{nicht umkehrbar} open process $\cal P$,
say, again,
from $s_i$ to $s_f$. He assumes that it can be closed into a cycle by some
{umkehrbar} process $\cal R$. from $s_f$ to $s_i$. For the cycle thus
obtained he uses the result (\ref{3}):
 \[  \oint
\frac{\dstreep Q}{T}
  = {\int_{s_i}^{s_f}}_{\cal P}
\frac{\dstreep Q}{T}
+ {\int_{s_f}^{s_i}}_{\cal R} \frac{\dstreep Q}{T}
\leq 0.\]
 For the reversible piece  $\cal R$ of the cycle one has
 \begin{equation}
 \int_{s_f}^{s_i} \frac{\dstreep Q}{T} = S(s_i) - S(s_f) \label{3c}.
 \end{equation}
Thus, for the \emph{nicht umkehrbar}  process $\cal P$%
\forget{from
 $s_i$ to $s_f$} one gets:
 \begin{equation} \int_{s_i}^{s_f} \frac{\dstreep Q}{T} \leq S(s_f) -
S(s_i) \label{4}.
 \end{equation}If  this  process is  adiabatic, i.e.\ if there is no
 heat exchange with the environment, we have $\dstreep Q =0$ for the
entire duration of the process and it follows that
 \begin{equation} S(s_f)
\geq S(s_i)\label{5}. \end{equation}
Hence we obtain:
\begin{quote}\small {\sc The Entropy Principle} (Clausius' version)
For every \emph{nicht umkehrbar} process in an adiabatically
isolated system which begins and ends in an  equilibrium state,
the entropy of the final state is greater than or equal to
 that of the initial state.  For every {umkehrbar} process in an
adiabatical system, the entropy of the final state is equal to
that of the initial state.  \end{quote} This is the first instance
of a formulation of  the second law as a statement about entropy
increase.
    Note that only the `$\geq$' sign is established for \emph{nicht
umkehrbar} processes.  One often
reads the stronger view that for irreversible processes the strict
inequality, i.e. with the `>' sign in (ref{5}), holds but this
has  no basis in Clausius' work.
 Note  also  that, in contrast to the
common view that the entropy principle obtains for isolated systems,
 Clausius' result applies to \emph{adiabatically} isolated systems.

Clausius concludes\begin{quote}\small \selectlanguage{german} Der
zweite Hauptsatz in der Gestalt, welche ich ihm gegeben habe, sagt
aus, dass alle in der Natur vorkommenden Verwandlungen in einem
gewissen Sinne, welche ich als den positiven angenommen habe, von
selbst, d.h.\ ohne Compensation, geschehen k\"onnen, dass sie aber
im entgegengesetzten, also negativen Sinne nur in der Weise
stattfinden k\"onnen, dass sie durch gleichzeitig stattfindende
positive Verwandlungen compensirt werden.  Die Anwendung dieses
Satzes auf das gesammte Weltall f\"uhrt zu einem Schlusse, auf den
zuerst W. Thomson aufmerksam gemacht hat [\ldots] Wenn n\"amlich
bei allen im Weltall vorkommenden Zustands\"anderungen die
Verwandlungen von einem bestimmten Sinne  diejenigen vom
entgegengesetzten Sinne an Gr\"osse \"ubertreffen, so muss die
Gesammtzustand des Weltalls sich immer mehr in jenem ersteren
Sinne \"andern, und das Weltall muss sich somit ohne Unterlass
einem
Grenzzustande n\"ahern.%
\footnote{
`The second law in the form  I have given it says that
all transformations  taking place in nature go by themselves
in a certain direction, which I have denominated the positive
direction. They can thus take place without compensation.
They can take place in the opposite direction, that is, the
negative, only when they are compensated at the same time by
positive transformations. The application of this law to the
universe leads to a conclusion to which W. Thomson first
called attention [\ldots]
namely, if in all changes of state in the universe the transformations in
one direction surpass in magnitude  those taking place in the opposite
direction, it follows that the total
state of the universe will
change continually in that direction and hence will
inevitably approach a limiting state.' }
\cite[p.~42]{Clausius67}\end{quote}
He  next notes that his theory is still not capable of treating the
phenomenon of heat radiation. Therefore, he `restricts himself'
---as he puts it--- to an
application  of the theory to the universe:
\begin{quote}\small
\selectlanguage{german}
  [\ldots] man [kann] die den beiden Haupts\"atzen der mechanischen
W\"armetheorie
entsprechenden Grundgesetze des Weltalls in folgender einfacher Form
aussprechen:\\
1.) Die Energie der Welt ist constant.\\
2.) Die Entropie der Welt strebt einem Maximum zu.\footnote{%
`One can express the fundamental
laws of the universe that correspond to the two main laws of
thermodynamics
in the following simple form:\\
1. The energy of the universe is constant.\\
2. The entropy of the universe tends to a
maximum.'
} (ibid.~p.~44)
\end{quote}
  These words of Clausius are among the most famous and most
often
quoted in the history of thermodynamics.
 Perhaps they are also the most controversial.  Even Planck, in many
regards a loyal disciple of Clausius, admitted that the entropy of the
universe is an undefined concept
 \cite[\S~135]{Planck}.
    For example, in order to define the entropy difference between two
states of a system we need the integral (\ref{3c}). But if that system is
the universe, it is unclear where the heat absorbed by the system might
come from.\nocite{VdWaalsKohnstamm}
 Van der Waals and Kohnstamm  (1927)
even argued that the universe
cannot be the subject of scientific study .
 Ironically, Clausius could have avoided this objection if he had not
`restricted' himself to the universe but generalised his formulation to
 an arbitrary adiabatically isolated system (but at least beginning and
ending in equilibrium).

 A more important objection, it seems to me, is that Clausius bases his
conclusion that the entropy increases in a \emph{nicht umkehrbar} process
on the assumption that such a process can be closed by an \emph{umkehrbar}
process to become a cycle.
 This is essential for the definition of the entropy difference between
the initial and final states. But the assumption is far from
obvious for a system more complex than an ideal gas, or for states
far from equilibrium, or for processes other than the simple
exchange of heat and work. Thus, the generalisation to `all
transformations occurring in Nature' is somewhat rash.

Another problem is what $T$ refers to in an \emph{nicht umkehrbar}
process. As noted above, in the integral (\ref{2}) this
temperature refers to the environment of the system (the
reservoirs with which it is in contact). In  an \emph{umkehrbar}
process the temperature of system and environment must be the
same, and one is allowed to consider $T$ as referring to the
system itself.
 But for arbitrary  processes we cannot take this
step. Moreover,
Clausius applies the integral to an adiabatically isolated system, i.e.\
one which does not interact with any reservoir.  Thus the $T$ in the
left-hand side of inequality (\ref{4}) is not properly defined.  This
paradox is somewhat mitigated by the fact that since $\dstreep Q
=0$, the value of $T$ does not matter anyway.

    On many occasions Clausius was criticised by
 his contemporaries.  I do not know if, in his own time, he was
criticised in particular for his famous formulation of the second
law
as the increase of the entropy of the universe.\footnote{%
         One can find some indications for this.
         Planck notes in his {\em Wissenschaftliche
         Selbstbiographie\/} \cite{Planck48} that
     prominent German physicists
     in the 1880s  rejected the application of the second law to
                       irreversible  processes.
         The book by \citeasnoun{Bertrand87} is
         also skeptical about the validity of the second
          law for irreversible (cyclic) processes:  `Je
          serai tr\'es bref sur les cycles irreversibles; les
          d\'emonstrations et les \'enonc\'es  m\^emes
       de leur
          propriet\'es me paraisent jusqu'ici manquer de
          rigueur et de precision'. (`I will be brief about irreversible
    cycles.  It appears to me that the demonstrations, and even the
    descriptions of their properties lack rigour and precision.' He
    discusses two favorable examples
           for the statement that
          `L'entropie de l'univers tend vers un maximum',
          but concludes: `Les examples [\ldots] n'autorisent pas a
          regarder le th\'eor\`eme general comme d\'emontr\'e. Il faudrai
          commencer par
          pr\'eciser l'\'enonc\'e, et, dans  beaucoup de cas,
          cela para\^\i t fort difficile.' (`Examples do not warrant
    regarding the general theorem as proved. One should start by
    making the statement more precise, and in many cases, that appears
    to be very difficult.'

     There are also more general
    complaints about the writings of Clausius.
          Mach writes: 
        `Die Darstellung von Clausius hat immer einen Zug von
         Feierlichkeit und Zur\"uckhaltung. Man weiss oft
         nicht ob Clausius mehr bem\"uht ist etwas
         mitzutheilen oder etwas zu
         verschweigen'~\cite{Mach}. (`The presentation by Clausius
always has a touch  of ceremoniousness and reservation.  One often does
not know whether Clausius is concerned more with communicating something
or with concealing  something.')
Maxwell too had
difficulty
        swallowing the work of Clausius:
         `My invincible ignorance of certain modes of thought has caused
        Clausius to disagree with me (in the digestive sense) so that I
        failed to boil him down and  he does not occupy the place in my book on
        heat to which his other virtues entitle him'
        (\cite[p.~222]{Garber}).
         }
 However, \citeasnoun[pp. 13-15, p.~260]{Kuhn} has pointed out
the remarkable fact that in the book \cite{Clausius76} he
eventually composed from his collected articles, every reference
to the entropy of the universe and even to the idea that entropy
never decreases in irreversible processes in adiabatically
isolated systems is deleted!  The most general formulation given
to the second law in this book, which may be regarded as the
mature presentation of Clausius' ideas, is again the relation
(\ref{3}), where the system is supposed to undergo a cycle,
 and entropy increase is out of the question.\footnote{%
        That is, of course, for the system itself.
        For the heat reservoirs this may be different.
        But since  Clausius' argument has the purpose
        of establishing the existence of the property to be
        called the `entropy' of the system,
        we cannot suppose without further ado that the
         reservoirs already possess entropies or even thermodynamical
         states.}

We must conclude that in the work of Clausius and Kelvin the connection
between the second law and irreversibility is extremely fragile.
 Kelvin
claimed that the irreversibility of all processes in nature is a
necessary consequence of his principle, but gave not a shred of
argumentation for this claim. His later versions of the second law
were even completely disconnected from the arrow of time.
 Clausius does give  argumentation, but it is so untransparent and
dependent on implicit assumptions that his famous general conclusion (all
processes in nature proceed in the `positive' direction, i.e.\ the
direction of entropy increase)  cannot be considered as established.

Further, we have noted that Clausius employs a definition of
`\emph{umkehrbar}' that largely coincides with  `quasistatic'.
This concept is very different from Kelvin's concept of irreversibility
(i.e the irrecoverability of the initial state).
  The question then arises whether the \emph{(Un)umkehrbarheit} of
processes (in the sense of Clausius)  has anything at all to do the arrow
of time. The deceptive nomenclature may make this seem self-evident. But
Clausius also explicitly draws such a connection:  in the quotation on
p.~\pageref{x} he claims that every \emph{umkehrbar} process can be
performed in the reversed direction, but a \emph{nicht umkehrbar} process
cannot, at least not under influence of the same forces.

 An example to the contrary was given
by~\citeasnoun[p.~17]{Sommerfeld}. Consider a charged condensor
which
is short-circuited by a resistance  submersed in a heat reservoir.
 When the resistance is very large the discharge will take place by an
arbitrarily small current, and negligible disturbance of electrostatic
equilibrium. Thus, such a process is \emph{umkehrbar} in the sense of
Clausius. The reverse process, however, is not allowed by the second law.

An example from relativistic mechanics shows that the converse is
also conceivable: an `\emph{unumkehrbar}' process of which the
reversal is allowed by the theory.
 In order to bring a rod into motion it must be  accelerated.
 In general this will bring about internal stress in the rod, depending on
its constitution, the point where the force is applied, etc. In order to
determine the relativistic length contraction, one considers a change of
velocity performed so slowly that at every moment the force remains
negligible, so that the rod remains almost in internal equilibrium.  This
is analogous to Clausius' criterion for an \emph{umkehrbar} change of
state. Further, the time reverse of this transformation (the Lorentz
transformation)  is again a Lorentz transformation.  Now consider the
analogue of what Clausius called an irreversible change:
   the measuring rod is suddenly put in motion, causing internal
deformations and shock waves in the rod. The length of this rod is not
described by the Lorentz transformation.  Still, the time reversal of this
process, i.e.~a sudden deceleration, is also dynamically possible, even
under the influence of the same external forces. Clearly the claim that
\emph{umkehrbar} processes can be reversed, but \emph{nicht umkehrbar}
 not, is not a tautological truth.

 At the same time it is undeniable that the idea of grounding the
irreversibility of processes in a law of nature is very suggestive
and attractive.
  Even if for Kelvin and Clausius the idea may have been nothing more than
a short flirtation, many later authors, starting with
\citeasnoun{Boltzmann72}
and
Gibbs (1875), have built upon Clausius' famous formulation of
the second
law as the tendency of the universe towards an entropy maximum.
 In particular  Planck  propagated the view  that  the essence of the
 second law lies in the principle of the increase of entropy.


An example of the confusion that entered thermodynamics as a result of the
confusing terminology is provided by the fate of a criticism by
\citeasnoun{Rankine}. Rankine denied the validity of Kelvin's dissipation
doctrine, in particular the claim that heat radiation is an example of an
irrecoverable process.
 His argument was straightforward. Imagine that mirrors are placed on a
huge sphere around the sun, that would reflect the solar radiation
billions of years after its emission. The radiation would reconcentrate
and reheat the sun to its original temperature, even after it had become
cold and extinct, and thus undo the apparently irreversible dissipation.

   A response to this objection came from
\citeasnoun{Clausius64a}.
Since the dissipation principle was seen as a necessary consequence of
the second law, Clausius understood the objection as an attack on
his own work.
  Clausius believed that Rankine wanted to propose a construction in which
a body which absorbs radiation could be made {\em hotter\/} than
the  bodies emitting the radiation.  This, of course, would be in
conflict with Clausius' claim about the natural behaviour of heat
flow and lead to a \emph{perpetuum mobile} of the second kind.  He
argued that such a construction is impossible.  But clearly
Clausius had not understood Rankine. The latter was concerned with
a \emph{recovery} of the original state,  not a \emph{perpetuum
mobile} of the second kind.
 That is, his intension was to make a radiating body \emph{just as hot} as
it originally was, by refocussing its own radiation.
  Clearly, Kelvin's dissipation principle and the second law in the form
of a \emph{perpetuum mobile} principle are not equivalent: the supposition
that the processes mentioned by Kelvin as examples of dissipation are in
fact reversible does not entail the possibility of a \emph{perpetuum
mobile} of the second kind.

\section{The attempt at clarification by Planck\label{Planck}}

If someone can be said to have  codified the second law, and
given it its definitive classical formulation,
that someone is Max Planck.
  His {\em Vorlesungen \"uber Thermodynamik\/} went through eleven
successive editions between 1897 and 1966 and represent \emph{the}
authoritative exposition of thermodynamics par excellence for the
first half of this century.\footnote{%
    But even the \emph{Vorlesungen} have not received unanimous
    acclaim. \citeasnoun[p.~328]{Truesdell68}
        describes the work as `gloomy murk',
        \citeasnoun[p.~142]{Khinchin} calls it an `aggregate of logical
and
        mathematical errors superimposed on a general confusion in the
        definition of the basic quantities'. \nocite{Khinchin}
        Still, apart from a review  by \citeasnoun{Orr}
    of the first English translation, I do not know of any attempt to
    analyse the arguments in this book in some detail.}
 It is no exaggeration to claim that all later writers on the topic have
been influenced by this book.  Planck puts the second law, the concepts of
entropy and irreversibility at the very centre of thermodynamics.  For
him, the second law says that for all processes taking place in nature the
total entropy of all systems involved increases, or, in a limiting case,
remains constant.
 In the first case these processes are irreversible, in the second case
reversible. Increase of entropy is therefore a necessary and sufficient
criterion for irreversibility.

Before Planck's work there were also alternative  views.
We have seen that Kelvin attributed irreversibility to
 processes involving special forms of energy conversion. This view on
irreversibility, which focuses on the `dissipation' or `degradation' of
energy instead of an increase in entropy was still in use at the
beginning of the century; see e.g.~\citeasnoun{Bryan}.   Planck's work
extinguished these views, by pointing out that mixing processes are
irreversible even though there is no energy being converted or degraded.

Because of the enormous historical influence of Planck's
formulation of the second law I will attempt to analyse his
arguments. However, this is not a simple task. The eleven editions
of the \emph{Vorlesungen} show considerable differences, in
particular in the addition of footnotes. Also, the last English
translation~\cite{Planck45} (of the seventh German edition of
1922) contains some unfortunate errors. Moreover, from the eighth
edition onwards, Planck replaced his argument with a completely
different one. Below, I will analyse the argument up to the
seventh edition, because this is presumably the most widely known
today.
 In section~\ref{revisit}, I will consider Planck's later revision of the
argument. Let us first distinguish the meaning of two concepts that Planck
uses.

\subsection{Planck's concept of \emph{Umkehrbarheit}}
We have already noted that the concept of  a reversible process was
used by Clausius and Kelvin with very different meanings.
In Planck's work we encounter a passage which is quite similar
 to  Clausius (1862), cited above on page~\pageref{x}:
 \begin{quote}\small
\selectlanguage{german}
Von besonderer theoretische Wichtigkeit sind
diejenigen thermo\-dyna\-mi\-schen Pro\-zesse, welche, wie man sagt,
unendlich langsam verlaufen, und daher aus lauter
Gleich\-ge\-wichts\-zust\"anden bestehen. W\"ortlich genommen ist
zwar diese Ausdrucksweise undeutlich, da ein Proze{\ss}
 notwendig Ver\"anderungen, also St\"o\-rung\-en des
Gleichgewichts zur Vor\-aus\-setzung hat. Aber man kann diese
St\"o\-rung\-en, wenn es nicht auf die Schnelligkeit, sondern nur
auf das Resultat der Ver\-\"an\-derungen ankommt, so klein nehmen
wie man irgend will, namentlich auch beliebig klein gegen die
\"ubrigen Gr\"o{\ss}en, welche im Zustand des betrachteten Systems
eine Rolle spielen. [\ldots ] Die hohe Bedeutung dieser
Betrachtungsweise besteht darin, da{\ss} man jeden ``unendlich
langsamen'' Proze{\ss} auch in entgegengesetzer Richtung
ausgef\"uhrt denken kann. Besteht n\"amlich  ein Proze{\ss} bis
auf minimale  Abweichungen aus lauter Gleichgewichtszust\"anden,
so gen\"ugt  offenbar immer eine ebenso minimal  passend
angebrachte \"{A}nderung, um ihn in entgegengesetzter Richtung
ablaufen zu lassen, und diese minimale   \"{A}nderung kann durch
einen Grenz\"{u}bergang
ebenso ganz zum verschwinden gebracht werden. (\S 71--73)\footnote{%
    `Of particular theoretical importance are those thermodynamical
    processes which, as one says, proceed infinitely slowly, and which,
    therefore, consist purely of equilibrium states. Strictly
    speaking, this terminology is unclear, because a process presupposes
    changes, i.e.\ disturbances of equilibrium. But when only the
    result of the changes matters, and not their speed, one can make
     these disturbances as small as one wishes, namely, arbitrarily
small compared with the other quantities which play a role in the
state of the considered system. [\ldots] The high significance of
this viewpoint is that one can think of every `infinitely slow'
process as being carried out in the opposite direction.  Indeed,
if a process consists purely of equilibrium states, then,
obviously, a minimal change, appropriately applied, will suffice
to make it proceed in the opposite direction, and this minimal
change can also be made to vanish by a limiting procedure.'}
  \end{quote} \selectlanguage{english}
Obviously,  Planck's `disturbance of equilibrium'
is intended to mean the same thing as
Clausius' `inequality of forces'.
In fact,
 Planck immediately proceeds to call his infinitely slow processes
\emph{umkehrbar}, just as  Clausius had done before.
Indeed, this name  turns out to be his favourite
and he uses the term  `infinitely slow'  only rarely.

On a closer reading there is a distinction between the passages
from Clausius and Planck. Consider a container filled with gas and
closed by a piston so tight that when it moves it experiences
friction with the walls of the container. When we compress the gas
extremely slowly, the force on the piston must be large enough to
overcome this friction;  but reversal of the process is not
physically possible under the same force, because friction always
opposes the motion. Such processes are not \emph{umkehrbar}
according to Clausius, even if they are performed infinitely
slowly, in contrast to the criterion of Planck (and Carnot).

However, one may wonder whether this distinction was intended by Planck.
His claim that  infinitely slow processes  can also be  performed in the
opposite direction with some suitable  minimal adaptions, which can be
as small as we wish,  suggests that the example just mentioned  would not
qualify as `infinitely slow'  or \emph{umkehrbar}.

Another difficult issue is how to judge when deviations of
equilibrium are small. How to compare e.g.\ a small variation in
the temperature of the whole system with a larger temperature
variation in a small part? It is obvious that there are many ways
to quantify the `disturbance' of equilibrium, and we cannot speak
of a well-defined limit by making the
disturbances smaller and smaller.\footnote{%
    For example,
    consider a container, half of which contains  gas, the other half
    being vacuum and partitioned by a large number $n$ of parallel membranes
    into tiny empty volumes. If one ruptures the membranes, one by one, one
    can let the gas expand  in $n$ steps, until it fills  the entire
    container. If we take $n$
    very large, so that at each rupture the gas expands over a tiny
        volume,
    and wait between ruptures until the gas attains equilibrium,
    there is a sense in which the disturbance from equilibrium is small at
    each step.  Still it would be undesirable to allow  this as an
    `infinitely slow' process; see e.g.~\citeasnoun[p.~99]{Callen}.}

\subsection{Planck's concept of reversibility}
 Planck  also defines  the term {\em reversibel}.
 \begin{quote}\small
 \selectlanguage{german}
  Ein Proze{\ss} der auf keine einzige Weise vollst\"andig
r\"uckg\"angig gemacht werden kann, hei{\ss}t ``irreversibel'', alle
andere Prozesse ``reversibel''. Damit ein Proze{\ss} irreversibel ist,
genugt es nicht, da{\ss} er sich nicht von selbst umkehrt,
 ---das ist auch bei vielen mechanische Prozessen der Fall, die nicht
irreversibel sind--- sondern es wird erfordert da{\ss} es selbst mit
Anwendung alle in der Natur vorhandenen Reagentien kein Mittel gibt, um,
wenn der Proze{\ss} abgelaufen ist,
 allenthalben genau den Anfangszustand wiederherzustellen, d.h.\ die
gesamte Natur in den Zustand zur\"{u}ckzubringen, die sie am Anfang des
Prozesses besa\ss.\footnote{%
    `A process which can in no way be completely undone is called
    ``irreversible'', all other processes ``reversible''. In order for
    a process to be irreversible, it is not sufficient that it does not
    reverse by it self, ---this is also the case for many mechanical
    processes, which are not irreversible--- rather, it is demanded
    that, once the process has taken place, there is no means,
     even by  applying all the
    agencies available in Nature, of restoring exactly  the
    complete initial state, i.e.\ to return the totality of Nature to
    the state which it had at the start of the process.'} (\S 112)
 \end{quote} \selectlanguage{english}
 `\emph{Reversibel}' denotes of possibility of
 undoing processes.
It deals with the recoverability of the initial state, and
is obviously closer to Kelvin's  (1852) `reversibility'
than  to Clausius' `\emph{Umkehrbarheit}'.
The unfortunate fact that  the English translation of
 Planck's work (but  also that of \citeasnoun{Clausius62})
uses {\em reversible\/} in both cases surely bears part of the
blame for the widespread confusion in the meaning of this term.
The English reader of Planck is faced with a curious text which
apparently ventures to define a term
 in \S 112 that has already been used on many previous occasions, but in a
different meaning.  In order to keep the distinction between these two
concepts, as well as with previous notions of irreversibility, I will in
the sequel denote Planck's concept of reversibility by his own phrase
`reversibel', and, for ease, treat it and its conjuncts as if it were an
English word.  (Thus I will also write `(ir)reversibelity', etc.)

Three remarks are in order.
  In the first place,
 Planck speaks about a complete recovery of the initial situation
in `die gesamte Natur'. This does not merely refer to the initial state
of the system. He emphasizes:
 \begin{quote}\selectlanguage{german}
 Die in dem Worte `vollst\"andig'
   ausgesprochene Bedingung soll nur die sein da{\ss}
   schlie{\ss}\-lich \"uberall wieder genau die bekannte
   Anfangszustand  [\ldots]
    hergestellt ist wozu auch notwendig geh\"ort da{\ss} alle etwa
   benutzten  Materialien und Apparate am Schlu{\ss} sich wieder
    in demselben Zustand befinden wie am Anfang, als man sie in Benutzung
      nahm.\footnote{%
    `The condition expressed in the word ``completely'' will be only
    this: that eventually the given initial state [\ldots] is
    restored everywhere, and this includes necessarily that all
    materials and
    apparatuses employed are returned to the same state as they
        occupied initially, before they were used.' } (\S 109)
 \end{quote}\selectlanguage{english}
 Obviously it is no mean feat  to restore the initial
 state everywhere in the `totality of Nature'. Suppose we
perform some process on a thermodynamical system.
In the meantime, the Earth
rotates, an atom on Sirius emits a photon, etc.
 Do we have to be able to undo all of this, before we can say that the
process is reversibel?  In that case Planck's completeness
condition would become grotesque.  It appears reasonable to assume that
the recovery is complete when we restrict the `gesamte Natur' to all
bodies that have interacted with the system in the original process. That
is to say, I will understand the clause mentioned by Planck (after `wozu
auch notwendig geh\"ort' in the above explanation of the completeness
condition as not only necessary but also sufficient.

Secondly, Planck emphasises that the way in which the initial
state is restored may be chosen freely; i.e.\ it is not necessary
that the system retraces every stage of the original process in
reverse order. Any procedure whatsoever that restores the initial
state will do.
 He says: `Was dabei an technischen Hilfsmitteln, Maschinen
mechanischer, thermischer, elektrischer Art verwendet wird, ist ganz
gleichg\"ultig'\footnote{%
    `Whatever technical resources, apparatuses of
    a mechanical, thermal, electrical nature are used here is
    completely indifferent.'} (\S 109).
 On many occasions \cite{Planck05,Planck48}, he emphasised the importance
of this aspect of his concept of {irreversibelity}. It implies that
the statement that a particular thermodynamical process is
{irreversibel} has consequences, not only for thermodynamical
processes, but for all types of interactions occurring in nature,
including even those not yet discovered. In this respect too the concept
`{reversibel}' differs from `\emph{umkehrbar}'!

In the third place, it should be noted that for Planck's criterion of
`{reversibelity}' it is necessary that recovery can be obtained with
 ``in der Natur vorhandenen Reagentien''.  That we might conceive of a
recovery process as in a thought experiment, (i.e.\ a process allowed by
the theory in some possible world) is not good enough for Planck.

 An example (given by Planck himself) of a {reversibel} process is
the motion of a harmonic oscillator. This system returns after
every period to its initial state without demanding any change
occurring anywhere else in nature. The motion is therefore
{reversibel}; it is not infinitely slow, because the deviation
from equilibrium (rest) is not
negligible.\footnote{%
      The harmonic oscillator can be realised as a thermodynamic system
       e.g.\ as a cylinder containing two ideal gases separated
      by an adiathermal frictionless piston. }

As another example, consider a non-periodic mechanical process, say a free
particle in uniform motion through empty space.  To bring it back to a
previous state, we must interfere with it, e.g.\ by means of a collision
with another body.  This will return the particle to its original
position. Then we must also reverse its velocity, e.g.\ by a collision
with a third body, to obtain the original mechanical state. But is this a
complete restoration?  The answer is not so obvious.  The body has gone
through at least two collisions, and thus exchanged momentum with two
other bodies in its environment.  In order to realize complete
restoration, this exchange must be undone.  One can do this, e.g., by
arranging a rigid connection between these two auxiliary bodies, e.g.\
they
are walls of a rigid box, so that the total momentum exchange vanishes.
Then, the particle simply bounces to and fro periodically, and we have
complete {reversibelity}.

Planck claims that all mechanical processes are, in fact,
{reversibel}.
But it is not clear to me whether they always comply with Planck's
condition of complete recoverability, especially if  one demands that
restoration should be achieved by `{in der Natur vorhandenen
Reagentien}'. It would not seem far-fetched to me,  if one argues
that, by Planck's criterion, the motion of the solar system is
{irreversibel}.

\forget{
Another example - not  given by Planck-  is from celestial mechanics.
Assuming the solar system
is not periodical, every constellation  of the planets occurs only once.
 To restore  some constellation, one has to disturb the system; e.g.\ by
collisions with eternal objects.
But then there is exchange of  (angular) momentum with these external
objects.

 Kelvin's idea from
 1874 to a uitdijende zwerm deeltjes terug te brengen tot zijn
initial configuration door ieder deeltje terug te kaatsen met een
`moleculair cricketbatje'. The feit dat het op deze wijze, mogelijk is,
de deeltjes exact terug te laten keren tot de
initial configuration, is vaak refered to als voorbeeld voor the reversibility
}

 These subtle aspects of Planck's concept of reversibelity have not
always been noticed.  The unfaithful English translation (although
sanctioned by the author)  surely contributed to the proliferation of
confusion.
  It is no surprise, therefore, to find Planck at the end of
his life complaining
about confusion on the true meaning of reversibelity:
\begin{quote}\small
 der Fehler, [\ldots] den ich mein ganzes Leben hindurch unerm\"udlich
be\-k\"ampft habe, ist, wie es scheint nicht auszurotten.
  Denn bis auf den heutigen Tag begegne ich statt der obigen
 Definition der Irreversibilit\"at der folgenden:
``Irreversibel ist ein Proze{\ss}, der nicht in umgekehrter
Richtung verlaufen kann.'' Das ist nicht ausreichend.
 Denn
von vornherein ist es sehr wohl denkbar, da{\ss} ein Proze\ss, der
nicht in umgekehrte Richtung verlaufen kann, auf irgendeine Weise
sich vollst\"andig r\"uckg\"angig machen l\"a{\ss}t.\footnote{`the
error which I have  battled against my entire life with tiring
appears to be inextinguishable. Untill this very day I  meet, in
stead of the above definition   of irreversibility the following:
``An irreversible process is one which cannot proceed in the
reverse order.'' This is insufficient. Indeed,  it is very well
conceivable that a process which cannot proceed in the reverse
direction can be fully undone in some other way.'}
 \cite[p.~10]{Planck48}
\end{quote}
\subsection{The  second law for ideal gases}

In part 3 of his book, Planck sets the aim of demonstrating that the
 second law in the form of the principle of increase of entropy follows
from Kelvin's principle.
 At this stage he has already announced that this proof
``bei dem heutigen Stande der Forschung nicht leicht sorgf\"altig
genug gef\"uhrt werden kann, da theils seine Allgemeing\"ultigheit
noch mehrfach bestritten, theils seine Bedeutung, auch von seinen
Anh\"angern, noch recht
verschieden beurtheilt wird.'\footnote{%
    `cannot easily be demonstrated
    carefully enough, at the present stage of research, partly because
    its general validity is sometimes denied and partly because even its
adherents interpret its meaning very differently.'} (\S 55).

This task is finally taken up in \S 106--136. For clarity, I have
organized the argument into a number of Lemmas.
 Consider  $n$ moles of ideal gas in a state of equilibrium,
characterized by the temperature $T$ and volume $V$.  Planck defines the
entropy of the gas straight away as a function of these equilibrium
states:
 \begin{equation}
 S(V,T) :=  n( c_V \log T  + R \log V  +  K)
 \label{6}
\end{equation}
 where $R$ is the gas constant and $c_V$ is the specific heat capacity at
constant volume.  The choice of the constant $K$ is arbitrary, as long as
it does not depend on $V$ and $T$.\footnote{However, $K$ may depend on $n$
and the units used for $V$ and $T$.}

  Planck shows
\begin{theorem}
 In every adiabatic \emph{umkehrbar} process performed on an ideal gas its
entropy $S$ remains constant. \end{theorem}
      Such a process can be approximated by a succession of equilibrium
states, and thus be represented as a curve in state space (i.e.\
$(T,V)$-diagram). For each infinitesimal element of such a curve one can
write
      $\dstreep Q =p dV + dU =0$.
      For an ideal gas one has, by definition,
      $pV =nRT$ and  $U= nc_V T$.
      Substitution gives: $\dstreep Q = n(RT \frac{dV}{V} + c_V dT)
      = TdS=0$, which proves the lemma.

   Next, Planck considers a system consisting of $N$ ideal gases in
separate containers.  Its state is characterised by the $2N$
variables: $s=(V_1, T_1, \ldots ,V_N T_N)$. The total entropy of
such a system is defined as
 \be
                    S_{\mbox{tot}}
(s) :=  \sum_i  n_i(c_{V i}
\log T_i+ R \log V_i+ K_i).
\ee
Planck shows (\S 121-- \S 123)

\begin{theorem}
In every adiabatic \emph{umkehrbar}
process performed on a
 system consisting of $N$ ideal gases,  which are connected by diathermal
walls and  remain in
thermal equilibrium, the total entropy $S_{\mbox{\rm tot}}$ remains
constant.
 \end{theorem}

This lemma is proven as follows: when the gases are connected by
diathermal walls, the condition of thermal equilibrium implies
that their temperatures are equal at each stage of the process:
$T_i= T_j=: T$. In an adiabatic process the gases can exchange
heat only with each other. If the amount of heat absorbed by gas
$i$ is $\dstreep Q_i$, one has
    \[ T\sum_i dS_i= \sum_i \dstreep Q_i= 0,\] which implies that
$S_{\mbox{tot}}$
 is constant.

  Combining the previous Lemmas, he then argues for
\begin{theorem}
 Every pair of states  $s,s'$ of a system consisting of $N$ ideal gases in
which the
total entropy is the same can be transformed into each other by means of
an adiabatic \emph{umkehrbar} process. \end{theorem}
 Proof: let $s = (V_1, T1,\ldots ,V_N, T_N)$ and $s'= (V'_1, T'_1\ldots
V'_N, T'_N)$ be two arbitrary states such that $ S_{\mbox{tot}} (s) =
S_{\mbox{tot}} (s')$. We first assume that each gas is adiabatically
isolated from the others. By \emph{umkehrbar} expansion or compression, we
can change the volumes $V_1,\ldots ,V_N$ to any desired set of values.
Since the entropies $S_i$ remain constant in such an expansion or
compression, the temperatures change and can also be made to attain any
desired set of (positive) values. In particular, we can perform a series
of adiabatic \emph{umkehrbar} expansions or compressions until all the
temperatures are equal.

   Next, one introduces a diathermal connection between the gases, while
the whole system remains adiabatically isolated.
Continuing
with \emph{umkehrbar} changes of volume, the gases will now exchange heat
and entropy,
while, according to lemma 2 the total entropy remains constant.
Perform such changes of volume until the entropies
 $S_i$ have attained the values $S'_i= S(V'_i, T'_i)$. At that point, one
removes the diathermal contacts, so that each gas becomes adiabatically
isolated as before. Finally we change the volumes again (adiabatically and
\emph{umkehrbar}) until they attain the values $V'_i$. Since the
entropies
$S'_i$ are conserved in this stage too (according to lemma 1), both the
volumes and the entropies of all gases are the same as in the state $s'$.
But then this holds for their temperatures too, and the final state is
identical to $s'$.  Thus we have constructed a series of adiabatic
\emph{umkehrbar} processes starting from $s$ and resulting in $s'$.

Up till here the development of the argument has been
straightforward. The only point worth mentioning is that the
argument is constructive and relies on the availability of
\emph{umkehrbar} adiabatic processes by which the  volume or  the
temperature can be made to attain any value desired. For the ideal
gas this assumption is of course unproblematic, but for more
general fluids it is not.

But now Planck argues (\S 122, 123:)
\begin{theorem}
All  processes considered in Lemmas 1, 2 and 3 are
{reversibel}.    \end{theorem}
 It is here that Planck's concept of reversibelity
enters into the argument. It is also here that the argument
becomes liable to confusion and misunderstanding.
   Planck's argument for this Lemma is exceedingly brief. Considering the
processes of
Lemma 2 (with $N=2$) he writes:
\begin{quote}\small
     Ein jeder derartiger mit den beiden Gasen ausgef\"uhrter Proze\ss{}
ist offenbar in allen Theilen reversibel, da er direkt  in
umgekehrter Richting ausgef\"uhrt werden kann, ohne in anderen
K\"orpern irchendwelche Ver\"anderungen zu hinterlassen.\footnote{
`Every process of this kind performed on the two gases is
obviously in all parts reversibel, because it can be performed
directly in the opposite direction without leaving any changes in
other bodies.'} (\S 122)
 \end{quote}
 The claim that such processes are `directly' and `in all
parts' {reversibel} obviously relies on the claim that every
`infinitely slow'  process can be performed in the opposite
direction after some minimal suitable adaptions.  But in order to
qualify  the process as {reversibel}, one needs a complete
restoration of the initial state of the system as well  as  its
environment.
 Planck's claim that the considered processes do not leave any changes in
other bodies is somewhat rash, because the argument up till now did not
pay any attention to the environment of the system.

 Perhaps worries about the environment of the system are most easily
expressed by formally assigning a state to the environment. We can then
denote the complete situation with a pair of states and represent a
process by a transformation (change of state)
  \begin{equation}
\langle s, Z \rangle \statechange{P} \langle s', Z' \rangle,
\label{c}
  \end{equation}
 where $s$ is the thermodynamical state of the system, and $Z$ the formal
state of the (relevant part of) the environment. A process $\cal P$ is
then {reversibel} just in case there exists a process $\cal P'$ which
produces the transformation: \be
 \langle s', Z' \rangle
\statechange{P'}
\langle s, Z \rangle.
\label{p'}
\ee
 Apparently, Planck assumed that the processes considered in the previous
Lemmas simply do not require any changes outside of the
system.\footnote{%
    It is clearly Planck's intention to consider such  interventions
        as
    the establishing or breaking of a diathermal connection as  operations
     requiring no or negligible effects on the environment.}
 That is, one can put $Z = Z'$ in (\ref{c}) and (\ref{p'}). In that case,
Lemma 4 would be an immediate consequence of the symmetry of Lemma 3 under
the interchange of $s$ and $s'$.

  However, the assumption is false.  The point is, of course, that an
adiabatic \emph{umkehrbar} change of volume involves work, and therefore
an exchange of energy with the environment. Something or somebody has
exchanged mechanical energy with the system and in order to call the
process {reversibel} there must be a restoration process which
returns that energy to its previous owner.

According to Orr (1904), it was Ogg,
 the translator of the first edition into English, who pointed this objection
out to Planck. In response, Planck included a couple of footnotes in the
second edition in which the matter is discussed further.
  \forget{He argues that the energy exchange can be achieved by means of
raising or lowering a weight. The change of height of this weight is to be
disregarded, because, as he says, it is not an internal change but only an
external one. Thus,
 }
Appended to the phrase `ohne in anderen K\"orpern irchendwelche
Ver\"anderungen zu hinterlassen'\footnote{`without leaving any
changes in other bodies'} of \S 122 quoted above, he adds the
footnote:
 \begin {quote}\small
Hier ist das Wort ``in'' zu beachten. Lagen\"anderungen starrer K\"orper
(z.B. Hebung oder Senkung von Gewichten) sind keine inneren \"Anderungen;
wohl aber der Tempe\-ra\-tur und der Dichte.\footnote{%
    `Here the word `in' must be emphasised.  Changes of place of rigid
    bodies (e.g.\ the raising or lowering of weights) are not internal
    changes; in contrast to changes of temperature or density.'}
 \cite[p.~89]{Planck2}
 \end{quote} and when the phrase
reappears one page later in the same paragraph we read the footnote:
 \begin{quote}\small
`Denn die Leistung der erforderlichen mechanischen Arbeiten
kann durch Heben oder Senkung von unver\"anderlichen Gewichten erfolgen,
bedingt also keine innere Ver\-\"ande\-rung.\footnote{5
    `Since  the mechanical  work
    needed here can be obtained by the raising or lowering of
    inalterable weights, this  does not presuppose any internal
changes.'}
 \end{quote} (In later editions the exact phrasing of these footnotes is
altered, but their essential content remains the same.) Clearly then,
Planck's strategy for avoiding the problem is to assume that any exchange
of work is done by means of weights, and that lifting or lowering weights
is not a relevant change in the environment because it is not `internal'.

  I want to make three remarks about this manoeuvre.
  First, it does not completely save Lemma 4, because the assumption
  is obviously special.  One
can also obtain  work by means of an electrical battery, by a combustion
engine, by muscle, etc. In all these cases the {reversibelity} of the
process is at least doubtful.
 Thus, the claim that \emph{every} adiabatic \emph{umkehrbar} process in a
system of ideal gases is {reversibel} is not proven.

Secondly, coupling a thermodynamical system to a weight obviously requires
the presence of a gravitational field. This is often regarded as
undesirable in thermodynamics.\footnote{%
                An obvious problem is that an ideal gas in a
            gravitational field is  no longer homogeneous with
            respect to pressure and density, and therefore, strictly
            speaking, not a fluid. Some thermodynamicists even argue
            that the notion of adiabatic isolation is applicable
        only when gravity is excluded \cite[p.~5]{Pippard}. }
 For this reason, \citeasnoun{Giles} proposed to replace
    the weight by a flywheel, as an alternative
    mechanical `work reservoir'.  Of course, one may wonder whether
         a change of angular velocity of a flywheel would be considered by
        Planck as an `internal' change or not.

  But the most important remark is that the way out of the objection
chosen by Planck seems completely at odds with what he had written before.
Just a few pages earlier, in his explanation of the completeness
requirement in his concept of {reversibelity}, Planck had explicitly
discussed a process where work is done on a system by means of descending
weights and heat is exchanged with a reservoir.
 To  call that process {reversibel}, we need to achieve the following
conditions:
\begin{quote}\small
so m\"u\ss{}te, damit der Proze\ss{} vollst\"andig r\"uckg\"angig
wird, dem Reservoir die empfangene W\"arme wieder entzogen und ferner
\emph {das Gewicht auf seine urspr\"ungliche  H\"ohe gebracht werden},
ohne
da\ss{} anderweitige  Ver\"anderungen  zur\"uckbleiben\footnote{%
     `in order for the process to become completely undone, the
    reservoir should give back the heat it received and \emph{the
    weight should be returned to its original heigth.}'}
(\S~110, emphasis added).
\end{quote}
   \forget{
    ``Verfahren [\ldots ] dass den Anfangszustand in der ganzen Natur genau
    wiederherstellt, d.h. \emph{die Gewichte wieder auf die ursprungliche
    H\"ohe schafft} [\ldots]und sonst keine Ver\"anderungen
    zur\"uckl\"a\ss{}t.'' }
 If we now decide that lowering or raising of a
weight is not really a relevant change of state at all, it seems puzzling,
to say the least, why one should insist that it is undone in a recovery
process.

  This leaves two options. Either one understands Planck's footnotes as
intending that any discussion of changes of bodies in the environment,
including the explanation of the concept of {reversibelity},
is to be understood as restricted to internal changes
  This would mean that one no longer requires the restoration of work done
on or by the system. This interpretation of Planck's intention was adopted
by Orr, who accused Planck of effectively using a different definition
than the one he had stated:
 \begin{quote}\small   \label{orr}
     It appears, then, that the enunciation of the propositions should
     be amended by changing the phrase ``without leaving changes in other
     bodies'' into ``without interchanging heat with other bodies", and
     that there should be a corresponding change in the definition of
     ``reversibility''. The definition which is \emph{used} by Planck
    appears in
     fact to be this, that a process is reversible (``reversibel'') if it
     is possible to pass the system back from the final state to the
     initial state without interchanging heat with external bodies (Orr,
     1904, p.~511).  \label{Orr}
\end{quote}
   However, Planck's reply (1905a)\nocite{Planck05} makes clear that
he  rejected this reading of his work.
\forget{\footnote{%
     If, on the other hand, we are liberal on this point too, and no
    longer insist that the complete recovery includes
        undoing the displacement of weights,
   then it is obvious  (*???*)  that every adiabatic \emph{umkehrbar}
process can be performed reversibly
 This reinterpretation of the term
   reversibility has been  proposed by  \citeasnoun{Orr}.
    intention.\label{verworpen}} }

The other option is that one sticks to Planck's original
definition of reversibelity, but allows for an exception in the
formulation of the Lemmas, whenever the phrase `ohne
zur\"uckbleibende \"Anderungen in anderen K\"orpern' or similar
words appear. I will choose  this second option, but for clarity,
will insert the exception explicitly in the formulation. Instead
of `without leaving changes in other bodies' I will speak of
processes which leave no changes in other bodies except the
possible displacement of a weight.

In order to bring this out in the notation, I will add the height of the
weight to the total state.  Thus the state of the environment is from now
on specified by the pair $\langle Z, h\rangle$.
A process $\cal P$ can then  be represented
as   a transition
\be\langle s, Z, h \rangle
\statechange{P}
 \langle s', Z' ,h'
\rangle,
 \ee and $\cal P$ is reversibel just in case there is another process
$\cal P'$ such that: \be \langle s', Z', h'\rangle \statechange{P'}
\langle s, Z,h \rangle.
 \ee

Thus I read Planck as establishing the lemma:
\\[1.2\baselineskip]
{\bf Lemma $\bf 4'$} \emph{All processes considered in Lemma 1, 2 and 3
which
do not leave any changes in other bodies except the displacement of a
weight  are {reversibel}.}\\[1.2\baselineskip]
This lemma follows from  the assumption that these processes do not leave
any changes in other bodies except the displacement of a weight, i.e.\
they are of the form
\be
\langle s, Z, h\rangle
\statechange{P}
\langle
s',
Z,  h'\rangle, \label{p} \ee
and, as shown by  Lemma 1 and 2, they obey
\[S(
s) = S(s'). \]

 The existence of a restoration process $\cal P'$
with
\[ \langle s', Z,{h'} \rangle
\statechange{P'}
\langle s, Z,h
\rangle \]
is now no longer trivial on grounds of the
symmetry of the premise in lemma 3. (This only entails the existence of a
process with $\langle s, Z, {h''}\rangle$ as final state.) But the
proposition is still true due to the conservation of energy.
  That is, every process which restores the original energy to the system
must also bring back the weight to its previous position.
 Note however,
that it is crucial here that the energy is delivered by a single weight.
When  two or more weights are employed, or more generally, if  their are
more  mechanical degrees of freedom in the environment
than conservation laws, this argument  fails.

Planck concludes this stage of his argument with:
\begin{theorem}
 Every pair of states of a system consisting of $N$ ideal gases in which
the total entropy is the same can be transformed into each other by a
{reversibel} process, without leaving any change in the environment,
except
the displacement of a weight.
 \end{theorem} This conclusion follows by application of Lemma 3.
Indeed, that Lemma showed that every two states of equal entropy
can be transformed into each other by means of a \emph{umkehrbar}
adiabatic process. When  this process is assumed to be of the form
(\ref{p}), Lemma $4'$ shows it is {reversibel}.

   Planck now (\S 118 and 124)  appeals to Kelvin's principle for the next
step in the argument:
 \begin{theorem}
 Adiabatic expansion of an ideal gas
without performance of   work
is an {irreversibel} process.
 \end{theorem} \label{pm}
 Adiabatic expansion without performance of work is a process in which
$T$ is constant and $V$ increases.
 \forget{It appears from  (\ref{6}) that  entropy increases in such a
process.}
 One can think of a gas expanding into a vacuum  after a partition has
been removed in a
two-chamber container. The process proceeds without requiring any change
in the environment.

 The lemma is arrived at by a reductio ad absurdum. Suppose the process were
{reversibel}. Then there is a process in which the expanded gas is
driven back into its initial volume, which similarly proceeds without
producing any changes in the environment.
  Planck argues that by means of this process one could construct a
\emph{perpetuum mobile} of the second kind.

 Let us represent the adiabatic expansion process without  performance
of work by
 \[ \langle s_i, Z,h \rangle  \statechange{P}
\langle s_f, Z,h
\rangle \]
 where  $s_i = (T_0,V_0)$, $s_f= (T_0, V_1)$ and $V_1 >V_0$,
and we have  assumed that
$\langle Z_i ,h_i\rangle=
\langle Z_f, h_f\rangle
=\langle Z, h\rangle$,
 i.e.\ the expansion occurs without any changes in the environment.
 Let the hypothesis be that this process is reversibel. Then there is
another process ${\cal P}'$ which produces the transition:
 \begin{equation} \langle s_f, Z,h \rangle
\statechange{P'}
\langle
s_i, Z, h \rangle.\label{7}\end{equation}
 The combination of
these two gives rise to a cycle: \[ \langle s_i, Z, h \rangle
\statechange{P}
\langle s_f, Z, h \rangle
\statechange{P'}
 \langle s_i,
Z, h \rangle \]
 which establishes complete recovery.  But this, of course, is not yet a
\emph{perpetuum mobile}.

  Planck's argument is therefore more subtle. He assumes that the same
hypothetical recovery process (\ref{7})  can also be combined with another
process, $\hat{\cal P}$ in which the gas expands isothermally \emph{with}
performance of
work and simultaneous heat transfer.
  This is a process in another environment, in which the gas is not
adiabatically isolated but rather in thermal contact with a heat
reservoir.
 Let the transition in this process be:
 \begin{equation} \langle
s_i,\hat{Z}_i, \hat{h}_i \rangle \statechangehat{P}
 \langle s_f, \hat{Z}_f,
\hat{h}_f
\rangle\label{8}. \end{equation}
 The final state of the environment $\langle \hat{Z}_f, \hat{h}_f\rangle$
differs from $\langle \hat{Z}_i, \hat{h}_i\rangle $ because the system has
absorbed heat from a heat reservoir and has done work by raising the
weight.
  In order to combine the
 process
(\ref{8}) with  the hypothetical process (\ref{7})
 into a cycle,
 the  final state of
 process (\ref{8}) must be equal to  the initial state of (\ref{7}).  We
 should  therefore assume that
$\hat{Z}_f = Z$ and
$\hat{h}_f = h$.
In that case, performing  the processes
(\ref{8}) and  (\ref{7}) one after another yields
 \[ \langle s_i, \hat{Z}_i , \hat{h}_i \rangle
\statechangehat{P}
 \langle s_f, \hat{Z}_f, \hat{h}_f \rangle =
 \langle s_f, {Z}, {h} \rangle \statechange{P'}
 \langle s_i, Z, \hat{h}_f\rangle =
\langle s_i,\hat{Z}_f ,\hat{h}_f \rangle
 \]
 and we have indeed constructed a \emph{perpetuum mobile} of the second
kind: the
system undergoes a cycle and the only effect on the environment is
conversion of heat into work.
 Thus, we see that a crucial assumption in the argument is that states of the
system and environment can be chosen independently.

 Planck argues next (\S 126) that: \begin{theorem}
 Every process in a system of gases in which entropy increases and which
does not leave any changes in the environment other than the displacement
of a weight is irreversibel. In other words, there is no process in which
the entropy of a system of ideal gases is decreased without leaving any
changes in the environment other than the displacement of a weight.
\end{theorem}

The argument again proceeds  by  a reductio ad absurdum.
Suppose there were two states $s$ and  $s'$ of the system
which could be joined by a process obeying the mentioned
conditions.
Thus, suppose there exists a process
 \[\langle s, Z,h \rangle\statechange{P'}
 \langle s', Z,{h'}
\rangle
\;\;\mbox{with}\;\;
S(s') < S(s). \]
 Let now $s''$ be a third   state of the system which
differs from $s$ only  in the sense that
one single gas has a smaller volume, and which has the same total entropy
as  $s'$.
That is, if the state $s$  of the system is:
 \[s= ( V_1,T_1, V_2, T_2, , \ldots
, V_N, T_N ),\]
the state $s''$  has the form, say,
 \[s''= ( V''_1, T_1, V_2, V_2 \ldots
, V_N, T_N),\] where \[ V_1'' =  V_1  \exp^{S_{\rm tot}(s') -
S_{\rm tot}(s))/ (n_1 R)},
 \]
so that $S_{\rm tot} (s'') =S_{\rm tot}(s)$.
According to  Lemma 3, there is
  an  adiabatic \emph{umkehrbar}
process which
 connects  $s'$ and  $s''$:
 \[ \langle s', Z,{h'} \rangle \statechange{Q}
\langle s'',
Z,{h''}
\rangle. \]
Performing $\cal P'$ and $\cal Q$  in succession
yields a process
\[ \langle s, Z,h\rangle \longrightarrow \langle s'', Z, h''\rangle\]
 in which the only changes are that a single ideal gas has reduced its
volume and a weight has been displaced.
  Since the energy of an ideal gas is independent of its volume, one
concludes $U(s) = U(s'') $ so that by energy conservation one has also $h
=h''$.
  This would be a process that brings about the complete recovery of the
adiabatic expansion of an ideal gas without performance of work.
The impossibility of this process has already been demonstrated by
Lemma 6.

His conclusion is now that equality  of entropy is not only a
sufficient but also a necessary condition for the reversibelity of
a  process if it proceeds without leaving changes
 in other bodies, except for the possible displacement of a weight.
\forget{
We see here that it is  crucial for the argument  that the state of  the
environment is represented by  a single
independent quantity.
If, for example,  there  were
two
weights, at  heights $h_1$ and  $h_2$,
we could only
conclude that $h_1+h_2= h''_1 + h''_2$, and  conservation of energy
would not
guarantee   a complete recovery
(i.e.: $h_1 =h''_1$ and
 $h_2 =h''_2$).
Natuurlijk is het dan mogelijk een veronderstelling toe te voegen dat
ieder pair of mechanische states of equal  total energy in
elkaar over te voeren, zonder verdere tussenkomst of thermodynamische
systems. But die assumption zelf is  al een groot deel of wat Planck
wil conclude
 nl.\ dat alle irreversibele processes are of a  thermodynamical of
nature.
}

\subsection{The second law for arbitrary systems}
 The above argument has still only yielded a formulation of the entropy
principle
for the ideal gas, whose entropy  was introduced by a conventional
definition.
 The question is then of course how to proceed for other systems. Planck
considers an arbitrary homogeneous system for which the thermodynamical
state is determined by {\em two\/} variables (say temperature and volume).
Such a system is often called a \emph{fluid}.
 By exchange of work and heat, the system can undergo cyclic processes,
either reversibelly or irreversibelly.
 Planck assumes that the heat exchange is
obtained by means of ideal gases, which act  as heat reservoirs.
 (He does not consider the exchange of work, but it is probably
easiest
to assume that it is again obtained by means of an auxiliary weight.)
 The relevant environment then consists, apart from the weight, only of
ideal gases, and this allows us, by means of definition (\ref{6}) to speak
about the entropy of the environment.

At the end of the cyclic process the fluid has returned to its initial
state; but the states of the heat reservoirs have changed:  at least one
of them has absorbed heat and another one has lost heat.  If $\dstreep Q$
is the amount of heat absorbed by the system during an infinitesimal
element of the cycle from the heat reservoir with temperature $T$ one has:
 \[ \oint \frac{dQ}{T} \leq 0\]
 In particular, if there are only two heat reservoirs involved, heat must
have flown from the hotter to the colder gas.

If the cycle is  \emph{umkehrbar} the  special case
\[ \oint \frac{\dstreep Q}{T} =0\]
obtains.
 This implies that $\dstreep Q/T$ is an exact differential, which we may
call $dS$.
 We can express  this differential in the state variables  of
the  fluid.  Since  the \emph{umkehrbar} process obeys
 $\dstreep Q = dU + pdV$, one obtains:
\be  dS = (dU + pdV)/T   \label{fluid}\ee
The function $S(T,V)$ cannot be written explicitly,
 if the
equation of state  for the fluid
(or rather: the equations  expressing
  $U$ and  $p$ as functions  of $V$ and  $T$) is unknown.
 But ---and this is the main point according to Planck--- one can still
conclude that for arbitrary fluids there exists some function $S$ with
properties analogous to (\ref{6})  for the ideal gas, which enable us to
repeat the proof of the previous Lemmas.\footnote{%
    Obviously, to extend the proof of Lemma 3 to an arbitrary fluid,
    one needs to assume that an ample choice of adiabatic
    \emph{umkehrbar} processes is available by which
    one can change its volume from any given value to any other
    desired value.  This is not self-evident.
     This tacit assumption is brought out explicitly in the
    formulation of \C{} (see section 9).}
 He is satisfied with stating the result:
\begin{quote}\selectlanguage{german} Es ist auf keinerlei Weise m\"oglich
die Entropie eines System von K\"orpern zu verkleinern, ohne da{\ss} in
andere K\"orpern Aenderungen zur\"uckbleiben.\footnote{`It is
in no way possible to decrease the entropy of a system of
bodies, without leaving changes in other bodies.'}  (\S 132)\end{quote}
This last
clause about other bodies is simply lifted
 by including these other bodies in the system. The conclusion is then:
\begin{quote}\small {\sc The Entropy Principle (Planck's version)}
\selectlanguage{german} \emph{Jeder in der Natur stattfindende
physikalische und chemische Proze{\ss} verl\"auft in der Art,
da{\ss} die Summe der Entropieen s\"amtlicher an dem Proze{\ss}
irgendwie betheiligten K\"orper vergr\"o{\ss}ert wird.} Im
Grenzfall, f\"ur reversible Prozesse, bleibt jene Summe
unge\"andert. Dies ist der allgemeinste Ausdruck des zweiten
Hauptsatzes der W\"arme\-theorie.  (\S 132)

[\ldots Es] ist hier ausdr\"ucklich zu betonen da{\ss} die hier
gegebene Form  [des zweiten Hauptsatzes] unter allen die einzige
ist, welche sich ohne jede Beschr\"ankung f\"ur jeden beliebige
endlichen Proze{\ss} aussprechen l\"a{\ss}t, und da{\ss} es daher
f\"ur  die Irreversibilit\"at eines Prozesses kein anderes
allgemeines Maass gibt als den Betrag der eingetretenen Vermehrung
der Entropie.\footnote{`\emph{Every physical or chemical process
occurring in nature proceeds in such a way that the sum of the
entropies of all  bodies which participate in any way in the
process is increased.} In the limiting case, for reversibel
processes, this sum remains unchanged.  This is the most general
expression of the second law of thermodynamics.\ldots  [It]  must
be explicitly emphasised that the formulation  [of the second law]
given here is the only one  of them all which can be stated
without any restriction, and that, therefore, there is no other
general measure for the irreversibelity of a process than the
amount of increase of entropy.'}
 ( \S 134) \end{quote}\selectlanguage{english}
Shortly  thereafter
 (\S 136)  Planck raises the question whether there are any restrictions
to the validity of the second law.  In principle, he recognises
two possible restrictions.  Either the starting point of his
argument could turn out to be false.  That is, a
 {\em perpetuum mobile\/} of the
 second kind  can be realised after all.
Or else, there might be a logical defect in his argumentation.
Planck  dismisses this last possibility light-heartedly.
It `erweist sich bei n\"aherer
Untersuchung als unstichhaltig'.\footnote{It `turns out, after closer
examination, to be untenable'.}
  On the former option only experience can give the final answer.  But
Planck is full of confidence.  He predicts that future
metaphysicians will assign the entropy principle a status even
higher than empirical facts, and recognise it as an {\em a
priori\/} truth.  The quotation from Eddington in
section~\ref{Edd} confirms that Planck was right about that.

\subsection{Evaluation}
Let us summarise the weak and strong aspects of Planck's argument.
 A good point is that Planck, by assuming that the heat reservoirs in the
environment of the system consist of ideal gases allows for an
explicit thermodynamical description  of their state.  Thus, in
contrast to previous approaches, it is now possible to conclude
that, at least in this case, if a system performs an
\emph{unumkehrbar} cycle, the entropy of its environment
increases.

Less good aspects are the following.
  When he wants to show that  one can assign an
entropy to arbitrary systems, Planck  restricts his  discussion to
`beliebige homogene K\"orper von der Art wie wir in \S 67 ff.\
betrachtet haben'\footnote{`arbitrary bodies of the kind we
considered in \S 67 and further.'} (\S 128). The text in \S 67
makes clear that this refers to
 fluids.
  In the course of the argument, this restriction is never mentioned again.
He simply refers to these systems   as  `\emph{K\"orper}'.
  But fluids are not to be confused with the arbitrary bodies mentioned in
the recurring phrase about `zur\"uckbleibende \"Anderungen in
anderen K\"orpern'\footnote{`remaining changes in other bodies'}.
These other bodies in the environment include heat reservoirs,
rigid bodies like stirrers and pistons, weights, and  `technische
Hilfsmittel, Maschinen mechanischer, thermischer, elektrischer
Art'\footnote{`technical devices, apparatuses of mechanical,
thermal, electrical nature'}  and maybe living creatures.

The step of regarding all such bodies in the environment simply as parts
of the thermodynamical system, ---without considering the question how
their entropy is to be defined--- does not appear very plausible:  when
these bodies are more complex than a fluid, or if they are not in
equilibrium
 or if their environment is more complex than a system of ideal gases,
  their entropy still remains undefined.%
\footnote{%
      Planck emphasises that the attribution of
      entropy is not restricted  to systems in equilibrium.
      But concrete applications of such a non-equilibrium entropy remain
      restricted to the remark that as long as a system is locally in
      equilibrium its total entropy can be identified with a sum or
      integral of local entropies.  But this is a meager harvest. For this
      application one still needs recourse to equilibrium states.
      For systems far from equilibrium this approach does not work.
 }

Another objection is that Planck's general formulation of the
second law states that the law is valid for arbitrary physical and
chemical processes. This is surprising. Only one page earlier
Planck  (rightly) emphasised that  the  expression (\ref{fluid})
for entropy could not be applied to chemical processes: `Denn von
\"Anderungen dieser Art [i.e.\  changes of
 mass or chemical composition] ist bei der Definition der Entropie nicht
die Rede gewesen (\S 131).'\footnote{`Because  changes of this
kind were not considered in the definition  of entropy'.}
 Indeed, the entropy
function (\ref{6}) is
defined up to a constant which may depend on the chemical nature of the
substance. How this restriction can suddenly be lifted remains
unclear.\footnote{In the seventh and later editions of the {\em
Vorlesungen\/} this problem is avoided by  simply dropping the
reservation about chemical processes!}

In my opinion, Planck's argumentation allows no more general
conclusion than the following statement.
 \begin{quote}\small
 For any system consisting of
fluids which are capable of exchanging work with the environment by means
of single mechanical coupling (a weight) and which can be placed at will
in a heat bath or in adiabatic isolation: if irreversibel
processes take place in the system whose final result is only a change of
volumes and/or temperatures of the system, leaving no changes in auxiliary
systems other than the displacement of the weight, its entropy increases.
And conversely, if entropy increases
 during such a process, it is irreversibel.\end{quote} There is no
argument that all natural processes are of this kind.  Examples mentioned
 by Planck such as mechanical friction are already outside this category.

Yet, there is another remarkable aspect of Planck's result worth
mentioning. His version of the entropy principle is not restricted
to adiabatically isolated systems, as in Clausius' version.
Instead, it applies to all processes performed by a system which
proceed under the condition that all auxiliary systems in the
environment which are employed during the process return to their
initial state, with the possible exception of a single weight.

 In one sense, this condition is much more general than the condition of
adiabatic isolation, because it allows for heat exchange between
the system and its environment. In another sense, it is more
restricted, because processes in adiabatic isolation may very well
proceed by interaction with auxiliary systems which do not return
to their initial state.  We shall see in a later section how Lieb
and Yngvason adopted Planck's condition to devise a new definition
of the term "adiabatic"

Conclusion: the goal of Planck's approach is to take the
phenomenon of irreversibility of natural processes as the
essential element of the second law. But his claim to have derived
a formulation of universal generality cannot withstand scrutiny.
Besides, this emphasis on irreversibelity and the universal
validity of the second law actually remains sterile in Planck's
own work.
 The final part of his book, which is devoted to applications of the
second law, only discusses equilibrium problems.

\section{Gibbs\label{Gibbs}}
 The work of Gibbs in thermodynamics (written in the years 1873-1878) is
very different from that of his European colleagues. Where Clausius,
Kelvin and Planck were primarily  concerned with processes, Gibbs
concentrates his efforts on a description of equilibrium states.
 He assumes that these states are completely characterised by a finite
number of state  variables like temperature, energy, pressure,
volume, entropy, chemical potentials etc.  He makes no effort to
prove the existence or uniqueness of these quantities from
empirical principles.

Gibbs proposes:
 \begin{quote}\small
{\sc  The Principle of Gibbs:}
 For the equilibrium of any isolated system it is necessary and sufficient
that in all possible variations of the state of the system which do not
alter its energy, the variation of its entropy shall either vanish or be
negative.  \cite[p.56]{Gibbs}
 \end{quote}
He writes this  necessary and sufficient condition as:
\[ (\delta S )_U  \leq 0\]

Actually, Gibbs did not claim that this statement presents a formulation
of the second law. But, intuitively speaking, the Gibbs principle, often
referred to as \emph{principle of maximal entropy}, does suggest a strong
association with the second law.
  Gibbs corroborates this suggestion by placing Clausius' famous words
(`Die Entropie der Welt strebt ein Maximum zu')  as a slogan above his
article.  Indeed, many later authors do regard the Gibbs principle as a
formulation of the second law.

\forget{\footnote{Note that Gibbs'
    principle, being restricted to isolated systems,
    is narrower than Clausius' entropy principle,
     which applies to all adiabatically isolated systems.}}

Gibbs claims that his principle can be seen as `an inference naturally
suggested by the general increase of entropy which accompanies the changes
occurring in any isolated material system'.  He gives a rather obscure
 argument for this inference.\footnote{\label{fg}%
        His argument that the principle is a sufficient condition for
        equilibrium contains the following passage:
      \begin{quote}
    Let us suppose [\ldots] that a system may have the greatest entropy
     consistent with its energy without being in equilibrium. In such a
     case, changes in the state of the system must take place, but these
     will necessarily be such that the energy and entropy will remain
     unchanged and the system will continue to satisfy the same condition,
     as initially, of having the greatest entropy consistent with its
     energy. Let us consider
     the change which takes place in any time so short that the change may
     be regarded as uniform in nature throughout that time. [\ldots] Now
     no change whatever in the state of the system, which does not alter
     the value of the energy, and which commences with the same state in
     which the system was supposed at the commencement of the short time
     considered, will cause an increase in entropy. Hence, it will
     generally be possible by some slight variation in the circumstances
     of the case to make all changes in the state of the system like or
     nearly like that which is supposed to actually occur, and not
     involving a change of energy, to involve a necessary decrease of
     entropy, which would render the change
     impossible.
       \end{quote}
   His argument that the condition (or actually the equivalent condition
    $(\delta U)_S \geq 0$) is necessary reads:
 \begin{quote}
     whenever an isolated system remains without change, if there is any
    infinitesimal variation in its state which would diminish its energy
   [\ldots] without altering the entropy [\ldots] this variation involves changes
   in the system which are prevented by its passive forces or analogous
   resistances to change. Now as the described variation in the state of the
   system diminishes its energy without altering its entropy, it must be
   regarded as theoretically possible to produce that variation by some
   process, perhaps a very indirect one, so as to favor the variation in
   question and equilibrium cannot subsist unless the variation is prevented
   by passive forces \cite[p.~59--61]{Gibbs}.\end{quote}}
 But fortunately there is no need for us to fret about the exact meaning
of his words, as we did in the case of Planck. His approach has
been followed by many later authors,
e.g.\.~\cite{Maxwell76,VdWaalsKohnstamm,Callen,Buchdahl}. We can
follow the lead of these Gibbsians and~\citeasnoun{Truesdell86}
about how the principle is to be understood.

The first point to note is that in this view the principle of
Gibbs is not literally to be seen as a criterion for equilibrium.
Indeed, this would make little sense because all states
characterisable by the state variables are equilibrium states.
Rather, it is to be understood as a criterion for {\em stable\/}
equilibrium. Second, the principle is interpreted as a variational
principle, analogous to other variational principles known in
physics such as the principle of least action, the principle of
virtual work, etc. Here, a `variation' is to be understood as a
comparison between two conceivable models or possible worlds
(i.e.\ states or processes). The variations are {\em virtual}.
That is to say, one should not think of them as (part of) a
process that proceeds in the course of time in one particular
world. Instead, the variational principle serves to decide which
of these possible worlds is physically admissible.

In the case of the principle of least action the compared worlds are
mechanical processes, and `admissible' means: `obeying the equations of
motion".  (The circumstance that the worlds are here themselves processes
of course immediate blocks the idea that variations could be considered as
processes too.)
 In the principle of virtual work, one considers mechanical states, and
`admissible' means: `being in mechanical equilibrium'. In the case of
Gibbs, similarly, the possible worlds are equilibrium states of a
thermodynamical system, and `admissible' means stable equilibrium.

According to this view, the principle of Gibbs tells us when a conceivable
equilibrium state is stable.  Such a proposition obviously has a modest
scope.
 In the first place,
 \forget{
 the statement does not apply to all states occurring
in nature.  Unstable equilibrium states can in fact be produced in the
laboratory (e.g.\ overheated liquids or undercooled vapours).
 Secondly,} Gibbs' principle is more restricted than previous statements
of the second law in the sense that it applies to systems which
are \emph{isolated} (i.e.\ no energy exchange is allowed) and not
merely adiabatically isolated.  But more importantly, of course,
it contains no information about evolutions in the course of time;
and a direction of natural processes, or a tendency towards
increasing entropy, cannot be
obtained from it.\footnote{%
    Another argument for the same conclusion is that Gibbs proposes
    another formulation of  his principle, which  he claims to be
    equivalent.  This is the principle of minimal energy, saying that in
    every variation which leaves the entropy of the system unaltered,
    the variation of energy should be positive or vanish:
      $(\delta U)_S \geq 0$. Does this express a tendency in Nature
    towards decrease of energy? }

To be sure, there is a long tradition in physics of regarding variational
principles as expressing a tendency or preference, or even purpose, in
Nature; see~\citeasnoun{YM}.
  For example, the principle of least action has often been explained as a
preference for efficiency.
 But even so it would be a mistake to interpret this as a statement
about evolution in the course of time.
 The principle of least action does not say that mechanical
processes tends to loose `action' during their course. Similarly, the
principle of maximal entropy is no basis for the idea that entropy will
increase as time goes by.

In fact, a description of processes is simply not available in the
approach of Gibbs.  Indeed, the resulting theory is sometimes called
\emph{thermostatics} (Van der Waals and Kohnstamm,
1927).\nocite{VdWaalsKohnstamm}
 Obviously, there are no implications for the arrow of time in the second
law as formulated by Gibbs.

Of course this view is not completely coincident with Gibbs' own
statements.  In some passages he clearly thinks of variations not
as virtual but as actual processes within a single world, as in
the quotation in footnote~\ref{fg}: `it must be regarded as
generally possible to produce that variation by some process'.
Some sort of connection between virtual variations and actual
processes is of course indispensable if one wants to maintain the
idea that this principle has implications for time evolutions.

\forget{In some physical problems these connections may seem almost
self-evident: if a system is not in stable equilibrium, say an undercooled
Van der Waals gas, isn't it natural to assume that the system will make a
transition to a more stable state?  Perhaps spontaneously, or else under
the influence of tiny perturbations from the environment, which one can
never completely prevent?
 However, the point is not whether such an assumption is natural, but
whether it is implied by the present formulation of the second law. One
can only conclude that this is not the case.
 }

 Probably the most elaborate
attempt to provide such a connection is the presentation by
\citeasnoun{Callen}. Here, it is assumed that, apart from its
actual state, a thermodynamic system is characterised by a number
of \emph{constraints}, determined by a macroscopic experimental
context. These constraints single out a particular subset $\cal C$
of $\Gamma$, consisting of states which are consistent with the
constraints. It is postulated that in stable equilibrium, the
entropy is maximal over all states allowed by the constraints.

 A process is then conceived of as being triggered by the cancellation of
one or more of these constraints. Examples are the mixing or
expansion of gases after the removal of a partition, loosening a
previously fixed piston, etc.  It is assumed that  such a process
sets in spontaneously, after the removal of a constraint.

Now, clearly, the set of possible states is always enlarged by the
removal of a constraint.  Hence, if we assume that the final state
of this process is again a stable equilibrium state, and thus
characterised by a maximum value for the entropy among all states
consistent with the remaining constraints, one concludes that
every process ends in a state of higher (or at best equal)
entropy.

I will not attempt to dissect the conceptual problems that this
view brings along, except for three remarks. First, the idea of
extending the description of a thermodynamical system in such a
way that, apart from its state, it is also characterised by a
constraint brings many conceptual problems. For if the actual
state is $s$, it is hard to see how the class of other states
contained in the same constraint set $\cal C$ is relevant to the
system. It seems that on this approach the state of a system does
not provide a complete description of its thermodynamical
properties.

Second, the picture emerging from Callen's approach is somewhat
 anthropomorphic. For example he writes, for the case
 that there are no constraints, i.e.\ ${\cal C} = \Gamma$, that `the
system is free to select any one of a number of states' \nocite{Callen}
(1960, p.~27). This sounds as if the system is somehow able to `probe' the
set $\cal C$ and chooses its own state from the options allowed by the
constraints.

Third, the established result that entropy increases in a process
from one equilibrium state to another, depends rather crucially on
the assumption that processes can be successfully modeled as the
removal of constraints. But, clearly, this assumption does not
apply to all natural processes. For instance, one can also trigger
a process by imposing additional constraints.  Hence, this
approach does not attain the universal validity of the entropy
principle, as in Planck's approach.

\section{Carath\'eodory \label{Car}}
 Constantin Carath\'eodory~\nocite{Cara09} was the first mathematician
to work on thermodynamics and to pursue its rigorous
formalisation. For this purpose he developed a new version of the
second law in 1909.
 Apparently, he had no revolutionary intentions in doing so. He emphasised
that his purpose was merely to elucidate the mathematical structure of the
theory, but that the physical content of his version of the second law was
intimately related to the formulation by Planck.  However, as we shall
see, his contribution was not received with a warm welcome, especially not
by Planck.

Before I consider this in more detail, I want to mention some further
merits of Carath\'eodory's work. In the first place, he is the first to
introduce the concept of `empirical temperature', before the treatment of
the first and second law.
 The empirical principle he proposed for this purpose was later baptised
as the zeroth law of thermodynamics (by Fowler). Also, \C's introduction
of the first law is superior to the flawed version by Planck (cf.\
footnote 3).
 Most modern textbooks use his formulation of these two laws, often
without mentioning his name. However, I will not discuss these aspects
of his work.

Carath\'eodory follows Gibbs in the idea that thermodynamics should be
construed as a theory of equilibrium states rather than (cyclic)
processes. A thermodynamical system is described by a space $\Gamma$
consisting of its possible states, which are represented by $n$ state
variables.
 It is assumed that this state space can be represented as a (subset of an)
$n$-dimensional manifold in which these thermodynamic state variables
serve as coordinates.
 Carath\'eodory assumes that the state space is equipped with the standard
Euclidean topology.
 However, metrical properties of the space do not play a role
 in the theory.  For example, it makes no sense to ask whether
  coordinate axes are orthogonal. Further, there is no preference for a
particular system of coordinates.\footnote{%
        Some authors~\cite{TH}, \cite[p.~118]{Truesdell86}
        raise the objection that \C's formulation  would demand the use
    of pressure and volume as coordinates for the state of a fluid.
        These are not always suitable.
        For example, water of about $4^o \rm C$ possesses physically
    distinct states with the same values of $(p,V)$.
        It is true that, at the beginning of his paper, \C{} chooses
    this pair of coordinates to represent the state of a system, but,
    as far as I can
    see, this is not essential to the theory. In fact he explicitly
        extends his treatment to general coordinates.  }

However the coordinates are not completely arbitrary.
 Carath\'eodory distinguishes between `thermal coordinates' and
`deformation coordinates'. (In typical applications, temperature or energy
are thermal coordinates, whereas volumes of the components of the system
are deformation coordinates.) The state of a thermodynamic system is
specified by both types of coordinates;  the `shape' ({\em Gestalt\/}) of
the system by the deformation coordinates alone.

 Although he does not mention this explicitly, it seems to be assumed that
the deformation coordinates remain meaningful in the description of the
system when the system is not in equilibrium, whereas the thermal
coordinates are generally defined only for equilibrium states. In any
case, it is assumed that one can obtain every  desired final shape
 from every initial state by means of an adiabatic process.

The idea is now to develop the theory in such a way that the second
law provides a characteristic mathematical structure of state space.
 The fundamental concept is a relation between pairs $(s,t)$ of states
that represents whether $t$ can be reached from $s$ in an adiabatic
process.\footnote{%
     A characteristic (but for our purpose not very important)
     aspect of the approach is that \C{} wishes to avoid the
    concept of `heat' as a primitive term.
     Therefore he gives a more cumbersome
     definition of the term `adiabatic process'.
     He calls a container adiabatic if the system  contained in it
     remains in equilibrium, regardless of
     what occurs in the environment, as long as the container is not moved
     nor changes its shape.
     Thus, the only way of inducing a process in a system contained in an
     adiabatic vessel  is by deformation of the walls of the vessel.
     Examples of such deformation are compression or expansion and also
     stirring  (the stirrer is also part of the walls).
       Next, a process is called adiabatic
     if it takes place while the system is adiabatically isolated,
     i.e.\ contained in an adiabatic container. }
 This relation is called \emph{adiabatic accessibility}, and I will
denote it, following \citeasnoun{LY}, by $s_1 \prec s_2$. This
notation of course suggests that the relation has the properties
of some kind of ordering. And indeed, given its intended physical
interpretation, such an assumption would be very natural. But \C{}
does not state or rely on this assumption anywhere in his paper.

In order to introduce the second law, \C{} starts from an
empirical claim: from an arbitrary given initial state it is not
possible to reach every final state by means of adiabatic
processes.  Moreover, such inaccessible final states can be found
in every neighbourhood of the initial state. However, he
immediately rejects this preliminary formulation, because it fails
to take into account the finite precision of physical experiments.
Therefore, he strengthens the claim by the idea that there must be
a small region surrounding the inaccessible state, consisting of
points which are also inaccessible.

 The second law thus receives the following formulation:
 \begin{quote}\small {\sc The Principle of Carath\'eodory:} In every  open
neighborhood $U_s \subset \Gamma$ of an arbitrarily chosen state
$s$ there are states $t$ such that for some open neighborhood
$U_t$  of $t$:
 all states $r$  within  $U_t$  cannot be reached adiabatically from
$s$.  Formally:
\be  \forall s \in \Gamma \; \forall  U_s \:
\exists  t \in U_s \mbox{ \& } \exists U_t \subset U_s   \:
\forall
r \in U_t \; : \: s
\nprec
r ,\label{cp}\ee
  where $U_s$ and $U_t$ denote open neighborhoods of $s$  and $t$.
 \end{quote}

 He then specialises his discussion to so-called `simple systems', obeying
four additional conditions.  In the first place, it is demanded that the
system has only a single independent thermal coordinate.  Physically
speaking, this means that the system has no internal adiabatically
separated subsystems since in that case it would have parts with two or
more independent temperatures.
 For a simple system the state can thus  be represented with coordinates
  $s=(x_0, \ldots, x_{n-1})$, where $x_0$ is, by convention, the thermal
coordinate.

 Secondly, it is demanded that for any given pair of an initial state and
final shape of the system there is more than one adiabatic process $\cal
P$ that connects them, differing in the amount of work done on the system
during the process.  For example, for a gas initially in any given state
one can obtain an arbitrary final value for its volume by adiabatic
expansion or compression.  This change of volume can proceed very slowly
or very fast, and these two procedures indeed differ in the amount of work
done.
 This assumption can also be found in the argument by Planck
 (see page~\pageref{pm}).

The third demand is that the amounts of work done in the processes just
mentioned form a connected interval.  In other words, if for a given
initial state and final shape there are adiabatic processes ${\cal
P}_1,{\cal P}_2$ connecting them, which deliver the work $W({\cal P}_1)$
and
$W({\cal P}_2)$ respectively, then there are also adiabatic processes
$\cal P$ with
any value of $W({\cal P})$, for $W({\cal P}_1) \leq W({\cal P}) \leq
W({\cal P}_2)$.

In order to formulate the fourth demand Carath\'eodory considers a
more special kind of adiabatic process.  He argues that one can
perform a process  starting in any given initial state and ending
with any given final shape, where the changes of the deformation
coordinates follow some prescribed continuous functions of  time:
 \be x_1(t), \ldots, x_{n-1}(t), \label{func}\ee
 Note that the system will in general not remain in equilibrium in
such a process, and therefore the behaviour of the thermal
coordinate $x_0$ remains unspecified.

Consider a series of such processes in which the velocity of the
deformation becomes infinitely slow, i.e.\ a series in which the
derivatives
 \[ \dot{x}_1(t), \ldots, \dot{x}_{n-1}(t).\]
 converge uniformly towards zero. Such a limit is called a
\emph{quasi-static change of state}.

For example, if the deformation coordinates  (\ref{func}) are
prescribed on the interval $0 \leq t \leq 1$, one can consider the
series of processes ${\cal P}_\lambda$, defined  on the time
intervals $[0, \lambda]$,  where the deformation coordinates
change as:
 \be x_1( \frac{t}{\lambda}), \ldots, x_{n-1}(\frac{t}{\lambda}),
\label{fnc}\ee
 with  $\lambda \rightarrow \infty$.

The fourth demand is now that in such a series of processes the
work done on the system converges to a uniquely determined value,
depending only on the given  initial state and  final shape, which
can be expressed as a time integral:
  \[  W =   \lim_{\lambda
\rightarrow \infty}
 W({\cal P}_\lambda)  = \int_{t_i}^{t_f} \dstreep W \]
 where
$\dstreep W$ denotes a differential form of the deformation coordinates:
 \[ \dstreep W = p_1 dx_1 + \cdots + p_n dx_n,\]
and $p_1, \ldots, p_n$
denote  some given functions on $\Gamma$, i.e.\ they may depend on $x_0,
\ldots,
x_{n-1}$.
 This value $W$ is the work done on the system in a quasi-static
adiabatic change of state.  Physically, this demand says that for
adiabatic processes, in the quasi-static limit, there is no internal
friction or hysteresis.

By means of \C's version of the first law (which I have not discussed
here), one can then show that
 \be  \dstreep W = dU, \label{c1}\ee
 and hence $W = U(s_f) - U( s_i)$, or in other words, the work done on the
system equals the energy difference between final and initial state.  This
means that for a quasistatic adiabatic change of state between a given
initial state and final shape the thermal coordinate of the final state is
also uniquely fixed. Since the choice of a final shape is arbitrary, this
holds also for all intermediate stages of the process.

 Thus, a quasistatic adiabatic change of state is represented by
a unique curve in $\Gamma$. That is to say, it is what Planck called an
`infinitely slow process'.  It represents a limit of processes performed
so slowly that the system can be considered as if it remains in
equilibrium for the whole duration of the process.
 \forget{\C{} does not
elaborate much on the nature of the perturbation, just like Planck. But he
is more specific by demanding that in this limit the derivatives of all
variables with respect to time vanish uniformly and that the amount of
work done during the process converges to a unique value.
}

With this concept of a `simple system'  he obtains:
 \begin{quote} {\sc  \C's Theorem:} For simple systems,
 Carath\'eodory's principle  is equivalent to the proposition that
the
differential form $\dstreep Q :
 =dU - \dstreep W
$ possesses an integrable
divisor, i.e.\ there exist functions $S$ and $T$ on the state space
$\Gamma$ such that
 \be \dstreep Q = TdS. \label{sc}\ee
 \end{quote}
 Thus, for simple systems, every equilibrium state can be assigned
 values
for entropy and absolute temperature. Obviously these functions are not
uniquely determined by the relation (\ref{sc}). Carath\'eodory discusses
further conditions to determine the choice of $T$ and $S$ up to a constant
of proportionality. However, I will not discuss this issue.

Because of \C{}'s first law, i.e.\ relation (\ref{c1}), the curves
representing quasi-static adiabatic changes of state are
characterised by the differential equation
 \[ \dstreep Q
 = 0, \]
and by virtue of (\ref{sc}) one can conclude that (if $T\neq 0$) these
curves lie on a hypersurface
\[ S(x_0, \ldots x_{n-1}) =\mbox{const}. \]
 Thus, for simple systems, the entropy remains constant in adiabatic
quasi-static changes of state.

 Next, \C{} argues that $T$ is suitable to serve as a thermal coordinate.
In such a coordinate frame, states with the same entropy differ only in
the values of the deformation coordinates, so that all these states are
mutually adiabatically accessible.

Before we proceed to the discussion of the relation of this
formulation with the arrow of time, I want to summarise a number
of strong and weak points of the approach. Undoubtedly, a major
advantage of the approach is that Carath\'eodory provides a
suitable mathematical formalism for the theory, and brings it in
line with other theories in modern physics.  The way this is done
is comparable to the development of relativity theory. There,
Einstein's original approach, which starts from empirical
principles like the invariance of the velocity of light, has been
replaced by an abstract geometrical structure, Minkowski
spacetime, where these empirical principles are incorporated in
local properties of the metric.
 Similarly, \C{} constructs an abstract state space where an empirical
statement of the second law is converted into a local topological
property. Furthermore, all coordinate systems are treated on the same
footing (as long as there is only one thermal coordinate, and they
generate the same topology).\footnote{%
    Indeed, as \LY{} have shown,  the analogy with
    relativity theory can be stretched even beyond this point.
    Let ${\cal F}_s = \{ t \;:\; s\prec t\}$ be the `forward cone' of
    $s$.
    This is similar to the definition of the future lightcone of a
    point $p$ in Minkowski spacetime which can similarly be
    characterised as the set of all points $q$ which are `causally
    accessible' from $p$.  \forget{This analogy might
                One would like, of course, that this set is a simply
        connected closed subset of state space.  If this condition is
        satisfied,}
        Thus, \C's principle implies that $s$ is always on the boundary of
    its own forward cone.}

Note further that the environment of
the system is never mentioned explicitly in his treatment of the theory.
This too is big conceptual advantage. Accordingly, nearly all attempts in
the subsequent literature to produce an axiomatic formalism for
thermodynamics take the work of Carath\'eodory as their point of
departure; e.g.~\cite{Giles,Boyling,Jauch,Hornix2,LY}.

It is also remarkable that in contrast to previous authors, \C{}
needs many special assumptions, which are packed into his concept
of a `simple' system, in order to obtain his theorem. The reason
for this distinction, is of course that \C aims to present a
formal theory, where the formalism decides  what is possible.
Thus, while Planck simply assumed without further ado that it is
possible to perform some required process, e.g.\ compressing or
expanding  a gas to any desired volume, this is because he took
`possible'  in sense (ii) of section 3 above.
 For him it suffices to observe that in the actual world provides the
means to do this. But for \C{} a process is possible if the
formalism allows it. For this purpose, the theoretical assumptions
which are needed to complete such arguments must be made explicit.
Again, this is an important advantage of \C{} approach.

But Carath\'eodory's work has also provoked less positive reactions among
thermodynamicists, in particular because of its high abstraction.  Many
complain that the absence of an explicit reference to a \emph{perpetuum
mobile} obscures the physical content of the second law. The complaint is
put as follows by Walter:
 \begin{quote}\small
 A student bursts into the study of his professor and
calls out: ``Dear professor, dear professor! I have discovered a perpetual
motion of the second kind!' The professor scarcely takes his eyes of his
book and curtly replies: ``Come back when you have found a neighborhood
$U$ of a state $x_0$ of such a kind that every $x\in U$ is connected with
$x_0$ by an adiabat (cited in \cite[p.~118]{Truesdell86}).\footnote{Note
    that Walter only states \C's preliminary version of his
    principle here.}
 \end{quote}
 The question has been raised (e.g.\ by \citeasnoun{Planck26}) whether the
principle of {\C} has any empirical content at all.  However,
\citeasnoun{Landsberg64} has shown that for simple systems Kelvin's
principle implies {\C}'s principle, so that any violation of the latter
would also be a violation of the former.

Other problems in {\C}'s approach concern the additional
assumptions needed implicitly or explicitly to obtain the result
(\ref{sc}). In the first place, we have seen that the result is
restricted to simple systems, a restriction which involves four
additional auxiliary conditions.  Even the definition of
quasi-static changes of states is confined to simple systems
alone.
 \citeasnoun{FJ}\label{FJ}
objected that the division of these five assumptions into four pertaining
to simple systems and one `Principle', intended to express a general law
of nature, seems ad hoc.  Indeed, the question whether \C's principle can
claim empirical support for non-simple systems still seems to be open.

  Secondly, there is an implicit assumption that thermodynamic state
variables can be used as differentiable coordinates on $\Gamma$. For
systems that possess phase transitions or critical states this assumption
is too strong. (This objection can obviously also be raised against other
approaches.) A generalization of Carath\'eodory's work, encompassing
certain non-simple systems (namely, systems composed of simple subsystems)
is given by \citeasnoun{Boyling}. A different elaboration, avoiding
assumptions of differentiability has been given by~\citeasnoun{LY}. This
is discussed in section 11.

In the third place, \citeasnoun{Bernstein} has pointed out technical
defects in the proof of \C's theorem. What \C{}'s principle actually
implies for simple systems is merely the \emph{local} existence of
functions $S$ and $T$ obeying (\ref{sc}). That is, for each state $s$
there is some environment $U_s$ in which one can find such functions.
$S_s, T_s$ 
 But this does not mean that there exists a single pair of functions, defined
globally on $\Gamma$, that obey (\ref{sc}). In fact, a purely local
proposition like \C{}'s principle is too weak to guarantee the existence
of a global entropy function.

\begin{figure}[t]
 \begin{center}
\begin{picture}(288,190)(10,20)
 \epsfig{file=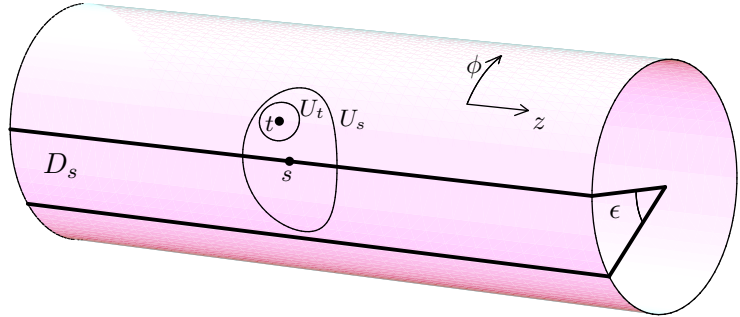}
\put(-180,137){\footnotesize $s$}
\put(-186,156){\footnotesize $t$}
\put(-158,158){\footnotesize $U_s$}
\put(-173,162){\footnotesize $U_t$}
\put(-85,157){\small $z$}
\put(- 110, 178){\small $\phi$}
\put(-56,123){$\epsilon$}
\put(-270,140){$D_s$}
\end{picture}
  \end{center}\vspace{-1 cm}
 \caption{%
\small \it
Carath\'edory's principle can be valid in a cylindrical phase space, even
though there is no global continuous entropy function.
           }
\end{figure}

        As a simple-minded counterexample, consider
the case where $\Gamma$ is the surface of a cylinder (see Figure
2),
        with coordinates $(\phi, z)$, $0\leq \phi<2\pi$. Let
    $z$ represent the deformation coordinate.
        For every point $s\in \Gamma$ let $D_s$ denote a narrow strip
         of points  just below $s$:
     $D_{s } := \{ t\in\Gamma \; :\;  \phi(s)-\epsilon < \phi(t)
    \leq\phi(s) \; (\mbox{mod } 2 \pi) \}$
     where $ \epsilon$ is a positive  number less than
     $2\pi$. Now  define the relation $\prec$ by:
    $s \prec t$ if and only if $ t\in D_s$. This relation
     obviously satisfies the principle  of \C.  Moreover
    the theorem of \C{} is also satisfied:
    adiabatic quasi-static changes  obey $ d\phi
    =0$, and this differential equation is integrable.
    Further,  for every point $s$,  one can find a
  continuous differentiable  coordinate function (namely $\phi$ itself,
   plus, if necessary, an appropriate constant)  such that locally,
    in a small  environment of $s$:
    $ s\nprec t$  if and only if $\phi(t) < \phi(s)$.
     But there is no continuous function that does this globally.
    In fact, the relation $\prec$  is  intransitive in this example, and
    every point can be reached from every other point by a \emph{series}
    of adiabatic changes of state.
   \citeasnoun{Boyling68}) has given a more sophisticated example of a
   two-dimensional  state space with ordinary (contractible)
    topology, which satisfies the principle of \C{}, without a global
   entropy function.
 \forget{(* Is $\prec$ in this case  transitive?*) }

 For the purpose of this essay, of course, we need to investigate
whether and how this work relates to the arrow of time.  We have
seen that \C{}, like Gibbs, conceives of \TD{} as a theory of
equilibrium states, rather than processes. But his concept of
`adiabatic accessibility' does refer to processes between
equilibrium states.  The connection with the arrow of time is
therefore more subtle than in the case of Gibbs.

\forget{%
 At some points in his argument  he uses  the term
`reversible'.  He writes  about the task a quasi-static curve in
state space to construct `Die Physikochemiker nennen es: a
Process reversible machen'' Here, the meaning of the term  is
closely
related to Planck's `omkeerbaar' or `infinitely slow'. As
argued before,  this  has little to do with  the arrow of time.
}

In \S 9 of \citeasnoun{Cara09}, he gives a discussion of the notorious
notion of
irreversibility. Consider, for a simple system, the class of all final
states $s'$ with a given shape $(x'_1, \ldots, x'_{n-1})$ that
are adiabatically accessible from a given initial state
 $s =(x_0, \ldots x_{n-1}) $.
 For example, an adiabatically isolated gas is expanded  from
some initial state $(T,V)$
to some desired final volume $V'$. The expansion may take place by
 moving a piston, slowly  or more or less suddenly.
 The  set of final states that can be reached in this fashion differ only
in the values of their thermal coordinate $x'_0$.  For this
coordinate one may  choose either energy $U$, temperature $T$  or
entropy $S$. For simple systems, (due to demand 3 above)  the
class of accessible final states constitutes a connected curve,
parameterised by an interval on the  $U$-axis. \C{} argues  that,
for reasons of continuity,  the values of $S$  attained on this
curve will also constitute a connected interval. Now among the
states of the considered  class there is the  final state, say
$t$,  of a quasi-static adiabatic change of state starting from
$s$. And we know that
 $S(s) =S(t)$.
He then claims that the entropy value $S(s)$ cannot be an internal
point of this interval. Indeed, if it were an internal point, then
there would exist a
 small interval  $( S(s) -\epsilon ,   S(s) +\epsilon )$
such that the corresponding states on the curve would all be
accessible from $s$. Moreover, it is always assumed  that we can
change the deformation coordinates  in an arbitrary fashion by
means of adiabatic state changes. By quasi-static adiabatic
changes of state we can even do this with constant entropy But
then, all states in a neighborhood of $s$ would be adiabatically
accessible, which violates  \C's principle.

Therefore, all final states with the final shape $(x'_1, \ldots,
x'_{n-1})$ that can be reached from the given point $s$ must have an
entropy in an interval of which $S(s)$ is a boundary point.  Or in other
words, they all lie one and the same side of the hypersurface $S=
\mbox{const}$.  By reasons of continuity he argues that this must be the
same side for all initial states.
 Whether this is the side where entropy is higher, or lower than that of
the initial state remains an open question. According to \C, a further
appeal to empirical experience is necessary to decide this issue.

 He concludes:
 \begin{quote}\small
\selectlanguage{german} [Es] folgt aus unseren Schl\"ussen,
da\ss{}, wenn bei irgend einer Zustands\-\"anderung der Wert der
Entropie nicht Konstant geblieben ist, keine adiabatische
Zustands\"anderung gefunden werden kann, welche das betrachtete
System aus seinem End- in seinem Anfangszustand \"uberzuf\"uhren
vermag.

\emph{Jede Zustand\"anderung, bei welcher der Wert
der Entropie variiert, ist ``irreversibel''}\footnote{%
    `[It] follows from our conclusions that, when for any change of
    state
    the value of the entropy has not remained constant, one can find
        no adiabatic change of state, which is capable of returning the
    considered system from its final state back to its initial state.
    \emph{Every change of state, for which the entropy varies is
    ``irreversible''.'}}\nocite{Cara09} (\C~1909, p.~378).
\end{quote} \selectlanguage{english}
Without doubt, this  conclusion sounds pleasing in the ears of anyone
who believes that irreversibility is the genuine  trademark of
the  second law.\footnote{%
    \C{} argues that  his formulation of the second
    law imply Gibbs' principle.  I'd rather not go into this.
 }
  But a few remarks are in order.

  `Irreversibel' here means that the change of state cannot be undone in
an {\em adiabatic\/} process. This is another meaning for the term,
different from those of Carnot, Kelvin, Clausius and Planck.  In fact,
this definition is identical to the proposal by Orr, discussed on
page~\pageref{orr}.
  \forget{`It appears, then, that the enunciation of the
     propositions should be amended by changing the phrase `without leaving
     changes in other bodies' into `without interchanging heat with other
     bodies' and that there should be a corresponding change in the
     definition of `irreversibility'.'' \cite[p.~512]{Orr}. Planck
    explicitly rejected this proposal\nocite{Planck05}.}
 The question is then of course whether changes of states that cannot be
undone by an adiabatic process, might perhaps be undone by some other
process. Indeed, it is not hard to find examples of this possibility:
  consider a container of ideal gas in thermal contact with a heat
reservoir. When the piston is pulled out quasi-statically, the gas
does work, while it takes in heat from the reservoir.  Its entropy
increases in this process, and the process would thus qualify as
irreversible in \C's sense.
 But Planck's book discusses this case as an example of a reversibel
process.  Indeed, when the gas is recompressed equally slowly, the heat is
restored to the reservoir and the initial state is recovered everywhere,
i.e.\ for both system and environment.   Thus, \C's concept of
`irreversibility' does not coincide with  Planck's.

The obvious next question is  whether such counterexamples can be
avoided by restricting the proposition to all \emph{adiabatic}
changes of state of a simple system in which the entropy varies.
But this does not solve the problem. Planck says explicitly in his
criterion for reversibelity that in the recovery process, any
auxiliary systems available may be employed. The system certainly
need not remain in an adiabatic container.  Even if the original
process were adiabatic, it would  remain reversibel as long as a
non-adiabatic recovery process can be found.
 There seems to be no guarantee that something like that is excluded in
\C's approach.

There is also another way to investigate whether \C's approach
captures the content of the second law \`a la Clausius, Kelvin or
Planck,
 namely by asking whether the approach of \C{} allows models in which
these formulations of the second law are invalid.
       An example is obtained by applying the formalism to a fluid while
swapping the meaning of terms in each of the three  pairs `heat
/work', `thermal/deformation coordinate' and `adiabatic'/`without
any exchange of work'.  The validity of \C's formalism is
invariant under this operation, and a fluid remains a simple
system. Indeed,  we obtain, as a direct analog of (\ref{sc}):
$\dstreep W =p dV$ for all quasi-static processes of a fluid. This
shows that, in the present interpretation, pressure and volume
play the role of temperature and entropy respectively.
Furthermore, irreversibility makes sense here too. For fluids with
positive pressure, one can increase the volume of a fluid without
doing work by expansion into a vacuum, but one cannot decrease
volume without doing work on the system.  But still, the analogues
of the principles of Clausius of Kelvin are false in this
application.  A fluid with low pressure can very well do positive
work on another fluid with high pressure by means of a lever or
hydraulic mechanism.

The next point worth remarking is that the conclusion of
Carath\'eodory is formulated as a time-symmetric statement: not
only an increase of entropy, but also a decrease cannot be undone
in an adiabatic process! As we shall discuss in section 10, Planck
criticised the approach by pointing out that a world where the
time reverse of Kelvin's principle holds, also obeys the principle
of \C{}.
  Although this does not mean that the principle of \C{} itself is time
symmetrical (that would mean that the time reversal of \emph{every}
possible world obeying the principle of \C{} obeys the same
principle\footnote{%
      In order to judge the time-symmetry of the theory of \C{}
      according to the criterion on page~\pageref{traf} it is necessary to
      specify a time reversal transformation $R$.  It seems natural to
      choose this in such a way that $Rs =s$ and $R( \mbox{$\prec$} )=
      \mbox{} \succ \mbox{} $.  (That is to say:  $s\prec t $ in ${\cal
      P}^*$ if $t\prec s $ in $\cal P$.)
      Then the theory is \emph{not} time-symmetric. Indeed, the
      principle of \C{} forbids that state space contains a `minimal
      state'
      (i.e.\ states $s$ for which $\exists U_s \, \forall t \in U_s\, :\,
      s\prec t $. It allows models where  state space possesses a
      `maximum', i.e.\ a state $s$ for which
      $\exists U_s \forall t \in U_s\,:\, t\prec s$.
      Time reversal of such a model is in conflict with the principle of
      \C.  However, this time-asymmetry manifests itself
    only in rather pathological cases.  (For a fluid, this would
      mean a local maximum for its temperature and volume.)
      If we exclude the existence of such maxima, \C's theory becomes time
      symmetric. Indeed a modern
      variation of the theory~\cite{Giles} has been given that is
      manifestly time-symmetric. (Giles calls this the
      `principle of duality'). The same goes for the formulation by  Lieb
         and Yngvason (see section~\ref{LY}).
      \label{R}}),
  according to Planck it is still not enough to characterise the direction
of irreversible processes. In fact \C{} admitted this point~\cite{Cara25}.
He stressed that an additional appeal to experience is necessary to
conclude that changes of entropy in adiabatic processes are always
positive (if $T>0$).
 In other words, in \C's approach this is not a consequence of the second
law.

Finally it is remarkable that the converse statement (i.e.\ that every
irreversible process in a simple system is accompanied by a change of
entropy) is not expressed. In this respect too the formulation of \C{} is
less far-reaching than Planck.

   \forget{30 juni}

\section{The debate between Born, Ehrenfest-Afanassjewa and Planck
\label{debat}}

  \C's work did not immediately have much impact on the physics
community.  Only twelve years later, when Max Born (1921)
formulated a popularised version of this work and explicitly
presented it as a critique of the traditional formulation of
thermodynamics, did the attention of the physicists awaken.

I first mention some of the  simplifications  introduced in this paper.
 In the first place, Born's formulation of \C's principle
is different:
 \begin{quote}{\sc  \C{}'s Principle  (Born's version)}:
            In every neighborhood of each state
         there are  states  that are inaccessible
         by means of adiabatic changes of state.
          In other words \[
\forall s \in \Gamma \; \forall \; U_s \:\;:\:\;
 \exists  t\in U_s \;  s \nprec t.\]
\end{quote}
In fact, this is the formulation of the principle which
    Carath\'eodory considered as a preliminary version,
 and then  rejected in favour of (\ref{cp}).
Nevertheless,
Born's formulation has since been adopted generally as {\em the\/}
statement of Carath\'eodory's principle. This is unfortunate because it
is evident that the statement by itself is inadequate for the derivation
of the result
(\ref{sc}).
      Indeed,  one's first association,  when reading that every
      neighborhood of a point contains points of another kind,
      is about the way in which rational numbers are imbedded
      in the real line.  In fact, if we call a real number $p$
      `adiabatically inaccessible' from number $q$ just in case
      $p-q$ is irrational, Born's version of Carath\'eodory's principle is
       satisfied for  $\Gamma = I\!\!R$.
      But clearly such a model is not intended at all.
      So  the formulation by Born does not suffice to obtain
      an interesting second law.  It presupposes additional tacit
      assumptions about the continuity of $\prec$.

Next, Born does not mention the restriction to simple systems, and
the subtle assumptions involved states bluntly that the approach
is applicable, without any problems, to `ganz beliebige Systeme,
wie sie die Thermodynamik zu betrachten
pflegt'\footnote{%
    `completely arbitrary systems, such as usually
    considered in thermodynamics'.}\cite[p.~162]{Born21}

  Further, instead of using \C{}'s definition of irreversibility, Born
calls a process \emph{reversibel} iff it is quasi-static:
\begin{quote} \small
Man leitet den Proze\ss{} unendlich langsam, derart, da\ss{} der
Zustand in jedem Momente als Gleichgewicht angesehen werden kann.
Man sollte solche Vorg\"ange \emph{quasi-statische} nennen, aber
man gebraucht gew\"ohnlich das Wort \emph{reversibel}, weil sie im
allgemeinen  die Eigenschaft haben umkehrbar zu sein. Wir wollen
hier auf die Bedingungen, unter denen das der Fall ist nicht
n\"aher eingehen, sondern annehmen da\ss{} sie erf\"ullt sind, und
werden beide Bezeichnungen als
synonym verwenden.\footnote{%
    `One conducts the process infinitely slowly,
      in such a way that the state at every moment can by regarded as an
equilibrium.   One should call such processes \emph{quasi-static},
but one usually employs the word \emph{reversible}, because, in
general, they have the property that they can be reversed. We do
not want to discuss the conditions under which this is the case,
but rather assume that they are fulfilled, and use both terms as
synonymous.'} \cite[p.~165]{Born21}
 \end{quote} That is, he
employs the term in the sense of Clausius' and Planck's
\emph{umkehrbar}.

The most striking point of Born's article is his claim that every
differential form defined on a two-dimensional state space has an
integrating divisor.  This provides a strong and elegant objection against
Planck's presentation of the second law, because it implies that the
existence of an entropy function for fluids, a topic which occupies a
substantial part of Planck's laborious argument, is in fact trivial; i.e.\
an empirical justification of this result is not needed at all!  Planck's
work thus appears to be an `attempt to crash through an open
door'~\cite[p.~207]{Kestin}.
  But this conclusion is not completely correct.\footnote{%
    It is true that differential forms in two dimensions always
       have integrating divisors. But these can still attain the value
      zero at some points.
      In such singular points the integral curves
       (i.e.\ the adiabats) can intersect.  (An example is
      the differential form $ydx - xdy$ in $I\!\!R^2$.)
         Kelvin's principle disallows
       the intersection of adiabats globally.
  The approach of Planck is thus not empirically empty, even for fluids. }

A different analysis was given by \TEA~(1925). Her interpretation of
\C{}'s work also contains a few curious alterations. For example, she
calls a process adiabatic in case there is no \emph{net} exchange of
heat with the environment. Thus, in this sense, a non-simple system, e.g.\
a composite system of gases in separate containers connected by
adiabatic pistons, undergoes an adiabatic process if one part absorbs
heat from a reservoir, while another ejects the same amount of heat to
some other reservoir.

This choice of terminology leads her to the view that \C's principle is
violated for non-simple systems.  All states in a neighborhood of a given
initial state can be `adiabatically' accessible for such a system.
Obviously, this view can be maintained consistently, and it inevitably
leads to the conclusion that \C's principle does not qualify as a law of
nature, but only as a special assumption for simple systems.
 The student mentioned in the quotation from Walter would not have to
search very far in order to surprise his professor!
 But since \C{} used the term `adiabatic' with a different meaning this
should not count as an objection against his work.

More important is that she correctly pointed out that \C{}'s principle
is also valid
in models where the time reverse of the principles of Kelvin or Clausius
hold, or in worlds where only reversible processes occur.
 She saw this as an important advantage:  it means one is able to
introduce the concepts of entropy and absolute temperature
and the equation (\ref{sc}), without being committed to a principle of
universal entropy increase or appealing to the existence of irreversible
processes.

Finally, Planck, the main representative of the tradition
criticised by Born, also responded \cite{Planck26}.
 He denied that Carath\'eodory's version of the second law could serve as
an adequate replacement of the principle of Kelvin and rejected it as a
`k\"unstliche und unn\"otige Komplikation
(an artificial and unnecessary  complication)'.  He advanced two
main objections.\footnote{%
    A third objection voiced by Planck is that in contrast to that of
    Kelvin, \C{}'s principle would have to be drastically reformulated
    when transposed into a statistical mechanical framework. He does
    not substantiate this claim, however.}

The first is that by speaking about arbitrarily small neighborhoods,
Carath\'eodory appeals to matters beyond the reach of observation. We
cannot possibly know, with our finite experimental faculties, whether
every neighborhood always contains adiabatically inaccessible states.
Hence the principle is speculative, and conflicts with the empiricist
guidelines along which thermodynamics ought to be developed, according to
Planck's point of view.

I don't think this objection is fair. It may be true that the connection
with experience is less manifest in Carath\'eodory's approach than in
that of Planck. But to conclude that it is therefore more speculative or
less reliable seems incorrect.  Planck too freely uses differential
calculus for thermodynamical quantities, presupposing that the state space
has the topological properties of a differential manifold.
  This is equally speculative: perhaps future experiments will teach us
that, on a very small scale, state space is discrete, or that it has some
other weird topology. The fact that Planck and previous authors ignored
these questions should not be mistaken for a sign of superiority by
empiricist standards.

The second objection is more important for our purpose. Planck writes:
   \forget{He claims that Carath\'eodory fails to draw a connection with
irreversibility of natural processes:
 \forget{As already noted by \TEA{}, Carath\'eodory's principle is
satisfied also in a world in which only processes occur in which entropy
decreases or remains constant.}  }
 \begin{quote}\small \selectlanguage{german}
 \ldots das
Prinzip [von \C] spricht nur von der Unerreichbarkeit gewisser
Nach\-bar\-zu\-st\"ande, es gibt aber kein Merkmal an, durch
welches die erreichbaren Nachbarzust\"ande von der unerreichbaren
Nachbarzust\"ande zu unterscheiden sind.  Mit anderen Worten: nach
dem Prinzip von \C{} k\"onnte es sehr wohl m\"oglich sein, W\"arme
ohne Kompensation in Arbeit zu verwandeln. Dann m\"u{\ss}te nur
der umgekehrte Vorgang, die kompensationslose Verwandlung von
Arbeit in W\"arme, als unm\"oglich angenommen werden. Oder es
k\"onnte auch sein, da\ss{} beide Arten von Verwandlung
unm\"oglich w\"aren.  Es ist selbstverst\"andlich, da\ss{} auf
einer solchen Grundlage der zweite W\"armesatz nicht vollst\"andig
aufgebaut werden kann und da\ss {} f\"ur diesen Zweck dem durch
das \C{}sche Prinzip ausgedr\"uckten Axiom noch ein zweites von
jenem unabh\"angiges Axiom, das sich auf irreversible Vorg\"ange
bezieht, hinzugef\"ugt werden
mu\ss\ldots.\footnote{%
    `\ldots  the principle [of \C{}]  speaks only of the
    inaccesibility of certain neighbouring states, but it provides no
    mark  by  which  the accessible states can be distinguished from the
    inaccessible states.  In other words, according to \C's principle
    it could very well be possible to  transform heat into work
    without compensation.  One only needs to assume that the reverse
    process, i.e.\ the compensationless transformation of work into
heat were
    impossible.  It is obvious that the second law cannot be built
    completely on this foundation and that for this purpose one needs the
    addition of a second, independent axiom which refers to irreversible
processes.'}
  \cite[p.455]{Planck26}
 \end{quote}
The  observation that Carath\'eodory's principle is neutral with
respect to the irreversibility of natural processes had  already
been noted by \TEA\@. But whereas   she thought of  this as a
major advantage, in the eyes of Planck it is a serious defect. He
diagnoses the theory of Carath\'eodory as suffering from
\emph{Erg\"anzungsbed\"urftigkeit} (need of completion).

It is worthwhile to dwell on the   exact nature of Planck's criticism.
  The first point  (viz.\ that \C's principle is also valid in a
time reverse of our world)
\forget{does not mean  that the principle does not
imply irreversibility.  This objection}
 is rather mild: both worlds could contain irreversible processes;  \C{}
only fails to provide a `Merkmal' to indicate the direction in which
they proceed.
 The second point is more serious:  a principle
that allows worlds in which only
 reversible processes occur
does not imply the existence of irreversible processes.
 This point is correct;  but it can also be raised  against
Kelvin's  or Clausius' formulations of the second law.

Planck then presents a new proof of the second law which, he
claims, shares the advantages of \C's approach (namely: that no
reference is made to the ideal gas or cyclic processes), but also
hinges essentially on irreversibility. In the eighth edition of
the {\em Vorlesungen\/} \cite{Planck26}, this proof replaces the
`careful' proof discussed in section~\ref{Planck}.

\subsection{Planck revisited\label{revisit}}

I will discuss Planck's new proof only briefly.
 The main difference with the argument discussed in section~\ref{Planck}
is that one does not start with  the ideal gas.
 Instead, the existence of a positive integrating divisor for the inexact
heat differential $\dstreep Q$ of a fluid is accepted unquestioningly.
  Thus, one writes immediately \[ \dstreep Q = TdS, \] where the entropy
$S$ and temperature $T$ are state variables of the fluid.

He then introduces the statement `friction is an irreversibel process',
which he considers as a formulation of Kelvin's principle. This view may
need some explanation, because, at first sight, this statement does
not seem to address cyclic processes or the \emph{perpetuum mobile} at
all. But
for Planck, the statement is equivalent to the proposition that there
exists no process which `undoes' the consequences of friction, i.e., a
process which produces no other effect than cooling a reservoir and doing
work.  The condition `no other effect' here allows for the operation of
any type of machinery that operates in a cycle.

He then considers an adiabatically isolated fluid which can exchange
energy with its environment by means of a weight.
 Planck asks whether it is possible to reach a state $s'$ of the system
 from a given initial state $s$, in a process which brings about no
changes in the environment other than the displacement of the weight.
 Let us represent this as
 \[ (s,Z,h)
\statechange{?}
(s',Z,{h'} ).\]
 He argues  that, by means of
`\emph{reversibel}-adiabatic'\footnote{%
    Apparently, Planck's pen  slipped here.
    He means: \emph{umkehrbar}-adiabatic.}
 transitions, starting from the state $s$, one can always reach a state
$s^*$ in which the volume   equals that of state
$s'$ and the entropy equals that of $s$.
 That is,  there is a change of state
 \[ (s,Z,h) \statechange{}
 (s^*,Z,{h^*} ),\]
with
 \[ V(s^*) = V(s')  \;\mbox{
and  }  \; S(s^*) =S(s).\]
 Whether the intended final state $s'$ can now be reached from the
intermediate state $s^*$ depends on the value of the only independent
variable in which $s^*$ and $s'$ differ.
For this variable one can either choose the entropy $S$, energy $U$ or
temperature $T$.

 There are three cases:\\
(1)  $ h^* = h'$. In this case,  energy conservation implies
$U(s^*)= U(s')$.
Because the coordinates
$U$ and $V$ determine the state completely,
$s^*$ and $s'$  must coincide.\\
(2)   $ h^*> h' $.  In this case, the state   $s'$  can be reached from
$s^*$  by letting the weight perform work on the system, e.g.\ by means of
friction,
until the weight has dropped to height $h'$. According to the above
formulation of Kelvin's principle,  this process is irreversible.\\
(3) $h^* < h'$. In this case the desired transition is impossible.
  It would be the
reversal of the irreversible process just mentioned in
  (2),  i.e.\ produce work by cooling the system and  would thus  realise a
\emph{perpetuum mobile}
 of the second
kind.\footnote{%
    Note how much Planck's application of the \emph{perpetuum mobile}
    differs from Carnot and Kelvin.
    The latter authors considered the \emph{engine}, i.e.\  the
device  which performs the cycle,
    as the system of interest  and the reservoir as part of
    the environment. By contrast, for Planck,  the
\emph{reservoir} is the thermodynamical
    system, and the engine  performing the cyclic process belongs to
    the environment.
    Related to this switch of perspective is the point that the
    reservoir is now assumed to have a finite energy content. Thus,
    the state of the
    reservoir can change under the action of the
    hypothetical \emph{perpetuum mobile} device.  As a consequence,
          the withdrawal of energy from the reservoir need not be
        repeatable. This is in contrast to Carnot's analysis (see
    section~\ref{Carnot}).  Indeed,  there is
       nothing `perpetual' about  Planck's present construal of the
    \emph{perpetuum mobile}.
}

Now, Planck argues that in all three cases, a transition from $s^*$ to
$s'$ is possible by means of heat exchange in an umkehrbar process in
which the volume remains fixed. For such a process one can write
  \[ dU = TdS. \]
 Using the assumption that $T>0$, it follows that, in the three cases
above, $U$ must vary in  the same sense as $S$. That is, the cases
$U(s^*)  < U(s') $, $U(s^*)  = U(s') $ or $U(s^*)  > U(s') $, can
also be characterised as
 $S(s^*)  < S(s'),
S(s^*)  = S(s')  $ and
$S(s^*)  > S(s')  $  respectively.

For a system consisting of several fluids the argument is analogous.
Planck argues that, here too, starting from a state $s$, a state $s^*$ can
be reached by means of quasi-static-adiabatic processes in which all
variables except one are equal to the values of the variables in state
$s'$, while the entropy has remained constant, etc.\footnote{%
      This is comparable to  \C{}'s assumption that from every
      initial state one can reach all values of the deformation
      coordinates by an adiabatic process.
}

Just as in earlier editions of his book, Planck generalises his
conclusions (without a shred of proof)  to arbitrary systems and
physical/chemical processes:
     \begin{quote}\small
 \selectlanguage{german}
 Jeder in the Natur
stattfindende Proze{\ss} verl\"auft in dem Sinne, da{\ss} die
Summe der Entropien aller an dem Proze{\ss} beteiligten K\"orper
vergr\"o{\ss}ert wird. Im Grenzfall, f\"ur einen reversibeln
Proze\ss, bleibt diese Summe unge\"andert. [\ldots] Damit ist der
Inhalt des zweiten Hauptsatzes der
Thermodynamik ersch\"opfend bezeichnet\ldots.\footnote{%
    `Every process occurring in nature proceeds in
    the sense in which the sum of the entropies of all bodies taking
    part in the process is increased. In the limiting case, for
    reversible processes this sum remains unchanged. [\ldots] This
    provides an exhaustive formulation of the content of the second
    law of  thermodynamics'}
\cite[p.~463]{Planck26}
 \end{quote}

 The argument just presented is Planck's definitive formulation of the
second law.  Although in some respects clearer and  simpler than the earlier
proof,  I do not believe it gives a substantial improvement.
 First, the assumption that for every fluid there always exist functions
$S$
and $T$
(with $T>0$) such that $\dstreep Q =TdS$  is
problematic. (Although
understandable, being a concession to
 Born). Secondly the generalization to arbitrary
 processes in arbitrary  systems  remains as  dubious as it was in the earlier
 versions.  There is nothing in Planck's argument that
indicates that the argument is valid beyond the simple systems
of Carath\'eodory.

I conclude that Planck has not succeeded in his attempt to show
that the theorem of \C{} is nothing but `an artificial and
unnecessary complication'. All he shows is that by adding Kelvin's
principle to that of \C{} (which is still necessary to guarantee
the relation
 $ \dstreep Q =T dS$), one obtains a time asymmetric statement that no
longer admits both entropy increases as well as decreases in adiabatically
isolated systems. Further, although it seems natural to understand the
proposition `friction is an irreversibel process' as intended to imply
that friction processes occur in our world, there is no need to assume the
actual existence of irreversibel processes in the argument.  Thus,
Planck's formulation also allows models in which all processes are
reversible, and does not repair this defect which he diagnosed in \C's
work.

\section{Lieb and Yngvason\label{LY}}
 It goes without saying that I cannot treat all the numerous
reformulations of the second law that have been attempted in the past 75
years.
 But this article would remain incomplete if I did not deal with a very
recent contribution by~\citeasnoun{LY}.  These authors provide a new
attempt to clarify the mathematical formulation and physical content of
the second law.  However, I cannot do justice to this important work in
the context of this article: the paper is 96 pages long and employs no
less than 15 axioms in order to obtain the second law. I note, however,
that these elaborate ramifications are partly due to the fact the authors
not only wish to obtain the second law in the form of an entropy principle
but also the result that entropy is an additive and extensive function and
(up to additive and multiplicative constants)  unique, and numerous other
results. Moreover, they wish to achieve most of these results without
assuming differentiability of the state space.

On the formal level, this work builds upon the approaches of
\citeasnoun{Cara09} and \citeasnoun{Giles}. (In its physical
interpretation, however, it is more closely related to Planck, as
we will see below.)  A system is represented by a state space
$\Gamma$ on which a relation $\prec$ of adiabatic accessibility is
defined. All axioms mentioned
 below are concerned with this relation.  Further, Lieb and Yngvason
introduce a formal operation of considering two systems in state $s$ and
$t$ as a composite system in state $(s,t)$, and the operation of `scaling'
a system, i.e.\ the construction of a copy in which all its extensive
quantities are multiplied by a positive factor $\alpha$. This is denoted
by a multiplication of the state with $\alpha$. These scaled states
$\alpha s$ belong to a scaled state space $\Gamma_{(\alpha)}$.

The main axioms of Lieb and Yngvason  apply to all states $s \in
\cup_{\alpha} \Gamma_{(\alpha)}$  (and compositions of such states). They
read:
\begin{quote}\sc
\item[\sc A1. Reflexivity:]     \be s \prec s  \ee
\item[\sc A2. Transitivity:]
\be  \label{trans}
s \prec t \;\mbox{~and~}    \; t \prec r \; \mbox{~imply~} \;
s\prec r \ee

\item[\sc A3. Consistency:]
\be    s \prec  s'\; \mbox{~and~} \:   t\prec t' \;  \mbox{~implies~}
 \; (s,t) \prec (s', t')   \label{combi}
 \ee
\item[\sc  A4. Scale invariance:]
\be  \mbox{  If  } \;  s\prec t \;\mbox{  then } \; \alpha s  \prec
\alpha t   \mbox{ for all  $\alpha >0$}
\ee
\item[\sc A5. Splitting and recombination:]
\be
\mbox{For  all }\;  0< \alpha  <1  : \;
 s \prec (\alpha s, (1 -\alpha )s) \;
\mbox{ and } \;
 (\alpha s, (1 -\alpha )s) \prec s \ee
\item[\sc A6. Stability:]
If
 there are  states $t_0$ and $t_1$ such that
 $(s, \epsilon t_0 ) \prec  (r
,\epsilon t_1)$  holds for  a sequence
of $\epsilon$'s   converging to zero, then   $s\prec r$.
\end{quote}

The meaning of these axioms is, hopefully, largely self-evident.
Axiom A1 and A2 demand that adiabatic accessibility is a
pre-ordering.
 Axiom A3
says that if subsystems of a composite system can each go through certain
 adiabatic changes of state,
 it is also possible to achieve these changes of states adiabatically in
the composite system. Axiom A4 expresses an analogous statement for
inflated or shrunken copies of the system.
 Axiom A5 says that separating and recombining subsystems are adiabatic
processes. One can think of the introduction or removal of a partition in
a fluid.  The stability axiom A6 expresses, roughly speaking, the
idea that if two
states $s$ and $r$ of a system are adiabatically  accessible whenever the
system is expanded by a negligibly small second system, e.g.\ a dust
particle, these states  themselves
 must also be adiabatically accessible.

The axioms above seem intuitively plausible and physically acceptable for
thermodynamical systems.
  This is not to say that one must see them as the expression of
empirical principles. Some  seem to follow
    almost immediately from the intended meaning
    of the relation, and have little empirical content;
    others seem very well
        capable of violation by arbitrary  physical  objects.
       (Consider the application of Axiom 5 to near-critical masses of
        plutonium.)
        It seems reasonable, however, to regard the axioms as an implicit
    definition of a `thermodynamical system'.
\forget{%
  In any case  the correctness of (largest part of) the above
 axioms implicitly of explicitly presupposed  in
almost all treatises over orthodox thermodynamics, including
the work of Planck.}

After stating  the above axioms, Lieb and Yngvason formulate the following
  \begin{quote} \small {\sc
7. Comparability hypothesis:}
\be \mbox{For all states $s,t$ in the
same space
$\Gamma$:}~~s\prec t \mbox{ or }\; t\prec s.~\footnote{%
    The clause `in the same space $\Gamma$'  means that  the
    hypothesis is not intended for the comparison of states of scaled
    systems. Thus, it is not demanded that we can either adiabatically
    transform a state of 1 mole of oxygen into one of 2 moles of
        oxygen or conversely.}
 \ee \end{quote}
    The comparability hypothesis has, as its name already indicates, a
lower status than the axioms.  It is intended as a characterization of a
particular type of thermodynamical systems,
namely, of `simple' systems
and systems composed of such
`simple' systems.\footnote{%
     Beware that the present meaning of the term does not
    coincide with that used by \C.  For simple systems
    in \C's sense the comparability hypothesis need not hold.
 }
 A substantial part of the effort by Lieb and Yngvason is devoted to an
attempt to derive this hypothesis from further axioms for these `simple'
systems and their compositions. I will, however, not go into this.

The aim of the work is to derive the following result, which  \LY{} call
 \begin{quote} {\sc The Entropy Principle  (Lieb and Yngvason version):}
There exists a function $S$
 \forget{$\cup_{\alpha} \Gamma_{(\alpha)}$}
defined on all states of all systems
 such that
\begin{itemize}
\item[a.] when $s$ and $t$ are comparable then
\be
\label{ep}
  s \prec t \mbox{ if and only if } S(s) \leq
S(t) \label{Hw2LY}.\ee
\item[b.] When  $s$  and $t$ are states of (possibly different) systems
\be S ((s,t)) = S(s) + S(t)   \label{add} , \ee
\be S (\alpha s) = \alpha S(s).   \label{ext} \ee
\end{itemize}\end{quote}

 The relations (\ref{add}) and (\ref{ext}) express that
the entropy function is additive and extensive.
 For our purpose, it is relation
(\ref{Hw2LY}) that is particularly relevant. The authors interpret
the result (\ref{ep}) as an expression of the  second law: `It
says that entropy must increase in an irreversible process.' and:
`the physical content of [(\ref{ep})] \ldots[is that]\ldots
adiabatic processes not only increase entropy but an increase in
entropy also dictates which adiabatic processes are possible
(between comparable states, of course).' \cite[p.~19,20]{LY}).

The question whether this result actually follows from their assumptions
is somewhat involved.  They show that a special case of the entropy
principle follows from the assumption of axioms A1--A6 and the
comparability hypothesis.  In particular, special conditions are needed
which (physically speaking)  express that mixing and chemical reactions
are to be excluded. To extend the principle beyond this restriction, an
additional ten axioms are needed (three of which serve to derive the
comparability hypothesis). And even then, only a weak form of the above
entropy principle is actually obtained,
 where `if and only if'  in (\ref{Hw2LY}) is replaced by `implies'.

 Before considering the interpretation of this result more closely, a few
general remarks on this approach are in order.  This approach
combines mathematical precision with clear and plausible axioms
and achieves a powerful and remarkable theorem.  This is true
progress in the formulation of the second law.  Of course it still
holds that the result applies only for special kinds of systems;
but this is also the case for \C's approach  and, when stripped
from rhetorical claims, also for Planck's.

 It is remarkable that the theorem is obtained without appealing to
anything remotely resembling \C's principle.  This is undoubtedly an
advantage for those who judge that principle too abstract. In fact the
axioms and hypothesis used above allow models which violate the principle
of \C~\cite[p.~91]{LY}.  For example, it may be that all states are
mutually accessible, in which case the entropy function $S$ is simply a
constant on $\Gamma$.

However, there is an additional axiom in \LY's approach which makes for a
closer connection with \C's principle.  One of the special axioms invoked
to derive the comparability hypothesis reads:
  \begin{quote} {\sc S1:  Irreversible
Processes:} for all $s\in \Gamma$ there is a $t \in \Gamma$ such that $s
\prec
t$ and $t\nprec s $.  \end{quote}
 Here the prefix `S' denotes that it is the first of a series of axioms
intended to hold for simple systems only. We shall have more to say about
what this axiom has to do with irreversibility below.
 For the moment, I only note that this axiom is the closest resemblance to
\C's principle to be found in this approach: it says that for each state
there is another adiabatically inaccessible state.
 In fact, the authors prove that, in conjunction with other axioms, it
implies what they call  `\C's principle'.\footnote{%
    Here,  \LY{} employ a formulation
    of \C's principle which deviates from both \C' own statement as
    well as from Born's version. It reads: $ \forall s \in \Gamma,
\forall
    U_s \; \exists t \in
    U_s $ such that  $s\nprec t$ or $t \nprec s$.}

However, the present axiom is much more liberal than \C's
principle. First, of course, it no longer demands that the states
which are inaccessible from $s$ occur in every local neighborhood
of $s$. Thus, this axiom evades Planck's objection that we don't
have empirical access to arbitrarily small neighborhoods.  More
importantly, this axiom is only intended to characterise `simple'
systems, and actually serves as (part of) a definition of this
notion.  This is in sharp contrast to \C's principle, which was
presented as a general law of nature. Thus, \LY{} also evade the
objection of Falk and Jung (see p.~\pageref{FJ}). Moreover, note
that this axiom is not essential to the proof of the entropy
principle, but only to the attempt to derive the comparability
hypothesis.  Anyone who accepts this hypothesis as physically
plausible will obtain the above entropy principle without having
to bother with \C's principle.

 For the purpose of this paper, the pertinent question is whether there is
a connection with the arrow of time in this formulation of the second law.
As before, there are two aspects to this question: irreversibility and
time asymmetry.  We have seen that \LY{} interpret the relation (\ref{ep})
as saying that entropy must increase in irreversible processes.  At first
sight, this interpretation (and also the name of the last-mentioned axiom)
is curious.
 We have found in the discussion of section~\ref{Car} that adiabatic
accessibility is not the same thing as irreversibility. How then, can the
present axioms on adiabatic accessibility be interpreted as having
implications for irreversible processes?

This puzzle is resolved when we consider the physical interpretation which
\LY{}
propose for the relation $\prec$:
 \begin{quote}\small\selectlanguage{english}
 {\sc Adiabatic accessibility:} A state $t$ is adiabatically accessible
from a state $s$, in symbols $s\prec t$, if it is possible to change the
state from $s$ to $t$ by means of an interaction with some device (which
may consist of mechanical and electric parts as well as auxiliary
thermodynamic systems) and a weight, in such a way that the auxiliary
system returns to its initial state at the end of the process whereas the
weight may have changed its position in a gravitational field'
\cite[p.~17]{LY}.  \end{quote}
 This view is rather different from
 \C{}'s, or indeed, from anybody else's:  clearly, this term is not
intended to refer to processes occurring in a thermos flask.  As
the authors explicitly emphasise, even processes in which the
system is \emph{heated} are adiabatic, in the present sense, when
this heat is generated by an electrical current from a dynamo
driven by descending weight. Actually, the condition that the
auxiliary systems return to their initial state in the present
concept is strongly reminiscent of Planck's concept of
`reversibel'!

      This is not to say, of course,  that they are identical.
         Let $Z$ be the state of the auxiliary
        system and $h$ the height of  the weight.
        For Planck, a process  $\cal P$ which produces the transition
         $\langle s, Z, h\rangle \statechange{P} \langle s', Z',
h'\rangle$
        is reversibel iff there exists a recovery process $\cal P'$ which
    produces
         $\langle s', Z' , h'\rangle \statechange{P'} \langle s,
Z,h\rangle$.
        Here, the states  $Z$ and $Z'$ are allowed to be  different
    from each other.
         For \LY{}, a process $\langle s , Z, h\rangle \statechange{P}
\langle
       s', Z', h' \rangle$  is called  adiabatic iff  $Z= Z'$.
        However, we have seen in section~\ref{Planck} that in his argument
        to obtain the entropy principle, Planck always restricted his
    discussions to such reversibel processes `which leave no changes
        in other bodies', i.e.\ that obey the additional requirement $ Z =
        Z'$. These reversibel processes are always adiabatic in the
    present  sense.
        \forget{In fact, the meaning attached by Lieb and Yngvason
        to the term `adiabatic' is identical to the meaning for
        the term reversible which we
        discussed (and rejected) in footnote~\ref{verworpen}.}
 A major difference with the conventional meaning of the term is that, in
the present sense, it automatically follows that if a process $\cal P$ as
above is  adiabatic, any recovery process $\cal P'$ is also
adiabatic.

Thus, we can now conclude immediately that if an adiabatic state
changes is accompanied by an entropy increase, this change of
state cannot be undone, i.e., it is irreversibel in Planck's
sense.
 This explains why the result (\ref{ep}) can be seen as a formulation of
a principle of entropy increase, and why axiom S1 is interpreted as
stating the existence of irreversible processes.
 In fact, we can reason as follows:  assume $s$ and $t$ are states which
are mutually comparable, and that $S(s) <S(t)$. According to (\ref{ep}),
we then have $s\prec t$ and $t \nprec s$.
 This means that there exists no process from $t$ to $s$ which proceeds
without producing any change in auxiliary systems except, possibly, a
displacement of a single weight.  At the same time there exists a process
from $ s$ to $t$ (under the same condition). This process is irreversibel
in Planck's sense.\footnote{%
    This conclusion, obviously, is crucially
    dependent on the non-standard meaning given to the term
    `adiabatic'. It is somewhat surprising, therefore, that in the
          published version (\LY, 1999) of the  manuscript (1997), a
passage is
      included in
        which the authors argue that their interpretation coincides with
        the conventional meaning of this term after all.
        An example may show that this claim is misleading.
         Consider a compound system consisting of two simple systems, each
    with a one-dimensional state space.
        Assume that these two systems are adiabatically
        isolated from each other.  For example: take
        two quantities of an incompressible fluid contained in
    calorimeters (fitted with a stirring device). In the
    conventional sense, the only processes which can be called adiabatic,
    are  (i) stirring  and (ii) heat exchange  among the two systems  by a
    temporary diathermal connection.  Under this interpretation
    the compound system does not obey the comparability hypothesis.
       For example, if the temperatures of $s$ and $t$ differ, then
     $(s, t \nprec (t, s)$ and $(t,s)  \nprec (s,t)$;
       cf.~\cite[p.~38]{Boyling}.

      However, in the interpretation of Lieb and Yngvason, adiabatic
       accessibility depends on which other systems are available as
    auxiliary devices.
      Suppose there is another system (say an ideal gas) capable of
      performing a Carnot cycle.  By means of this system, operated as a
     heat pump, we can transfer entropy from one subsystem to the other,
     and thus increase entropy of the one at the expense of the other.
       This process would be adiabatic by
    \LY's criterion.  (The auxiliary system goes through a cycle and
    hence returns to its original state;  the work needed to drive the
    heat pump can be provided by lowering a weight.)  Thus we have $(s, t)
    \prec (t, s)$ and $(t,s)  \prec (s,t)$.

    In fact, in this example of a world in which one-dimensional
    systems and normal fluids coexist the conventional definition of
    `adiabatic' does not obey the axioms A1--A6. The argument given by
    Lieb and Yngvason in order to conclude that the two interpretations
    coincide, which assumes the validity of these axioms, is therefore not
    applicable. (I acknowledge clarifying personal communications with Jakob
    Yngvason on this point.)}
 Thus we have at last achieved a conclusion implying the existence of
irreversibel processes by means of a satisfactory argument!

However, it must be noted that this conclusion is obtained only for
 systems obeying the comparability hypothesis and under the exclusion of
mixing and chemical processes.  The weak version of the entropy principle,
which is derived when we drop the latter restriction, does not justify
this conclusion.
 Moreover, note that it would be incorrect to construe (\ref{ep})  as a
characterisation of \emph{processes}. The relation $\prec$ is
interpreted in terms of the \emph{possibility} of processes.  As
remarked in section~\ref{Car}, one and the same change of state
can very well be obtained (or undone) by means of different
processes, some of which are adiabatic and others not.
 Thus, when $S(s) < S(t)$ for comparable states, this does not mean that
\emph{all} processes from $s$ to $t$ are irreversibel, but only
that there exists an adiabatic irreversibel process between these
states. So the entropy principle here is not the universal
proposition of Planck, even if we restrict ourselves to systems
for which the comparability hypothesis holds and exclude mixing
and chemical processes.

The next question is of course about the time-(a)symmetry of this
approach. There are two ways in which one may analyse this
question.  The first is to consider all structures $\langle
\cup_{\alpha} \Gamma_{(\alpha)},  (\cdot, \cdot), \prec \rangle$
as candidate models, and look upon the axioms as singling out a
class of possible worlds.
 In that case it is easy to show, using the implementation of time reversal
proposed in footnote~\ref{R}, that the six general axioms, as well as the
comparability hypothesis, are completely time-symmetric!%
\footnote{%
     This conclusion cannot be extended to the complete set
of axioms proposed by
Lieb and Yngvason. For example,  axiom S1 (cited above)
        is  already time-asymmetric. However, the
        time-asymmetry introduced  by this axiom is only temporary.
        In the course of their presentation, axiom S1 is subsumed by
    a stronger axiom
        (called `Transversality') which restores time-symmetry.
    (Transversality  entails that for all $s$
      there is also a state $r$ such  that $r\prec s$ and $s\nprec r$.)
    Yet, there  are two other axioms  (called A7 and
T1)  which address mixing and equilibration  processes.
These axioms are explicitly time-asymmetric.  (I thank Jakob Yngvason for
pointing
this out to me.)
  Note, however, that as far as the
 entropy principle is concerned,
these axioms are  needed only in the derivation of the (time-symmetric)
comparability
hypothesis.
}

Another way of analyzing the question is to start from the interpretation
proposed by the authors for the relation $\prec$ and note that it invokes
the term `possible'.  One may regard this as a modal relationship,
to be understood in terms of a `possible worlds' semantics. On this
reading, the statement `$s\prec t$' does not express a manifest property
of one single world, but rather commits one to the existence of possible
worlds in which the state $s$ can be transformed into $t$ without leaving
changes in auxiliary systems except the displacement of a weight.  It does
not, however, commit us to the existence of a possible world in which $t$
is transformed into $s$ under the same conditions.  Thus, the class of
possible worlds allowed by such a statement is time-asymmetric.\footnote{
    In this view, the role of the axioms would then be to characterise
    a kind of second-order possibility, namely, to determine which
    relations between possible worlds are possible (allowed by the
    theory).}

Therefore, the answer to the question whether this approach is
time-symmetric or not depends on whether one analyzes the question from
the point of view of the formalism or its interpretation.  Nevertheless,
the fact that it is not necessary to introduce time-asymmetry into the
formalism to obtain the second law, is very remarkable.

As I have said, the interpretation these authors give to the term
`adiabatic' is much wider than that of \C{}.
 For the mathematical formalism, this is of course irrelevant; but not for
its physical meaning.  The wider the interpretation of the
relation $\prec$, the stronger is the empirical content of the
postulates. This raises the question whether the proposed
interpretation is not, as we saw in the case of \TEA, perhaps
already so wide that the axioms conflict with experience. As far
as I can see, this is not the case.  Of course, the main point
responsible for this difference from \C's approach is that the
present axioms are, in certain aspects,  much weaker.

However,  this question leads immediately to one problematical aspect of
the proposed physical interpretation.
 It refers to the state of auxiliary systems in the environment of the
system. Thus, we are again confronted by the old and ugly
question, when shall we say that the state of systems in the
environment has changed, and when  are we  fully satisfied that
their initial state has been recovered.  As noted before
(footnote~\ref{tr}),  this question is rather intractable from the
point of view of thermodynamics, when one allows arbitrary
auxiliary systems whose states are not represented by the
thermodynamical formalism.  Thus, the question when the relation
$\prec$ is applicable cannot be decided on the basis of the
formalism itself.

\forget{
 Consider e.g.\ the axioms for  transitivity
(\ref{trans})
and   combination
(\ref{combi}).
 The validity of these is not so self-evident as it seems.  In general,
the simultaneous performance of two adiabatic processes in the premises of
(\ref{trans}) and (\ref{combi})  will yield a process have in which
\emph{two} weights are displaced.

For the validity of the axioms it seems necessary to assume that all
employed weights can then exchange energy such that eventually only one
has been displaced.
  Thus, if $(U_1, \ldots U_n)$ and $(U_1' ,\ldots U'_n)$ denote the energy
values of two states of a system consisting of $n$ weights, then it should
be possible to transform the former, without leaving any other changes in
auxiliary systems into $(U'_1, \ldots U'_{n-1}, U'_n\ + \Delta )$ where
$\Delta = \sum_{i=1}^{n} U_i -\sum_{i=1}^{n} U'_i $.
 This implies that changes of state of mechanical auxiliary systems must
be reversible without exception.  It is clear that this question is not
\emph{not} entailed by, and therefore also not decided by, the
time-symmetry of the mechanical equations of motion. A general theorem
guaranteeing the reversibility of mechanics in this sense is unknown to
me.
}

 \section{Summary and moral \label{moral}}
 What is the relation between the second law of thermodynamics and the
arrow of time? The deeper we go into this question, the more
remote a clear-cut relation appears to be. Nevertheless, I think
we can summarise this study by drawing several conclusions.
Moreover, I argue below that it may be more fruitful to abandon
the idea that time-asymmetry or irreversibility is essential to
the second law.

First of all, we have seen that a distinction should be made between
time-(a)symmetry and thermodynamical concepts of `(ir)reversibility'.
Time-asymmetry, in the sense in which we used this word, refers to a law
which allows some process (or possible world), while excluding its time
reversal.  In the stock philosophical literature, such processes are
called irreversible. But in thermodynamics a plethora of other meanings
are employed for this term.

The two most important of these are as follows. First, one can
understand `reversible processes' as processes which proceed so
slowly that the system always remains close to equilibrium.
Elaborating on conditions employed by Carnot and Kelvin (1851),
Clausius (1864)  and Planck (1897) defined the term
\emph{umkehrbar} in this sense. This concept is of crucial
importance to their formulations of the second law. In the physics
literature it is probably the most common usage of the term
`reversible', in spite of \C's proposal to use the better term
`quasi-static'. However, this concept is by itself irrelevant for
the arrow of time. That is to say, the claim that there exist
processes which are not reversible in this sense, or indeed, the
claim that some law implies that all processes in nature are
irreversible in this sense, does not imply time-asymmetry.

The second meaning of `reversible' is the notion of a process whose
initial state can be completely restored by some other process, using any
auxiliary device available in our world. This is Planck's notion of
`reversibel', which goes back to Kelvin (1852). This concept \emph{is}
relevant to the arrow of time, although it is not identical with the
notion of time-asymmetry.  Discussions on irreversibility and the second
law in the philosophy of physics seem to have largely overlooked this
distinction.

A second conclusion is that different presentations of classical
thermodynamics vary a great deal, both in their formulations of
the second law, and in their relationships with the arrow of time.
The main division here is between the work of Clausius, Kelvin and
Planck on the hand, and Gibbs and \C{} on the other. However, also
inside each of these `camps' there are significant distinctions.
 Perhaps the  table below  is  helpful:\\
\begin{table}[h]
 \begin{center}
\begin{tabular}{|l|c|c|c|c|c|}
\hline
\parbox{1.8cm}{
 Version of second law}
   &
\parbox{1.2cm}{Applies only to cycles?} &
\parbox{1.8cm}{Time-asym\-metric?}
 &
\parbox{1.6cm}{
 Allows irreversible\\ processes?}
& \parbox{1.7cm}{\mbox{} \vspace{-.4ex}\\
Implies  existence of  irreversible processes?\\[-0.8ex]} &
\parbox{1.9cm}{
Argues for universal
ir\-reversibility?}\\
\hline
Carnot's theorem   & yes   &  yes  &  yes   & no   &   no  \\
Clausius (1850)    & yes   &  no   &  yes   & no   &   no  \\
Kelvin (1851)      & yes   &  no   &  yes   & no   &   no  \\
Kelvin (1852)      & no    &  yes  &  yes   & yes  &   yes \\
Kelvin (1855)      & yes   &  no   &  yes   & no   &   no \\
Clausius  (1865)   & no    &  yes  &  yes   & yes  &   yes  \\
Clausius  (1876)   & yes   & yes   &  yes   &  no  &  no  \\
Planck  (1897)     & no    & yes   &  yes   &  yes &  yes \\
Gibbs    (1875)    & n.a.  & n.a.  &   yes  &  no  &   no    \\
\C  (1909)        &  no   &  yes  &    yes &  no  &   no\\
{Lieb \& Yngvason} (1999)
                   & no    & no    &  yes   & no  &   no
 \\
 \hline
\end{tabular}
\end{center}
\caption{\small \em Various aspects of the arrow of time for various
formulations of
the second law. Here, `irreversible' is taken in  Planck's sense, and
`n.a.' stands for `not applicable'.}
\end{table}

  In the tradition of Clausius, Kelvin and Planck, thermodynamics is a
theory about \emph{processes}.  That is to say, one considers the
evolution of a system in the course of time.
  To be sure, the role of evolutions is tiny (as the theory is mainly
restricted to cyclic processes), but the question whether the theory is
time-asymmetric or implies irreversibility (irrecoverability)  makes
sense.

 The answer depends on whether the second law is regarded as a statement
concerning cyclic processes only, or also about open
(i.e.~non-cyclic) processes. We have seen that the formulations
Clausius and Kelvin gave of the second law went through various
changes in this respect.  However, most of their formulations of
the second law only concerned cycles.  This is reflected in the
formulation known today as Kelvin's and Clausius' principle, or
the principle of the impossibility of the \emph{perpetuum mobile}
of the second kind. This formulation is time-asymmetric. The
`negative' character of these principles is no obstacle;
time-asymmetry is a characterization of a theory, and not of our
world.
 But, I have argued, they  do  not imply the existence  of
irreversible (irrecoverable) processes.

However, a few exceptional publications, notably Kelvin (1852) and
Clausius (1864, 1865), argued for a universal tendency of
processes to proceed in one direction only.  This view led, in
particular in the work of Planck, to a grand universal
generalization, according to which the second law  says that for
all processes in nature the total entropy of the systems involved
never decreases, and that therefore all processes (with the
exception of those in which the entropy remains constant) are
irreversible.  A convincing argument for this claim has never been
given.

 The versions of the second law developed by Gibbs, \C{} and Lieb and
Yngvason apply to \emph{equilibrium states}.  Here, the evolution of
systems in the course of time plays no role, and a connection with any
aspect of the philosophy of time is therefore much less prominent.

As to  Gibbs' version, i.e.\ the entropy maximum principle, I have
argued that such a connection is completely absent;  it rests, in
my opinion, on a confusion between virtual variations and
processes.
 In the case of \C, and authors that elaborated his approach, the
situation is more subtle. His formulation of the second law (\C's
principle) is a proposition intended to provide sufficient grounds for the
existence of entropy and temperature as functions of equilibrium states,
at least for simple systems.  There is no direct concern for
time-(a)symmetry or irreversibility here.

However, \C's principle employs a notion of adiabatic
accessibility between states, $s\prec t$, which is interpreted in
terms of the possibility of an adiabatic process which transforms
$s$ into $t$.  Here, time enters the picture, because the time
reversal of such a process obviously produces an adiabatic process
from $t$ to $s$.  I have argued that, if we construe the time
reversal of any model of \C's principle as one in which $\prec$ is
replaced by $\succ$, this theory is, strictly speaking,
time-asymmetric. However, this asymmetry is only noticeable in
rather pathological models. If the theory is applied to usual
systems (like ordinary fluids or systems composed of such fluids),
then models in which the second law according to the CKP tradition
holds, as well as the time reversals of these models, are allowed.
Thus, for this class no time-asymmetry emerges. The modern
extension of the formalism of \C{} by Lieb and Yngvason, is even
manifestly time-symmetric.

The connection with the entropy principle and irreversibility is
even more subtle. While \C{} gives a discussion aiming at the
conclusion that for all simple systems, adiabatic processes in
which entropy varies are irreversible, he only obtains this
conclusion by a redefinition of `irreversibility':  a process is
called irreversible if the change of state cannot be undone
\emph{adiabatically}.
 This result is obviously far removed from
Planck's universal entropy principle.  Even if we restrict
ourselves to simple systems, it is not guaranteed that entropy
increases in irreversible processes (in either Planck's or \C's
sense) nor does it follow that processes in which entropy
increases should be irreversible (in Planck's sense).

Also in the approach of Lieb and Yngvason, an entropy principle is
obtained that holds for a restricted class of `simple' systems (and
systems composed  of these). However, their notion of simplicity  does not
coincide with that of \C. In order to reach a
statement about irreversibility, this approach chooses the opposite
strategy: these authors redefine the concept of `adiabatic' in such a way
that it (almost)  coincides with Planck's concept of reversibility.  The
result is that for every two comparable states $s$ and $t$ with $S(s)<
S(t)$ there exists an irreversible process, beginning in $s$, and ending
in $t$.

It is striking that this version of the second law can be obtained without
invoking time-asymmetry at all.  However, the result does have
consequences in terms of irreversibility (in the sense of recoverability).
 But this consequence is rather mild:  it does not follow that \emph{all}
such processes from $s$ to $t$ are irreversible. Here too,
 the universal formulation of Planck has not been attained. One can even
ask whether the result is so interesting for the philosophy of time, or
threatening for the harmony between different parts of physics.  After
all, Hamiltonian mechanics also allows the existence of irreversible
processes, for example, the motion of a free particle in an otherwise
empty universe.

This summary leads to the question whether it is fruitful to see
irreversibility or time-asymmetry as the essence of the second law. Is it
not more straightforward, in view of the unargued statements of Kelvin,
the bold claims of Clausius and the strained attempts of Planck, to give
up this idea?
 I believe that Ehrenfest-Afanassjewa was right in her verdict that the
discussion about the arrow of time as expressed in the second law of the
thermodynamics is actually a \emph{red herring}.

The only way to evaluate such a proposal is by making up a balance-sheet.
What would we loose and what would we gain?
 It is clear that in fact all concrete applications of the second law in
classical thermodynamics, even in the work of the most outspoken
proponents of the claim that this law implies universal
irreversibility, are restricted to systems in equilibrium.  This
holds for Kelvin and Planck, but also more recent text books
(e.g.\ \cite{Becker}). A general opinion among thermodynamicists
is even that the theory is incapable of dealing with systems out
of equilibrium; (see the quotation from Bridgman on page
\pageref{Br}). Clearly, in terms of concrete applications, we
would loose very little.  What, then, do we gain with this
proposal? The main advantage is, to my mind, that the second law
would no longer represent an obstacle to the reconciliation of
different theories of physics. More specifically, attempts to
reduce thermodynamics to, or at least to harmonise it with, a
mechanistic world picture would get a new lease of life.

 The work of Boltzmann in kinetic gas theory is a particularly good
example of this problem. Boltzmann spent the main part of his career
 trying to find a mechanical underpinning of the second law. Essential for
this task was, in his opinion, finding a mechanically defined
function for isolated mechanical systems, which could exclusively,
or at least with very large probability, increase.  Every time he
believed himself to have succeeded in this task, e.g.\ in 1872
with the $H$-theorem for a dilute hard spheres gas, and in 1877
with his combinatorial argument for the ideal gas, objections to
his results emerged (viz.\ the famous \emph{Umkehreinwand} and
\emph{Wiederkehreinwand}). The problem of avoiding these
objections is still open.

But apparently there is another option.  If the second law does not
express time-asymmetry or irreversibility, it is not necessary to find a
mechanical quantity which can only increase and still achieve
reconciliation between thermodynamics and mechanics.

Among philosophers of science, the themes around the second law
have drawn a lot of attention;  (Reichenbach, Gr\"unbaum, etc.).
Sometimes, this discussion has taken a flight which seems far
removed from the original physical background. There are serious
discussions about the entropy of a footprint on the beach, or
about the question whether the second law can perhaps explain the
flow of time itself.  It seems to me that thess discussions can
only be understood if we construe terms like `entropy', `second
law' or even `thermodynamics' as metaphors that do not literally
refer to a actually existing physical theory. According to the
proposal such discussion can be avoided, or at least sharpened.

With this proposal I do not wish to suggest that there is no connection
between thermodynamics and the arrow of time. Therefore I conclude this
study by mentioning two areas in which the connection might be
analyzed with more success.

In the first place, a fundamental presupposition in classical
\TD{} is that isolated systems attain or approach an equilibrium
state, and, once they reach equilibrium, they remain there as long
as they are left to themselves.
 In fact, equilibrium is often defined as a state which \emph{will} not
change in the future, if the system is left to itself.
  Changes in the past, in contrast, are allowed or even explicitly
presupposed.  This gives a clear time-asymmetric character to
thermodynamics.

It is often said that this behaviour of thermodynamical systems (i.e.\ the
approach to equilibrium)  is accompanied by an increase of entropy, and a
consequence of the second law.
 But this idea actually lacks a theoretical foundation: for a
non-equilibrium state
there is  in general no thermodynamic entropy --or temperature-- at
all.  We get no further than where Clausius was in 1864 (see
page~\pageref{sys}): the second law cannot be seen as a statement about
the quantities of the system, but also involves its environment.
 \citeasnoun[ \S~112]{Planck} too emphasised that the approach to
equilibrium has nothing to do with the second law.  This aspect of
time-asymmetry is woven much deeper in the theory.

Although Boltzmann's $H$-theorem is not necessary to reconcile the
second law with mechanics, it can still be important to obtain a proper
description of the approach to equilibrium.
\citeasnoun{UF}
therefore associate the work of Boltzmann with a foundation of  what
they call the {\em
zeroth\/} law.\footnote{%
     This means: every isolated thermodynamical system
     reaches an equilibrium state in the long run.
     This terminology is  unfortunate since
     the term `zeroth law' is normally used to denote transitivity of
     thermal equilibrium.
     Perhaps it is better to speak of ` law $-1$' or even $`-\infty$'.}
 But clearly, for a mechanical explanation of the approach to equilibrium
it is not necessary to prove the monotonous increase of some mechanically
defined quantity.

There is another interesting remark to be made in this connection.
  There exists, apart from the works of Carnot, Clausius,
Kelvin, Gibbs and Planck, another classical tradition in the study
of heat phenomena, e.g.\ the heat equation of Fourier,
\forget{:\[
\frac{\partial u(x,t)}{\partial t} = \frac{\partial^2
u(x,t)}{\partial x^2} \].}
 This equation shows all the aspects one would like to associate with an
`arrow of time': it contains  time explicitly; the class of
solutions is not invariant when we replace $t$ by $-t$; they show
a clear unidirectional tendency to equalise temperature
differences, etc.  Similar remarks hold for the diffusion equation
of Fick, and other equations describing the macroscopic flow of
heat and matter (often collectively called `transport equations').

 Furthermore, transport equations form a \emph{bona fide} part of
classical physics. The question is then: what is the relation of this
tradition to thermodynamics?  The answer is rather surprising.
 Truesdell (1980)\nocite{Truesdell} observed that in one and a half
centuries of their coexistence, not a single work has appeared in
which the behaviour of heat, as described by the heat equation,
and as described by thermodynamics, are related to each
other.\footnote{However, \citeasnoun{Bertrand87} is an exception.}
One has to conclude that the heat equation and other transport
equations simply do not belong to classical thermodynamics!

However, since the Second World War, a lot of work has been done in
obtaining extensions of thermodynamics which could be applied to systems
out of equilibrium.  Such extensions, sometimes called `thermodynamics of
irreversible processes', would be able to describe the approach to
equilibrium, as illustrated by the heat equation;  see, e.g., de Groot
(1945); Prigogine (1955).\nocite{Groot,Prigogine55}
 Here, a more interesting connection
 with the arrow of time could result.  This work seems to have resulted in
a large number schools, and I can therefore say little about it.
It is  characteristic of this type of work that it is focussed on
applications and gives comparatively little attention to the foundations
and logical formulation of the theory.
 Usually, a time-asymmetric statement about entropy production is
postulated. The question how the entropy of a non-equilibrium
state is to be defined,  and the proof that it exists and is
unique for all non-equilibrium states, still seem to be largely
unexplored;\footnote{%
       A typical argument
      (de Groot, 1945; Yourgrau et al., 1966\nocite{Groot,Yourgrau})
    is that, on the one hand, orthodox
      thermodynamics is rejected on the grounds that
      it is not valid for irreversible processes,
      and on the other hand one  justifies
      the statement about positive entropy production
      for irreversible processes with
        an appeal to the (orthodox) second law.
        Another curious characterization
       of the foundations of this theory is
       by \citeasnoun{Callen}: `Irreversible thermodynamics is based on
       the postulates of equilibrium thermostatics
        plus the additional postulate
        of \emph{time reversal symmetry}'; }
    see also~\cite{Meixner,Lavenda}.

\section*{ Acknowledgements}
I have benefited a great deal from discussions with and comments
by Co Broeder, Craig Callender, Elliott Lieb, Janneke van Lith,
John Norton, Henk de Regt, Jakob Yngvason and two anonymous
referees. Thanks, all of you.

\end{document}